\documentclass[aps,showpacs,twocolumn,superscriptaddress,prb,floatfix,10pt,longbibliography]{revtex4-2}
\usepackage{graphicx}
\usepackage{subfigure}
\usepackage{amssymb,amsmath,cancel}
\usepackage{comment}
\usepackage{xcolor}
\usepackage{array}

\begin{document}

\title{Pseudogap phases, chiral anomaly and topological order with quantum loop entanglement}

\author{Predrag Nikoli\'c}
\affiliation{Department of Physics and Astronomy,\\George Mason University, Fairfax, VA 22030, USA}
\affiliation{Institute for Quantum Matter at Johns Hopkins University, Baltimore, MD 21218, USA}

\date{\today}

%
%
%

\begin{abstract}

A many-body quantum system whose topological defects are conserved, abundant and mobile is a correlated quantum liquid. Since topological defects can be classified by homotopy groups, each homotopy identifies a class of quantum liquids. Here we explore the quantum liquids based on the $\pi_3(S^2)$ homotopy group, i.e. Hopf fibration. Their topologically non-trivial dynamics emerges from the interlinking between magnetic flux or skyrmion loops in the charge and spin sectors respectively. We lay down a field theory foundation for analyzing such states by naturally incorporating the well-known framing regularization into the theory, and constructing the appropriate topological Lagrangian terms. We show that at least two strongly correlated phases of interlinked loops can exist in $d=3$ spatial dimensions at zero and low finite temperatures. These phases are closely related to the chiral quantum anomaly and do not have an obvious topological order, but they are distinguished from the trivial disordered phase by a generalization of the Wilson loop operator. In $d=4$ spatial dimensions, interlinked loops are able to produce topological order at zero temperature, featuring charge, angular momentum and braiding fractionalization. We discuss some possible experimental signatures of loop entanglement in the quantum noise of charge currents.

\end{abstract}

\maketitle

\section{Introduction}

Topological defects profoundly influence the dynamics because they are resilient to perturbations. The most intricate example is topological order, a non-local many-body quantum entanglement which produces a ground state degeneracy on topological manifolds without spontaneous symmetry breaking \cite{Wen1990a, Wen1990b, Wen1995a, WenQFT2004}. Topological order based upon charge current vortices is realized in two-dimensional fractional quantum Hall liquids, and exhibits charge and exchange statistics fractionalization. Other kinds of topological defects from charge or spin degrees of freedom can support topological orders in higher dimensions \cite{Nikolic2019}. A more subtle aspect of topological dynamics concerns the quantum fluctuations of a ``classical'' topological invariant on topologically trivial manifolds \cite{Nikolic2023a}. No smooth transformations of fields can alter the topological invariant formulated in the continuum limit, but quantum tunneling events, called instantons, can. Instantons may be strongly suppressed. The topological protection that defects enjoy normally amounts to a linear confining action potential between an instanton and an anti-instanton. This neutralizes the instantons and leaves only the topological index preserving fluctuations to shape the dynamics at large scales. However, even in that case, sufficiently strong thermal fluctuations can deconfine the instantons and cause a phase transition to a phase without any topological protection mechanism \cite{Nikolic2023a}.

In this paper, we study the topological dynamics associated with ``hopfions'', the topological defects classified by the $\pi_3(S^2)$ homotopy group \cite{Faddeev1975, Faddeev1976, Faddeev1997}. The simplest realization of such defects are interlinked loops and knots in three spatial dimensions \cite{StoneGoldbart2009}, but their mathematical structure and quantum fluctuations introduce various complications. The loops can be built with either charge or spin currents. The former are vortices and the latter are skyrmions. Both are theoretically described as interlinked quantized flux loops of a U(1) gauge field. One of our main results is that hopfions can exist in two entangled loop phases with a conserved Hopf topological invariant. The phase transition to a disentangled state occurs at a finite temperature \cite{Nikolic2023a}. There is no requirement that latent heat, spontaneous symmetry breaking or topological order be associated with such a transition; instead, the loop disentanglement is similar to the Berezinsky-Kosterlitz-Thouless transition \cite{Berezinsky1972, Kosterlitz1973, Kosterlitz1974}, but occurring in three spatial dimensions.

One strategy for the detection of loop disentanglement, which we explore here, is based on the quantum noise spectrum of electron currents. We will show that the chiral quantum anomaly \cite{Adler1969, Bell1969, Fujikawa1980, Nielsen1981, Nielsen1983, Peskin1995, ZinnJustin2001} is directly related to the Hopf index, so correlated Dirac and Weyl semimetals might provide a measurement platform. But, the correlation between the Hopf index and current fluctuations is more general. We expect that this correlation, and the transition itself, would be much more prominent with emergent U(1) gauge fields in solids instead of the natural gauge field. Vortex fluctuations in superconductors are one possible source of such a gauge field. There are experimental indications via Nernst effect \cite{Ong2001, Wang2006, Li2010} that vortices are present even in the pseudogap normal phase of cuprates, so they could live in an entangled phase. Related indications of incoherent Cooper pairing are found in other studies \cite{Corson1999, Armitage2007}. Another type of an emergent U(1) gauge field can arise from the magnetism in heavy-fermion materials or chiral magnets, especially quantum-disordered magnets \cite{Machida2007, Machida2010, Armitage2017b}. If local magnetic moments are thermally or quantum-mechanically disordered but retain fluctuating skyrmion loops, the emergent gauge field, which one would associate with the ``topological'' Hall effect, can carry a Hopf index and undergo the disentanglement topological phase transition at a certain temperature.

Our main goal is to develop a physical insight and theoretical formalism for exploring these possibilities. A particular challenge is that multiple types of topological defects can drive phase transitions, in addition to many conventional mechanisms. Monopole fluctuations in $d\ge 2$ dimensional compact U(1) gauge theories are understood fairly well \cite{Polyakov1975, Polyakov1977, Fisher1989a, Fisher1990, Nagaosa1993, Nagaosa2000, SubirQPT, herbut03, Herbut2004, Hermele2004, Nogueira2008}, but hopfions have been observed and considered so far only as topological defects without significant dynamics \cite{Babaev2002, Padgett2010, Mottonen2016, Sutcliffe2017, Zang2018, Smalyukh2018, Sutcliffe2018, Rybakov2019, Kiselev2019b, Voinescu2020, Fischer2021, Sugic2021, Khodzhaev2022, Burkowski2023}, or as a background Berry curvature for Hopf band insulators \cite{Moore2008, Duan2013, Duan2014, Duan2016, Hasan2017, Volovik2017, Schuster2019, Slager2019, Soluyanov2021, Trifunovic2021, Piechon2022, Goswami2023}. We emphasize here that hopfions in $d=3$ and $d=4$ spatial dimensions can drive independent topological phase transitions across which the flux loops survive (i.e. monopoles are suppressed) but lose a well-defined structure of interlinks. This requires at least some dedicated length scale in the problem other than the lattice constant. The lack of such a scale keeps the phase diagram simple and without any unconventional entangled states, as in the case of plain bosonic lattice models \cite{Fisher1990, SubirQPT}. If a quantum liquid has a conserved topological invariant, it can exhibit topological order \cite{Nikolic2019}. However, this is not implied. We will argue that the conserved Hopf index induces topological order in $d=4$ spatial dimensions, generalizable to $d=4n$ ($n\in\mathbb{N}$). In contrast, Hopf index conservation stabilizes a pseudogap phase in $d=3$, which is perhaps related to correlated symmetry-protected topological (SPT) phases \cite{Chen2011, Chen2013} by the likely ability to host gapless states on the system boundary.

The only topological defects of SU(2) spin configurations in $d=3$ spatial dimensions are hedgehogs and hopfions. They are classified by $\pi_2(S^2)$ and $\pi_3(S^2)$ homotopy groups respectively. Hedgehogs are equipped with a point singularity and enjoy topological protection against quantum fluctuations \cite{Nikolic2019, Nikolic2019b, Nikolic2023a}, while hopfions are non-singular configurations similar to $d=2$ skyrmions. Since electrons carry both charge and spin, their charge and spin currents are closely intertwined. If electron's spin current exhibits a topological structure, its charge current will reflect the same structure unless spin and charge are deconfined. The charge analogue to a hedgehog is a monopole. The binding of monopoles to hedgehogs is a ``topological'' magnetoelectric effect, and the equivalent phenomenon in two spatial dimensions, vortex-skyrmion binding, is a ``topological'' Hall effect. There are also charge current configurations analogous to hopfions. They are usually classified by the second Chern number of the U(1) bundle \cite{Peskin1995, StoneGoldbart2009, ZinnJustin2001}, and their association with the Hopf index of the $\pi_3(S^2)$ homotopy group is rarely mentioned \cite{Bandyopadhyay2012}. But, here we wish to explore the topological connections between the charge and spin dynamics, so we will adopt the classification of spin defects for the charge sector. 

The rest of the paper is organized as follows. Section \ref{secLoops} introduces the Hopf invariant and explains its quantization and topological protection in physical terms. We first review the construction of the Hopf index from gauge fields in Section \ref{secHopf}, and then connect it to the Gauss' linking number in Section \ref{secLinkNum} where it becomes apparent that matter must be coupled to the gauge field in order to resolve the framing regularization problem. Section \ref{secStimulate} considers generic Lagrangian density terms capable of stabilizing correlated quantum liquds with a conserved Hopf index. In Section \ref{sec3D} we study the quantum liquids of Hopf topological defects in $d=3$ spatial dimensions, which can be potentially realized in some correlated materials. We demonstrate in Section \ref{secHopfConservation} the conservation of the Hopf index despite the natural quantum processes which create, annihilate and link or unlink the flux loops. A further analysis of instanton events with renormalization group, presented elsewhere \cite{Nikolic2023a}, establishes the topological stability of a correlated entangled-loop phase at low temperatures. We in fact identify two Hopf indexes for every gauge field coupled to matter, and thereby two corresponding loop disentanglement transitions. In Section \ref{secFieldEq3D}, we formulate and explore the field equations for a prototype system that can host hopfions. This reveals the natural coupling between the charge and spin dynamics from which the ``topological'' Hall and magnetoelectric effects, as well as the chiral magnetic interaction, emerge. We also seek manifestations of topological order but find none, at least in the simple considered theory framework. Nevertheless, we argue in Section \ref{secHopfCorr} that certain signatures of the conserved Hopf entanglement could be observable in the quantum noise correlations and specific heat. We pay a special attention to the chiral quantum anomaly and explain in a simple manner how it is related to the Hopf index; from that point of view, the chiral anomaly might provide a window into the loop-entanglement transition in materials with Dirac or Weyl quasiparticles. By analogy with topological order, we briefly discuss a mechanism for anomaly fractionalization.

Four-dimensional topological order based on hopfions is finally explored in Section \ref{sec4D}. It features charge and angular momentum fractionalization (Section \ref{secFract}), fractional braiding statistics between particles and 2-spheres (Section \ref{secBraid}), topological ground state degeneracy (Section \ref{secTopDeg}), and a correlation between the charge and spin sectors (Section \ref{secTopSpinor}). All conclussions are again summarized in Section \ref{secConclusions}, and discussed in the view of remaining open problems. A generalization of the Hopf invariant to higher dimensions is presented in Appendices \ref{app2} for the spin sector and \ref{app3} for the charge sector.

In this paper, we use units $\hbar=c=1$ and Einstein's convention for the summation over repeated indices. Spatial directions are denoted by Latin letters $i,j,k,\dots \in \lbrace 1, \dots, d \rbrace$, and space-time directions by Greek letters $\mu,\nu,\lambda,\cdots\in \lbrace 0, 1, \dots, d \rbrace$. Indices of Levi-Civita tensors $\epsilon_{\mu\nu\lambda\cdots}$ inside manifold integrals always correspond to the directions locally tangential to the manifold.

\raggedbottom

\section{The topology of loop entanglement}\label{secLoops}

Thermodynamic phases of mobile topological defects are sharply distinct from fully disordered phases as long as a relevant topological charge (invariant) is conserved. Fractional quantum Hall liquids are examples of such phases. They can be generalized to higher spatial dimensions $d>2$ by utilizing $\pi_{d-1}(S^{d-1})$ homotopy groups where the relevant topological defects are monopoles and hedgehogs in the charge and spin sectors respectively.

Here we study the analogous unconventional dynamics of the topological defects characterized by the $\pi_3(S^2)$ homotopy group. The relevant topological invariant is the Hopf index. Physically, the quantum liquids in this class are made of entangled and interlinked flux loops. While a classical snapshot of such a state may look chaotic like a bowl of spaghetti (with each spaghetti strand wrapped into a loop), we will show that quantum dynamics combined with loop entanglement stabilizes unconventional strongly correlated topological phases which need not be classifiable as topological order.

\subsection{Hopf index}\label{secHopf}

Hopf index $N\in\mathbb{Z}$ is a topological invariant which characterizes configurations of a three-component unit-vector field $\hat{\bf n}({\bf x})$ in a three-dimensional space without boundaries, ${\bf x}\in S^3$. Since the set of $\hat{\bf n}$ vectors is equivalent to a unit-sphere $S^2$, and a simple $d=3$ space without boundaries is a 3-sphere $S^3$, the Hopf invariant distinguishes equivalence classes of the $\pi_{3}(S^{2})$ homotopy group. Two configurations $\hat{\bf n}({\bf x})$ with different Hopf invariants cannot be smoothly transformed into each other. We will construct the Hopf index and demonstrate its topological invariance by physical arguments in several stages.

Using Einstein's notation $\hat{n}^a$ for the components of the vector $\hat{\bf n}$, we define the chirality
\begin{equation}\label{Chirality}
J_{i}^{\phantom{x}}=\frac{1}{2}\epsilon_{ijk}^{\phantom{x}}\epsilon^{abc}\hat{n}^{a}(\partial_{j}\hat{n}^{b})(\partial_{k}\hat{n}^{c})
\end{equation}
and observe that its flux
\begin{equation}\label{FluxQuant}
\int\limits_{S^2} d^2x\, \hat{\eta}_i J_i = n\phi_0 \quad,\quad n\in\mathbb{Z}
\end{equation}
is quantized in the units of $\phi_0 = 4\pi$ on any closed surface $S^2$. The integer $n$ is the topological invariant of the $\pi_2(S^2)$ homotopy group, also known as Pontryagin index, and $\hat{\eta}_i$ is a unit-vector locally perpendicular to the surface $S^2$. A field $\hat{n}^a(x,y)$ which becomes uniform at large distances $r=(x^2+y^2)^{1/2}$ can carry a quantized Pontryagin index on every $z=\textrm{const.}$ plane because all points at infinite $r$ can be effectively identified, turning each plane into a sphere $S^2 = \mathbb{R}^2 \cup \lbrace \infty \rbrace$. Such a field configuration is a skyrmion line. We may smoothly deform the field $\hat{n}^a ({\bf x})$ to focus its non-zero flux (\ref{FluxQuant}) into a finite region $r<R$ and leave a uniform chirality-free configuration $\hat{n}^a\to \textrm{const.}$ at all points $r>R$; the skyrmion thickness $R$ can become arbitrarily small, yet the topological Pontryagin index cannot change under these smooth transformations. We may also arbitrarily reshape the skyrmion line. Generally, skyrmion lines can form closed loops; line terminations are not allowed because the ensuing Pontryagin index change across the termination point cannot be obtained from a smooth spatial variation of $\hat{n}^a$. 

In the first step of Hopf index construction, we specialize to the vector fields $\hat{n}^a$ which carry an arbitrary set of \emph{infinitely thin} skyrmions loops $C_l$. The corresponding chirality (\ref{Chirality}) is a singular function of coordinates
\begin{equation}\label{FluxFilament}
J_i({\bf x}) = \phi_0 \sum_l n_l \oint\limits_{C_l} dx_{l,i}\, \delta({\bf x}-{\bf x}_l) \ ,
\end{equation}
where $n_l$ are integers, ${\bf x}_l\in C_l$ is the set of points on the flux filament $C_l$, and $\delta$ is the Dirac $\delta$-function of a three-dimensional vector. The flux (\ref{FluxQuant}) is quantized with $n=n_l$ on any \emph{open} manifold $S^2\to A$ which intersects the filament $l$. Let us interpret the chirality as an ``electric current density''; the ``current'' carried by each loop $C_l$ is quantized, $I_l=n_l\phi_0$. The ``magnetic field'' ${\bf B}$ which obtains from all ``currents'' is given by Ampere's law (up to an unimportant factor):
\begin{equation}
{\bf J} = \boldsymbol{\nabla} \times {\bf B} \ .
\end{equation}
If the loops are interlinked, then the total current passing through the loop $C_{l}$ is
\begin{equation}\label{CrossCurrent}
\sum_{m}N_{lm} I_{m} = \int\limits_{A_l} d^2x\,\hat{\boldsymbol{\eta}} \cdot {\bf J} = \oint\limits _{C_{l}}d{\bf x}\cdot{\bf B} \ .
\end{equation}
Here, $A_l$ is a surface bounded by $C_l$, $\hat{\boldsymbol{ \eta}}$ is the unit vector locally orthogonal to it, and $N_{lm}$ is the number of times the loop $m$ wraps around the loop $l$. We allow $l=m$ on the left-hand side and denote by $N_{ll}$ the self-linking number; if $N_{ll}\neq 0$, then the loop $l$ is a knot. Note that inside the filament $l$ and along its contour $C_{l}$ we have $I_{l}d{\bf x} = {\bf J}dxda_{l}$, where $da_{l}$ is the infinitesimal area element of the $l$-filament's cross-section. So, after multiplying (\ref{CrossCurrent}) by $I_{l}$ and summing over all loops $l$, we get
\begin{equation}\label{HopfLink1}
\sum_{l,m}N_{lm} I_{l} I_{m}=\int d^{3}x\,{\bf B}\cdot{\bf J} \ .
\end{equation}
Finally, we recall the ``current'' quantization $I_l=n_l \phi_0$; the last equation becomes
\begin{equation}\label{Hopf0}
N = \sum_{l,m} N_{lm} n_l n_m = \frac{1}{\phi_0^2} \int d^{3}x\,{\bf B}\cdot{\bf J} \ .
\end{equation}
The number $N$ is evidently an integer.

From this point on, we will interpret the ``current density'' ${\bf J}$ as a curl of a gauge field ${\bf A}$:
\begin{equation}\label{JAgauge}
{\bf J}=\boldsymbol{\nabla}\times{\bf A} \ .
\end{equation}
The integer (\ref{Hopf0}) becomes the topological Hopf index
\begin{equation}\label{Hopf1}
N_A = \frac{1}{\phi_0^2}\int\limits_{S^{3}}d^{3}x\,{\bf A}\cdot(\boldsymbol{\nabla}\times{\bf A})=\frac{1}{\phi_0^2}\int\limits_{S^{3}}d^{3}x\,{\bf A}\cdot{\bf J}
\end{equation}
after showing that the flux of ${\bf J}$ need not be focused into singular filaments. The above simple argument about integer quantization of the Hopf index holds against all smooth transformations of $\hat{\bf n}({\bf x})$ which reshape the flux loops and turn them into thick flux tubes without introducing overlaps. Going beyond this restriction is less transparent. We will present a rigorous direct argument about the topological invariance of the Hopf index only in Section \ref{secTopSpinor} using SU(2) spinor representations. Until then, we will rely on a physical point of view. Let us regard an arbitrary gauge field ${\bf A}({\bf x})$ as the coarse-grained representation of the quantum superposition between many filament configurations (\ref{FluxFilament}). In this picture, quantum fluctuations deform and move the quantized flux filaments until the flux diffuses into a continuous and non-singular structure. We allow, in principle, all quantum processes. Some processes, which we call instantons, can convert a filament loop configuration into another one with a different Hopf index. We will see in Section \ref{secHopfConservation} that instantons cannot occur with smooth $\hat{\bf n}({\bf x},t)$ fields, and argue that they are ``confined'' at low temperatures, i.e. every instanton is accompanied by an anti-instanton no further than a certain finite space-time interval away. This dynamically protects the Hopf index at macroscopic scales against quantum fluctuations, and gives (\ref{Hopf1}) the meaning of a topological invariant.

We will work with two physical contexts in which the Hopf index is relevant. One context is the spin context we already introduced. The emergent gauge field ${\bf A}$ is derived from the chirality
\begin{equation}\label{ChiralGauge}
\epsilon_{ijk}\partial_{j}A_{k}=\frac{1}{2}\epsilon_{ijk}\epsilon^{abc}\hat{n}^{a}(\partial_{j}\hat{n}^{b})(\partial_{k}\hat{n}^{c})
\end{equation}
of the spin field $\hat{n}^a$. This is the effective gauge field of the ``topological Hall effect''. It captures the effective Lorentz force which spinful particles experience when moving through a topologically non-trivial magnetization background, assuming adiabatic limit (particle spin locked to the background magnetization). The topological quantization of the skyrmion number determines $\phi_0=4\pi$, so the Hopf index characterizes the interlinking of skyrmion loops. The second context is tied to charge fluctuations of a spinor field $\psi$. We extract the gauge field using
\begin{equation}
A_i \psi^\dagger\psi = -i \psi^\dagger \partial_i \psi \ .
\end{equation}
This gauge field captures a set of Dirac string loops in any coherent $\psi$ configuration, carrying $\phi_0=2\pi$ flux quantum. Fluctuations may continuously distribute this flux, but the Hopf index remains well defined and quantized as long as monopoles of ${\bf A}$ are not present.

The purpose of this paper is to explore unconventional correlated states of matter whose dynamics quantizes and topologically protects the Hopf index. The above demonstration of the Hopf quantization reveals the physical requirements on such states. The degrees of freedom must be able to support appropriate flux currents which carry a globally quantized flux. Monopoles must be absent, i.e. the flux must be closed into loops. We will show that stable non-trivial phases with a quantized Hopf index exhibit subtle correlation phenomena in $d=3$ spatial dimensions or topological order in $d=4$.

\subsection{Gauss' linking number, framing and twist}\label{secLinkNum}

\begin{figure}[!]
\subfigure[{}]{\includegraphics[width=1.2in]{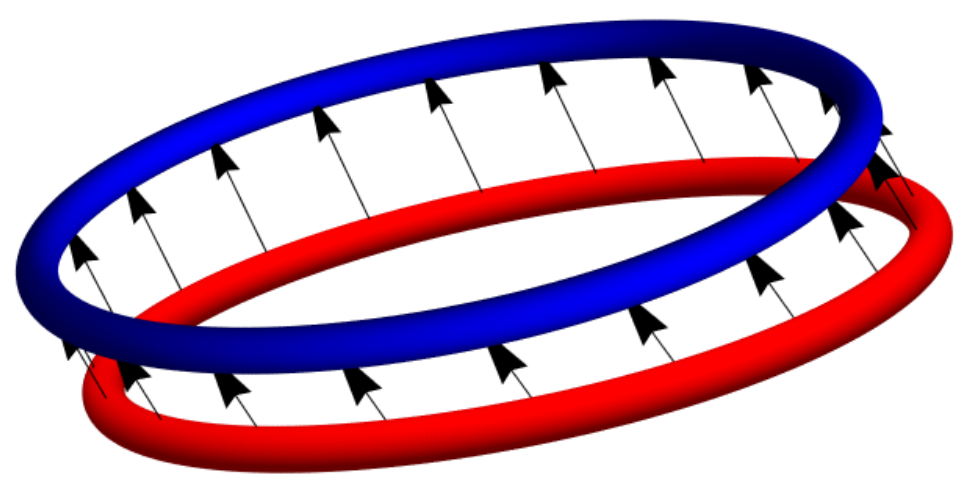}}
\hspace{0.2in}
\subfigure[{}]{\includegraphics[width=1.2in]{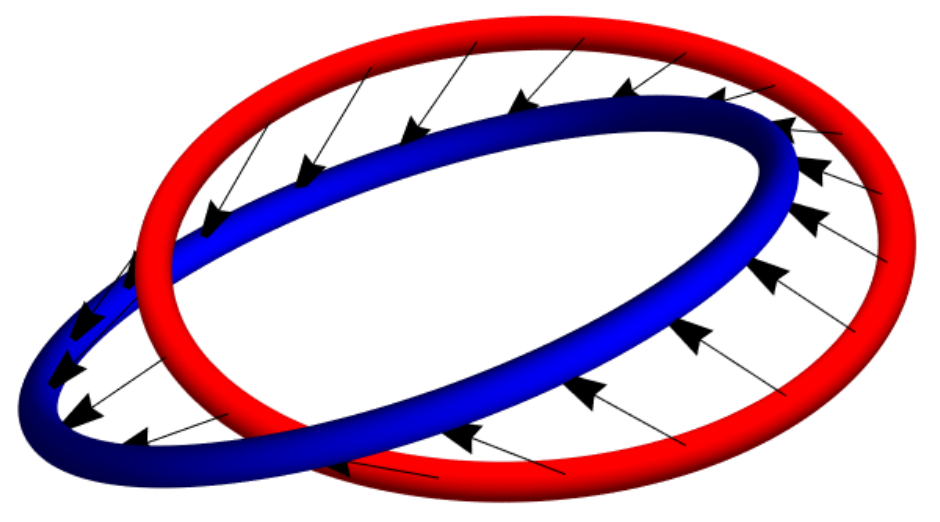}}
\caption{\label{Framing}Framing regularization: displace a loop by a framing field (black arrows), and compute the linking integral of the original (red) and the image (blue). The two shown configurations yield different linking integrals.}
\end{figure}

If the gauge flux ${\bf J}$ forms singular quantized loops $C_l$ with a fixed quantum $n_l=1$ according to (\ref{FluxFilament}), then the Hopf index (\ref{Hopf1}) is simply the sum of loops' linking numbers,
\begin{equation}\label{HopfLink}
N = \sum_{i,j}N_{ij}=\sum_{i}N_{ii}+2\sum_{i<j}N_{ij} \ .
\end{equation}
The linking number $N_{ij}$ of two loops (introduced by Gauss) obtains by substituting Biot-Savart law
\begin{equation}
{\bf A}({\bf x})=\frac{1}{4\pi}\int d^{3}x'\,\frac{{\bf J}({\bf x}')\times({\bf x}-{\bf x}')}{|{\bf x}-{\bf x}'|^{3}}
\end{equation}
into (\ref{CrossCurrent}) with interpretation ${\bf B}\equiv{\bf A}$:
\begin{equation}\label{GaussLink}
N_{ij}=N_{ji}=\frac{1}{4\pi}\oint\limits_{C_i}\oint\limits_{C_j}\frac{({\bf r}_{i}-{\bf {\bf r}}_{j})\cdot(d{\bf r}_{i}\times d{\bf r}_{j})}{|{\bf r}_{i}-{\bf {\bf r}}_{j}|^{3}} \ .
\end{equation}
Here, ${\bf r}_{i}$ represents the set of points on the loop $C_i$ and $d{\bf r}_{i}$ is the tangential infinitesimal displacement along $C_i$. Introducing a unit vector
\begin{equation}
\hat{{\bf m}}=\frac{{\bf r}_{i}-{\bf r}_{j}}{|{\bf r}_{i}-{\bf r}_{j}|}
\end{equation}
quickly demonstrates the quantization of the linking number
\begin{equation}
N_{ij}=-\frac{1}{4\pi}\oint\oint dx_{1}dx_{2}\,\hat{{\bf m}}\cdot\left(\frac{\partial\hat{{\bf m}}}{\partial x_{1}}\times\frac{\partial\hat{{\bf m}}}{\partial x_{2}}\right)\in\mathbb{Z} \ .
\end{equation}
There is a problem, however, when one uses (\ref{GaussLink}) to compute the self-linking number $N_{ii}$ of a knot. If $C_i=C_j$, then a singularity at ${\bf r}_i={\bf r}_j$ is integrated along the knotted loop. The regularization of this singularity is known as framing, see Fig.\ref{Framing}. One constructs an auxiliary image $C'_i$ of the knotted loop $C^{\phantom{x}}_i$ and displaces it everywhere from the original by a framing vector field ${\bf f}({\bf r}_i)\neq 0$. The self-linking number $N_{ii}$ is then computed as the well-defined interlinking number of the image and the original. The framing field can be infinitesimal and it is chosen, for concreteness, to be locally perpendicular to the original loop. As good as it gets, this procedure unavoidably introduces an ambiguity in the definition of self-linking. The framing field can complete $n\in\mathbb{Z}$ full rotations about the string as we traverse the loop. This wraps the image $n$ extra times around the original loop, and hence affects the computed inter-linking number. Introduced purely by the regularization, $n$ has nothing to do with the shape of the original loop. Physically, the framing field must be rooted in some microscopic degrees of freedom which we might have not encountered before.

Framing finds a very simple and natural representation in field theory. The gauge field ${\bf A}$ featured in the Hopf index is a dynamical degree of freedom with a certain Maxwell Lagrangian density $(\boldsymbol{\nabla}\times{\bf A})^2$. If we couple it to a matter field $\chi$ of charge $q$, we also introduce a gradient term $(\boldsymbol{\nabla}\chi+q{\bf A})^2$ into the Lagrangian density. For simplicity, we regard $\chi$ as an angle-valued field in an XY model context. Now, gauge invariance suggests an adaptation of the expression (\ref{Hopf1}) for the Hopf index
\begin{equation}\label{Hopf2}
N_\chi = \frac{1}{q\phi_0^2}\int\limits_{S^{3}}d^{3}x\,(\boldsymbol{\nabla}\chi+q{\bf A})\cdot(\boldsymbol{\nabla}\times{\bf A}) \ .
\end{equation}
Suppose we want to calculate the Hopf index of a self-linked flux loop $C$. The gauge field of a quantized flux filament has the curl given by (\ref{FluxFilament}), so that
\begin{equation}\label{Hopf3}
N_\chi = \frac{1}{q\phi_0}\oint\limits_{C}d{\bf r}'\cdot(\boldsymbol{\nabla}\chi+q{\bf A}) \ .
\end{equation}
The $\chi$ angle is free to wind $n$ times along the loop's path $C$, contributing $n$ to the Hopf index when the charge is quantized by $q\phi_0 = 2\pi$. This is different than the previously considered contribution of the pure gauge field, and requires $\chi$ to carry a singular vortex line somewhere in space, passing through $C$. In the charge context, the field $\chi$ is simply the quantum mechanical phase of the physical charged matter field. The microscopic origin of $\chi$ is not obvious in the spin context, but skyrmions are certainly complex objects which have enough structure to support another degree of freedom. The construction (\ref{Hopf3}) gives $\chi$ the role of a ``twist'' field which captures the internal twisting of the flux filament about its own axis as it forms a loop. The framing field ${\bf f}$ describes the same, so we can relate it to $\chi$ with
\begin{equation}
\hat{{\bf f}}\cdot\left(\frac{d\hat{{\bf f}}}{dr}\times d{\bf r}\right)=d\chi \ .
\end{equation}
In a coarse-grained field theory, $\chi$ captures the microscopic flux filament twist which cannot be resolved at macroscopic length scales.

It should be pointed out that (\ref{Hopf2}) is invariant under smooth gauge transformations ${\bf A}\to {\bf A} + \boldsymbol{\nabla}\theta$. However, it is not invariant under singular gauge transformations in which $\theta$ contains vortex singularities. This is essential for capturing the twist. It will become apparent shortly that quantum dynamics protects (\ref{Hopf2}) better that the pure gauge-field Hopf index (\ref{Hopf1}). Furthermore, the Hopf index is invariant under time-reversal but changes sign under mirror inversion transformations.

\begin{figure}
\subfigure[{}]{\includegraphics[width=1.3in]{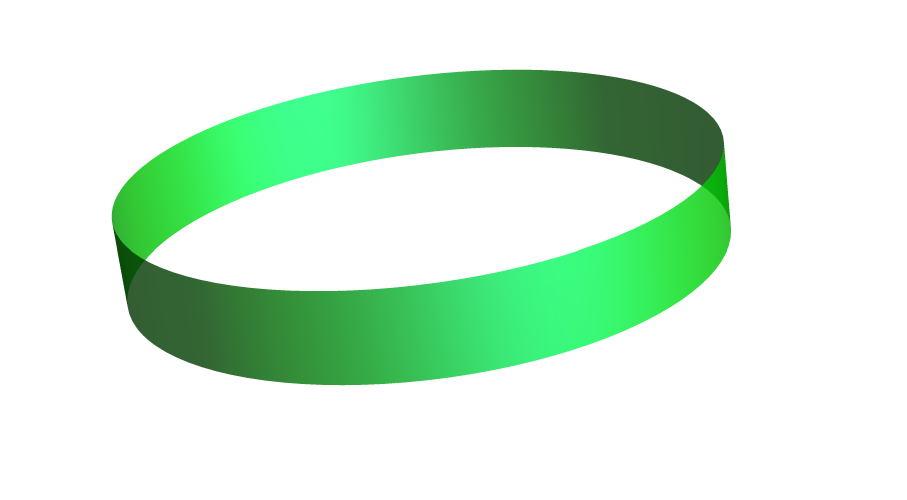}}
\subfigure[{}]{\includegraphics[width=1.3in]{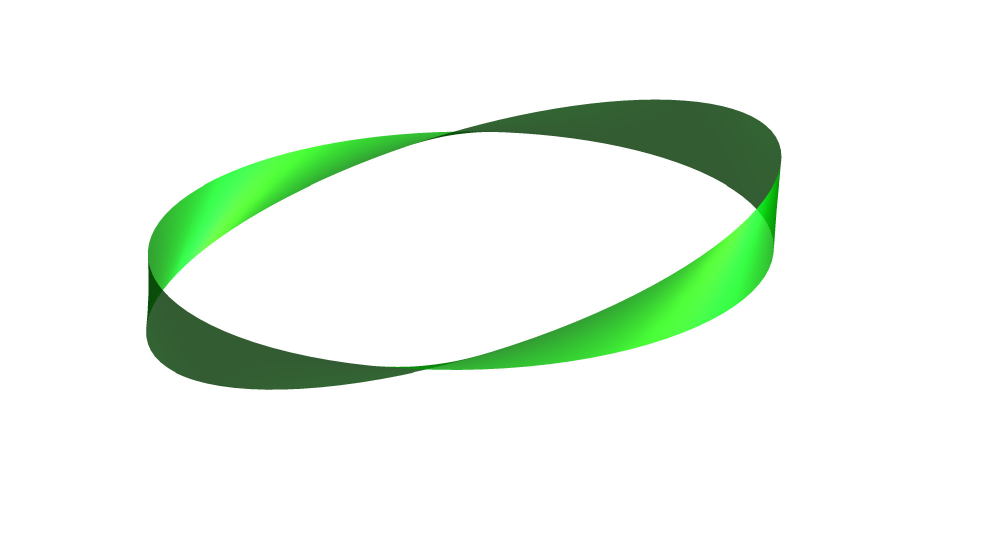}}
\subfigure[{}]{\raisebox{0.3in}{\includegraphics[width=1.3in]{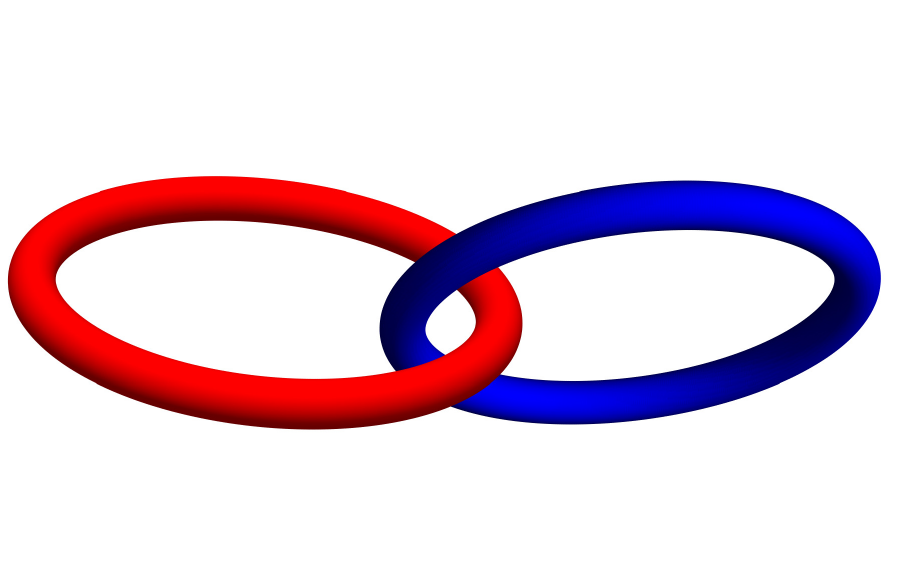}}}\hspace{0.2in}
\subfigure[{}]{\includegraphics[width=1.3in]{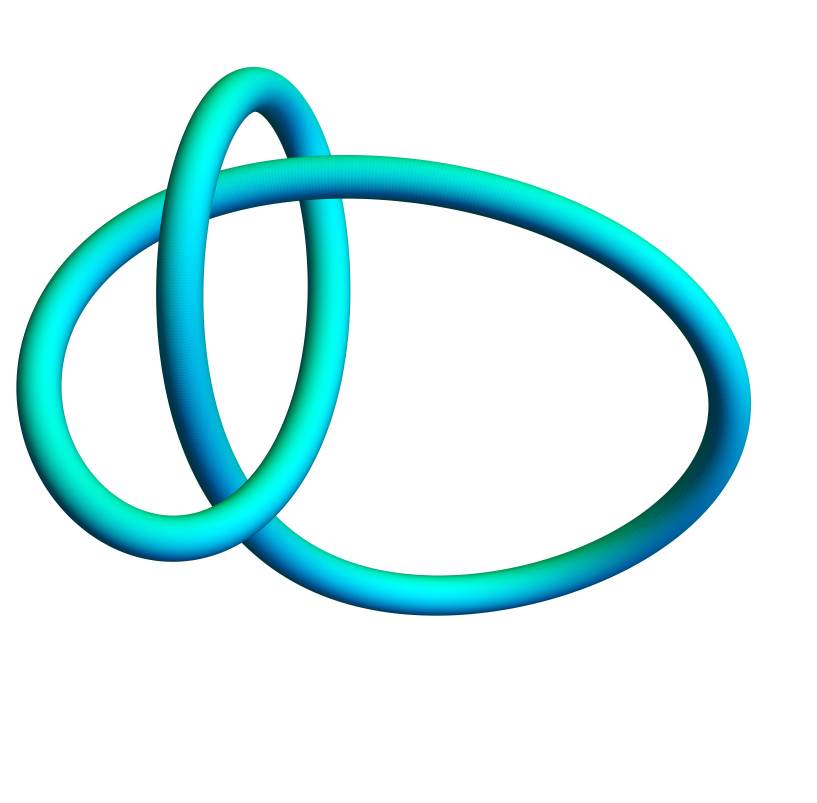}}
\caption{\label{Hopfions}Simplest Hopf topological defects: (a) simple loop $N=0$, (b) twisted loop $N=1$, (c) linked simple loops $N=2$, and (d) simplest (trefoil) knot $N=3$. Loops are oriented by the direction of the flux they carry. Even though time reversal changes the direction of flux, it does not affect the Hopf index. Mirror inversions change the sign of the Hopf index.}
\end{figure}

From now on, we will call a ``hopfion'' any field configuration that carries a non-zero Hopf index (\ref{Hopf2}). Hopfions are topological defects classified by the $\pi_3(S^2)$ homotopy group. The simplest hopfion configurations are shown in Fig.\ref{Hopfions}.

\subsection{Stimulating and conserving hopfions in the ground state}\label{secStimulate}

What kind of Lagrangian or Hamiltonian terms can stimulate the appearance of hopfions in the ground state of a system? Let us exploit the analogy with quantum Hall liquids to propose a generic answer. The topological defect behind quantum Hall liquids is a vortex, and the topological charge of vortices can be obtained from the curl of a gauge field, $\oint dx_i A_i = \int d^2x\, \epsilon_{ij} \partial_i A_j$. We used Stokes' theorem and the convention that the Levi-Civita tensor's indices live on the two-dimensional integration manifold whose boundary is the loop in the first integral. The Lagrangian density which nucleates a finite density of vortices is $\mathcal{L}_{\textrm{v}} \propto (\epsilon_{\mu\nu\lambda}\partial_\nu A_\lambda - B_\mu)^2$, where $B_\mu = B_0 \delta_{\mu,0}$ represents an external magnetic field.

In the case of hopfions, the expressions (\ref{Hopf1}) and (\ref{Hopf2}) are analogous to $\oint dx_i A_i$ of the quantum Hall effect. Consider
\begin{equation}
N_A = \oint\limits_{S^3} d^3x\, \epsilon_{ijk}A_i\partial_jA_k =
  \int\limits_{B^4} d^4x \, \epsilon_{ijkl}\partial_iA_j\partial_kA_l \ .
\end{equation}
written in Einstein's notation, with a convention that the Levi-Civita indices are locally tangential to the integration manifold. We assumed that the 3-sphere $S^3$ is embedded in a 4-dimensional space, where it acts as a boundary of a 4-dimensional ball $B^4$, so that we could apply Stokes-Cartan's theorem. It is now evident that a Lagrangian density term
\begin{equation}
\mathcal{L}_{\textrm{H,4D}} \propto (\epsilon_{\mu\nu\lambda\alpha\beta}\partial_\nu A_\lambda \partial_\alpha A_\beta - \mathcal{B}_\mu)^2
\end{equation}
with a ``magnetic field'' $\mathcal{B}_\mu = \mathcal{B}_0 \delta_{\mu,0}$ should stimulate a finite density of hopfions in the ground state of a hypothetical world with four spatial dimensions. The double curl in this formula is equivalent to $\epsilon_{\mu\nu\lambda\alpha\beta}(\partial_\nu A_\lambda) (\partial_\alpha A_\beta)$ in the absence of monopoles ($\epsilon_{\mu\nu\lambda\cdots}\partial_\mu\partial_\nu A_\lambda=0$), and each curl factor is related to magnetic flux lines or spin chirality (\ref{ChiralGauge}). We will explore the ensuing emergence of $d=4$ topological order analogous to fractional quantum Hall states in Section \ref{sec4D}.

The above exercise is also useful for the real three-dimensional world. However, since only four space-time indices are available, we can construct only a ``crippled'' Lagrangian density
\begin{equation}
\mathcal{L}_{\textrm{H,3D}} \propto (\epsilon_{\mu\nu\alpha\beta}\partial_\mu A_\nu \partial_\alpha A_\beta-\mathcal{B})^2 \ .
\end{equation}
A hypothetical non-zero scalar field background $\mathcal{B}\neq 0$ is not natural here because it would need to change sign under time reversal, while the Hopf index stays invariant. Instead, we should set $\mathcal{B}=0$ and let $\mathcal{L}_{\textrm{H,3D}}$ limit the confinement length $\lambda$ of Hopf instantons. If $\mathcal{L}_{\textrm{H,3D}}$ achieves $\lambda<\xi$, where $\xi$ is the correlation length of the matter field, then a disordered ground state can exist as a correlated quantum liquid of conserved hopfions \cite{Nikolic2023a}. Note that $\mathcal{L}_{\textrm{H,3D}}$ is not a priori required for $\lambda<\xi$. Other more realistic mechanisms, including the kinetic energy of the matter field, should also be able to limit $\lambda$ below $\xi$. Any one of them is sufficient for the discussion in this paper.

The means to generate a finite density of hopfions in $d=3$ systems are currently unclear beyond Hopf insulators \cite{Moore2008, Duan2013, Duan2014, Duan2016, Hasan2017, Volovik2017, Schuster2019, Slager2019, Soluyanov2021, Trifunovic2021, Piechon2022, Goswami2023} where delocalized hopfions are effectively imparted on electron wavefunctions.


\section{Hopfions in three dimensions: Instanton confinement and quantum anomaly}\label{sec3D}

\subsection{Hopf index conservation}\label{secHopfConservation}

Hopf index is topologically protected from all smooth transformations of the classical field configuration. However, quantum tunneling introduces local abrupt changes of the classical states (in a regularized theory). A high action cost of such tunneling doesn't necessarily make it statistically impossible, so quantum events, which we will refer to as instantons, are able to change the classical topological invariant. The question of topological protection, then, becomes the question of instanton confinement. If every instanton event is immediately followed by a nearby anti-instanton, we say that the instantons are confined into neutral dipoles and the topological index is globally conserved. A thermodynamic instanton confinement-deconfinement transition can occur at finite temperatures \cite{Nikolic2023a}.

\begin{figure}[t]
\includegraphics[height=1.0in]{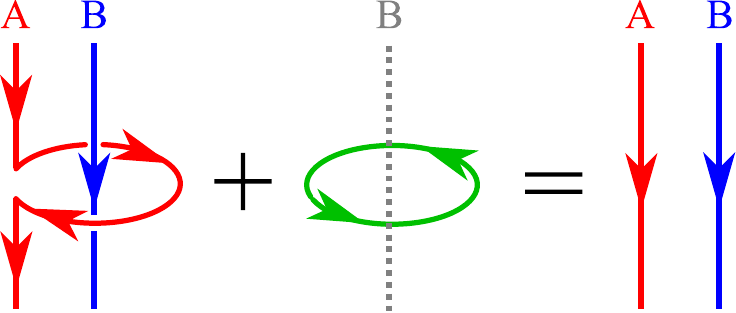}
\caption{\label{Unlink}Quantum and thermal fluctuations of small flux rings (green) can unlink and re-link flux loops (A and B). Such a ring has lower energy $(\boldsymbol{\nabla}\chi+q{\bf A})^2$ when it is twisted, because it needs a winding twist field $\chi$ to compensate the circular gauge field ${\bf A}$ of the B string. Its net Hopf index is zero, so the unlinked loops A and B have the same Hopf index as when they were linked; unlinking imparts a twist on the A string, as expected from the electromotive force induced by Faraday effect. Alternatively, if the small ring excitation is untwisted, then it costs more energy and carries a non-trivial Hopf index by being interlinked with B; the resulting untwisted unlinked A and B have a different Hopf index than before. The second unlinking event, in which the Hopf index changes, is an instanton.}
\end{figure}

\begin{figure}[t]
\subfigure[{}]{\includegraphics[height=0.4in]{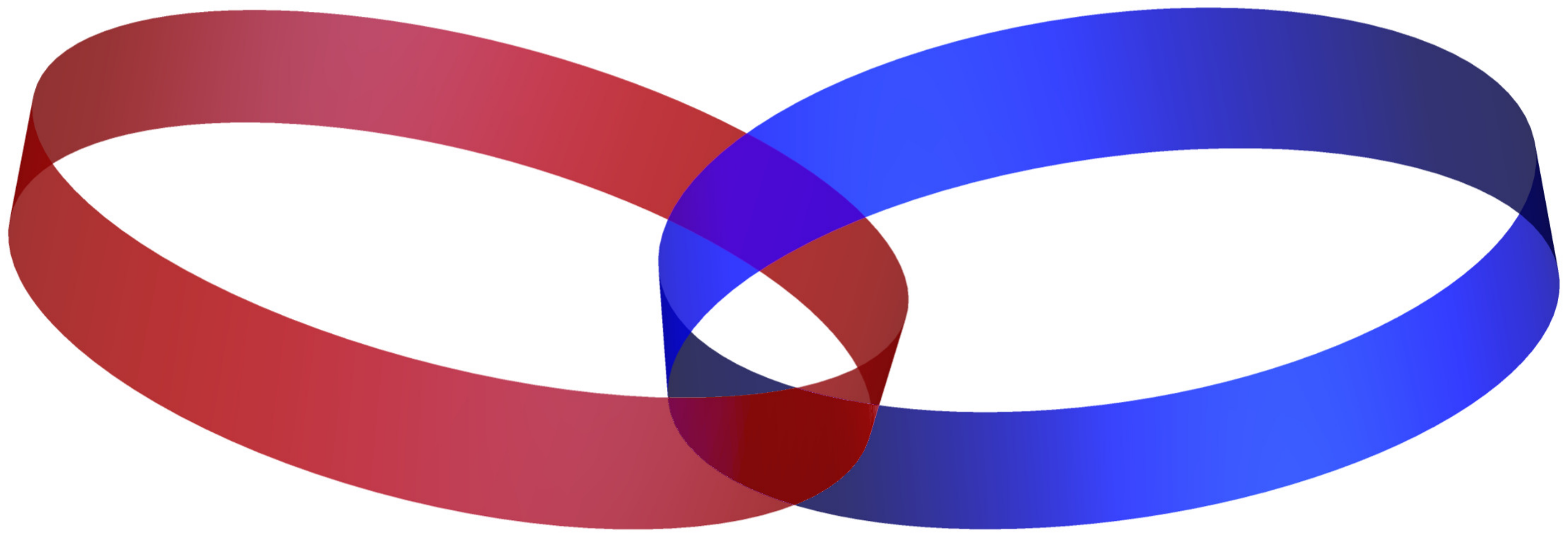}}
\hspace{0.2in}
\subfigure[{}]{\includegraphics[height=0.4in]{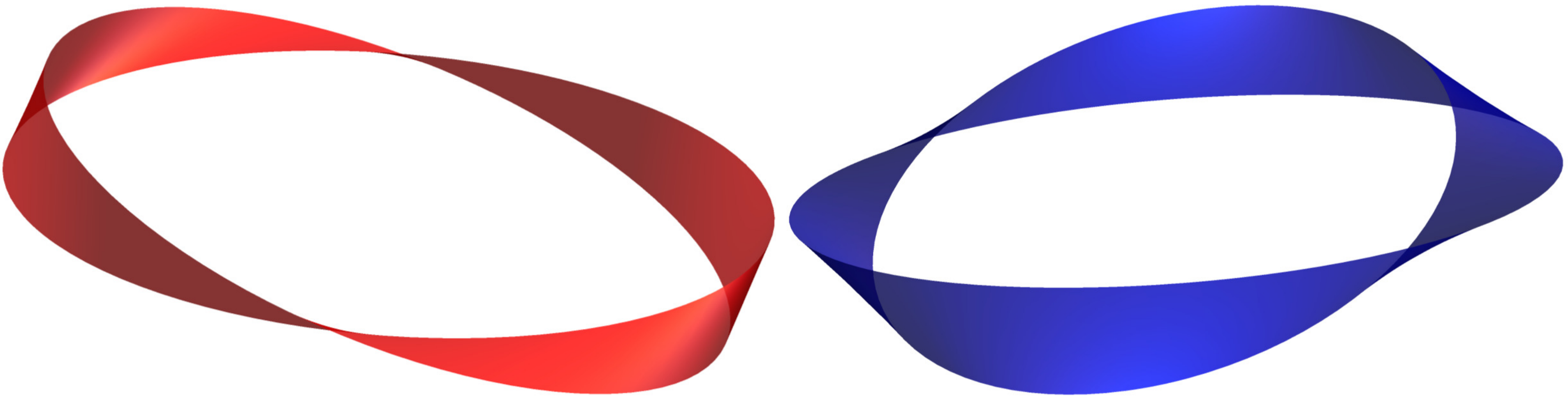}}
\subfigure[{}]{\includegraphics[height=1.1in]{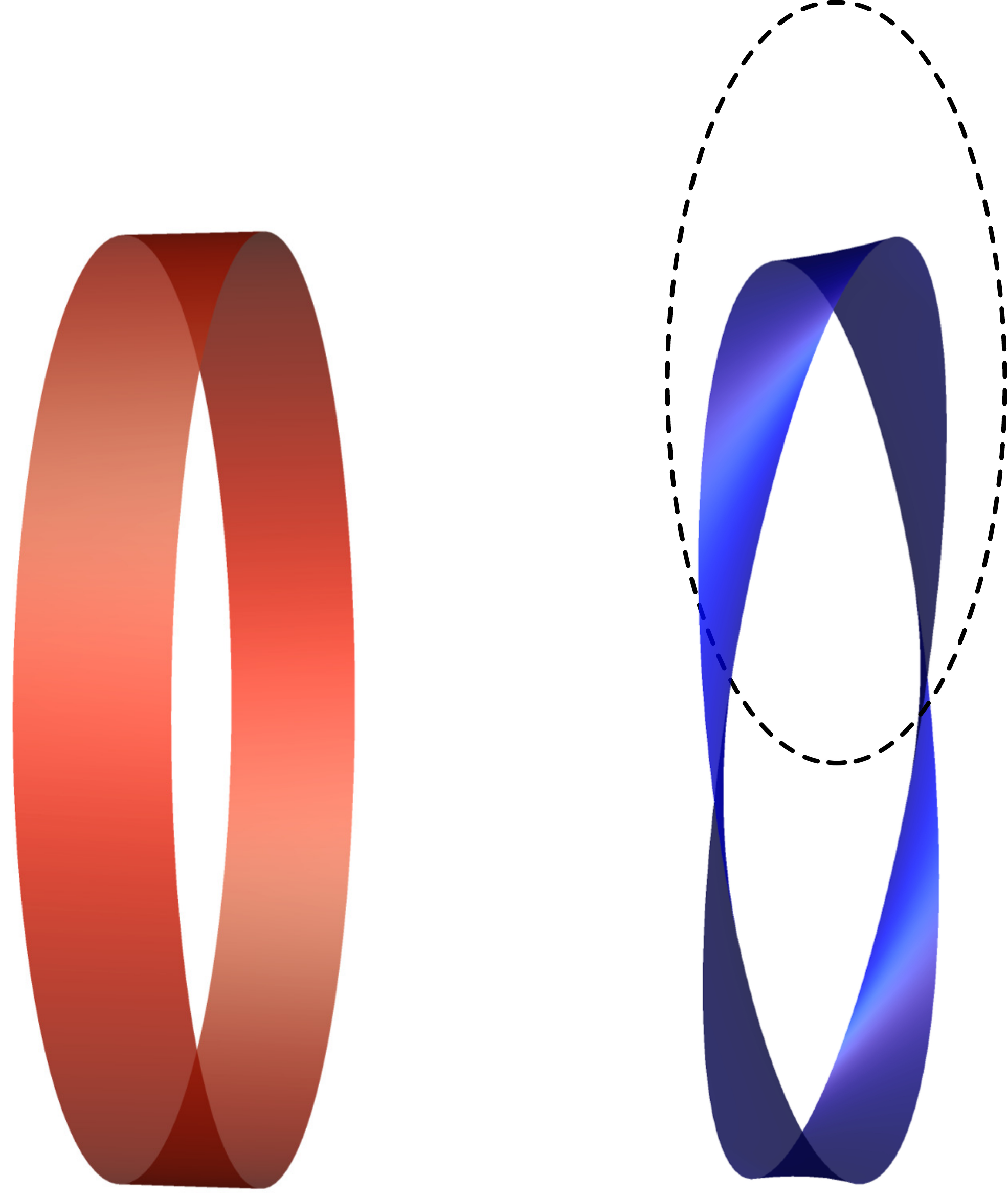}}
\hspace{0.7in}
\subfigure[{}]{\includegraphics[height=1.1in]{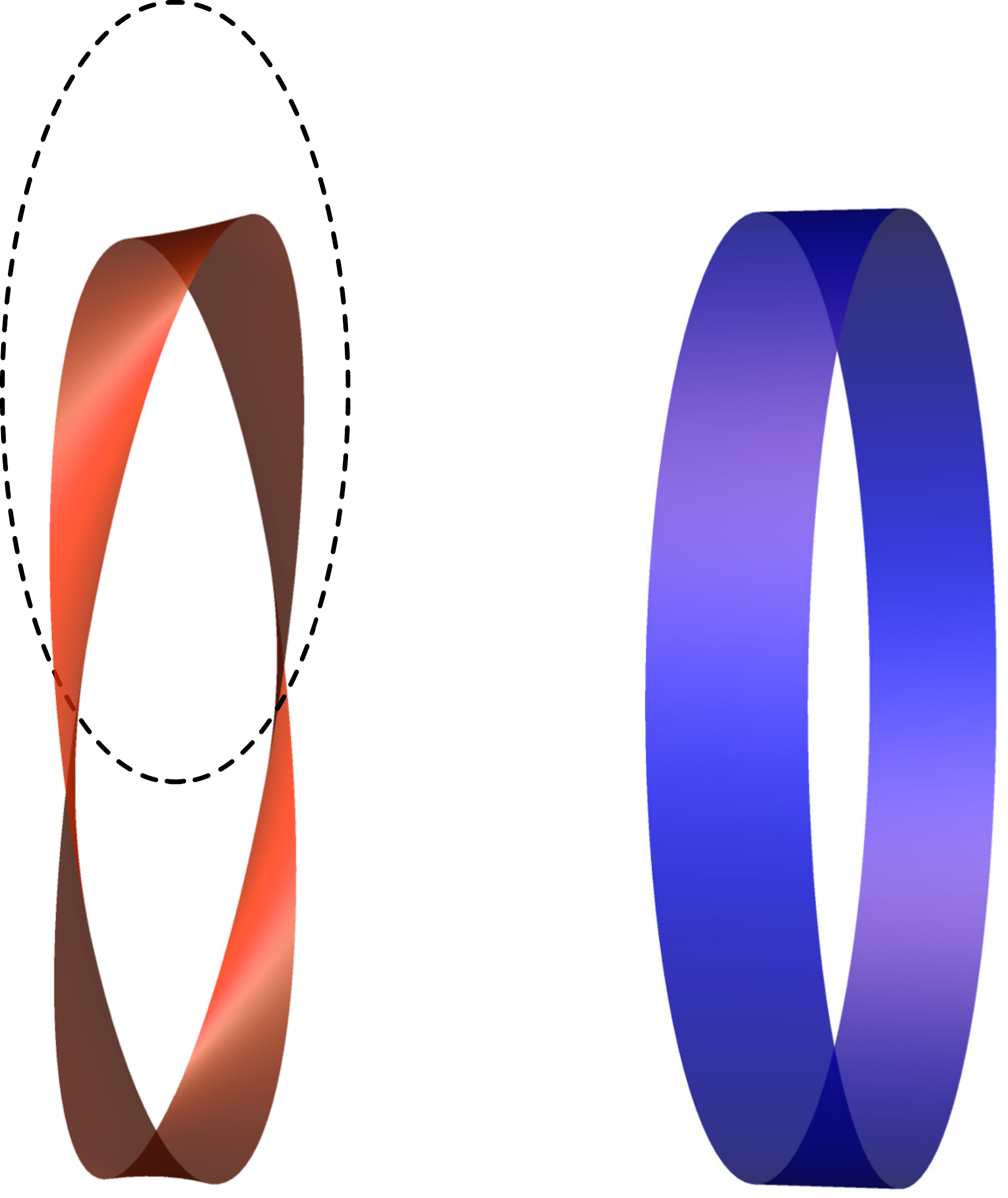}}
\caption{\label{HopfPreservation}Local processes which preserve the Hopf index (\ref{Hopf2}). (a)-(b) A symmetric unlinking and re-linking of two loops compensated by twist. (c)-(d) An exchange of twist between two nearby loops. If the $\chi$ field is not condensed (as we assume), then its gradient is not strictly tied to the gauge field ${\bf A}$. This makes it possible for the loop-singularity of $\boldsymbol{\nabla}\chi$, represented by the dashed ring, to tunnel from one flux loop to another without a classical energy penalty. Note that the untwisted loop can continuously shrink to a point. Therefore, flux loops can be created and annihilated, then linked and unlinked, without changing the Hopf index.}
\end{figure}

Quantum dynamics routinely generates virtual low-energy fluctuations of neutral topological defect clusters. These are vortex-antivortex pairs in two-dimensional superconductors, and small vortex loops in three-dimensional superconductors. An isolated flux loop is also the simplest neutral topological cluster of hopfions. Cheap fluctuations of small Hopf-neutral rings can gradually reshape the segments of a larger loop and move it through space. The interaction between loops is a simple superposition of their quantized fluxes; two flux filaments cancel out if they overlap exactly but have opposite directions. Importantly, fluctuating rings can also unlink and re-link two loops, as shown in Fig.\ref{Unlink}. The events which do not significantly modify the gradient energy
\begin{equation}
E_\chi \propto \int d^3x\, (\boldsymbol{\nabla}\chi+q{\bf A})^2
\end{equation}
preserve the Hopf index (\ref{Hopf2}). The examples illustrated in Fig.\ref{HopfPreservation} include twist-compensated unlinking and twist exchange. Other events which alter the Hopf index of the system are instantons. Interlinked and self-linked loops can be mathematically characterized by various other topological invariants \cite{Alexander1928, Jones1985, Jones1987, Kauffman1987, Witten1989, Livingstone1996, Morishita2012, Purcell2020}, but these cannot be conserved in the presence of low-energy unlinking fluctuations.

Hopf index can be protected only if instantons and anti-instantons attract each other with sufficiently strong ``forces'' that grow with the distance. Let us first analyze the protection of the pure gauge-field Hopf index $N_A$ given by (\ref{Hopf1}). Instantons must be local events in order to minimize action, so consider a unit instanton at the origin of space-time; this corresponds to a change of the Hopf index by $1$ between the times $t<0$ and $t>0$. Then, set up a 3-sphere manifold $S^3$ centered at the origin in the $D=4$ dimensional space-time. The Hopf index $N_A$ calculated on this manifold is $1$, the number of enclosed instantons (easily seen by stretching the manifold into an infinite slab with one flat boundary at $t<0$ and another at $t>0$). If $S^3$ has a finite space-time radius $r$, then the naively optimal gauge field configuration which realizes $N_A=1$ qualitatively behaves as $A_\mu \propto 1/|{\bf r}|$, with a dimensionless proportionality constant and a non-zero curl spread evenly across $S^3$. The fixed value of $N_A$ in the $r\to 0$ limit cannot be obtained without a singularity of ${\bf A}$, so, by (\ref{ChiralGauge}), an instanton corresponds to a singular configuration of $\hat{\bf n}({\bf x},t)$. This constitutes the topological protection of the Hopf index under smooth transformations of $\hat{\bf n}$. The imaginary-time action cost of the instanton's ${\bf A}$ field up to the distance $R$ from the origin is at best
\begin{eqnarray}\label{InstLog}
V(R) &\propto& \int\limits_{|{\bf x}|<R}\!\!\! d^4 x \, \left\lbrack \left(\frac{\partial{\bf A}}{\partial t}-\boldsymbol{\nabla}A_0 \right)^2 + (\boldsymbol{\nabla}\times {\bf A})^2 \right\rbrack \nonumber \\
&=& \int\limits_{|{\bf x}|<R}\!\!\! d^4 x \, \frac{1}{2}(\epsilon_{\mu\nu\alpha\beta}\partial_\alpha A_\beta)^2 \nonumber \\
&\sim& K \int\limits_0^R dr\, r^3 \left(\frac{1}{r^2}\right)^2 = K \log\left(\frac{R}{R_0}\right) \ .
\end{eqnarray}
We neglected the cost due to the gradient Lagrangian density $(\boldsymbol{\nabla}\chi+q{\bf A})^2$ because $\chi$ can compensate the gauge field of the instanton (at expense of becoming singular). When we introduce an anti-instanton at a distance $R$ from the origin, the behavior $A_\mu \sim 1/r$ becomes neutralized beyond $r>R$ as qualitatively anticipated in the integral. Therefore, $V(R)$ estimates the weakest possible interaction potential between two instantons a space-time distance $R$ apart.

Renormalization group analysis \cite{Nikolic2023a} shows that the logarithmic interaction is marginally confining in $D=4$. If such an interaction applies to every pair of instantons, then the partition function of the instanton gas suffers from an infra-red divergence. As a consequence, instanton fluctuations can proceed only if the interaction reconstructs into an exclusive form at large distances, where every instanton interacts with exactly one anti-instanton, and vice versa. The gauge flux $\boldsymbol{\nabla}\times{\bf A}$ spreads evenly only near the instantons, and then focuses into narrow tubes which connect the opposite-charge instantons across larger distances. The reconstructed potential is eventually linear, $V(R)\propto R$, and instanton confinement is guarantied at zero temperature. The gauge-field Hopf index (\ref{Hopf1}) is globally conserved below a finite critical temperature $T_{\textrm{c}A}$ for instanton deconfinement \cite{Nikolic2023a}.

Now we turn our attention to the enhanced Hopf index $N_\chi$ in (\ref{Hopf2}) which takes the framing regularization into account. Instantons of $N_\chi$ require an uncompensated gauge field and cost an extra gradient energy $(\boldsymbol{\nabla}\chi + q{\bf A})^2$. The previous ``best-case'' scenario $\partial_\mu\chi+qA_\mu \sim 1/r$ yields an extremely large gradient energy at large distances,
\begin{equation}\label{InstQuad}
V'(R) = \int\limits_{|{\bf x}|<R}\!\!\! d^4 x \, (\boldsymbol{\nabla}\chi + q{\bf A})^2
\sim \int\limits_0^R dr\, r^3 \frac{1}{r^2} \propto R^2 \ .
\end{equation}
Again, there is a better way. The disturbance of the gradient focuses into tubes and reduces the instanton interaction to a more manageable linear form, $V'(R)\propto R$. The Hopf index $N_\chi$ is topologically protected below a finite critical temperature $T_{\textrm{c}\chi}$ for instanton deconfinement.

The protection of $N_A$ implies the protection of $N_\chi$, but not the other way round. The more dramatic large-distance behavior of (\ref{InstQuad}) relative to (\ref{InstLog}) suggests that instantons of the twist field have a shorter confinement length than the instantons of the gauge field. Hence, we may expect two deconfinement phase transitions with $T_{\textrm{c}\chi} > T_{\textrm{c}A}$. The lower-temperature confined phase preserves the flux loop entanglement and possibly conserves some knot invariants \cite{Alexander1928, Jones1985, Jones1987, Kauffman1987, Witten1989, Livingstone1996, Morishita2012, Purcell2020} other than the linking number $N_A$. This is the only confined phase of Hopf instantons in a pure gauge theory, realized in the continuum non-compact limit. If matter is coupled to the gauge field, then the conservation of $N_A$ requires at least a short-range Higgs mechanism across a coherence length scale $\zeta$ which exceeds the instanton confinement length $\lambda_A$. This insight comes from the renormalization group \cite{Nikolic2023a}. The flow into the fixed point $\zeta>\lambda_A, \lambda_\chi\to 0$ indicates instanton confinement even in the absence of long-range order ($\zeta<\infty$), according to a generalized Wilson loop correlation. We suspect that the conservation of $N_A$ is a necessary (probably not sufficient) condition for the stability of topological orders defined by loop braiding \cite{Walker2012, Wang2014a, Jiang2014, Jian2014, Furusaki2019, Wen2019a, Wen2019, Burkov2020a, Burkov2020b}.

The higher-temperature confined phase in the temperature range $T_{\textrm{c}\chi} > T > T_{\textrm{c}A}$ is possible only when the gauge field is coupled to an independent matter field. It preserves a single knot invariant, the Hopf index $N_\chi$. There is no surviving loop coherence. Note that (\ref{Hopf2}) is not the same as (\ref{Hopf1}) with a redefined gauge field ${\bf A} \to q^{-1}\boldsymbol{\nabla}\chi+{\bf A}$; the curl of $\boldsymbol{\nabla}\chi$ is not included in the Hopf index definition because we assume that an underlying lattice regularization renders the quantized $\chi$-vortex filaments physically unobservable. This enables $N_\chi$ to acquire value $1$ from the twist of a single loop, while $N_A$ without framing is always even. 

Incompressible quantum liquids with topological order are a broad class of phases whose confined instantons cannot ruin the conservation of delocalized topological charge. The realizations of topological order in all $\pi_n(S^n)$ homotopy groups are based on singular topological defects (monopoles and hedgehogs). The number conservation of mobile singularities is captured by a topological Lagrangian term in the effective Lagrangian density, i.e. a Chern-Simons or background-field coupling that arises from some microscopic Berry phase mechanism. This implements particle-singularity attachment, a fractional quantization of charge and exchange statistics, as well as ground state degeneracy on topological manifolds at the level of classical field equations. The dynamics of hopfions is fundamentally different because the field configurations of hopfions do not have singularities in $d=3$ spatial dimensions. A full topological Lagrangian term for hopfions can be constructed only in $d=4$ spatial dimensions, leading to topological order which we discuss in Section \ref{sec4D}. Its remnant in $d=3$ dimensions is only a coupling that regulates the dynamics of instantons. The instanton term can indirectly drive instanton confinement transitions by controlling the coupling constants in (\ref{InstLog}) and (\ref{InstQuad}), but the absence of symmetry breaking hides this transition in the classical field equations. Furthermore, there is no mechanism for fractionalization because this depends upon the ability of particles with a conserved number to bind singular topological defects with a conserved number.

We can compare hopfions to skyrmions in $d=2$, which are also classical topological defects without singularities. Skyrmions are the only topological defects of the three-component unit-vector field $\hat{\bf n}$ in $d=2$. While the skyrmion number is a topological invariant given by (\ref{Chirality}) and (\ref{FluxQuant}), it is not protected against instantons which take form of $\hat{\bf n}$-hedgehogs in the $D=3$ space-time. The interaction between a hedgehog and anti-hedgehog is given by the Coulomb potential, $V(R)\sim 1/R$ in $D=3$. This is not a confining potential, and the renormalization group \cite{Kosterlitz1977} reveals that instantons are unavoidably deconfined. Therefore, the skyrmion number necessarily fluctuates in quantum magnets. A physical consequence is that the topological Hall effect shaped by skyrmions cannot be topologically ordered and fractional, and there is no ``pseudogap'' state with confined instantons below a finite critical temperature.

\subsection{Field equations for hopfion dynamics}\label{secFieldEq3D}

Here we study the classical dynamics of coupled charge and spin degrees of freedom. We will find that the dynamics of hopfions is left out and entirely expressed in quantum processes. Without having to worry about the particle exchange statistics, we will derive the classical equations of motion from a simple bosonic XY-like model. A basic Hamiltonian density of charged particles with field operator $\psi=\psi_0 e^{i\theta}$ and local magnetic moments (spins) $\hat{\bf n}$ is:
\begin{eqnarray}
\mathcal{H} &=& \frac{\kappa_{c}}{2}\left(\partial_{i}\theta\!+\!a_{i}\!+\!qA_{i}\right)^{2}+\frac{\kappa_{s}}{2}(\partial_{i}\hat{n}^{a})^{2}-\mu b_a \hat{n}^a \nonumber \\
&& +\frac{\kappa_{t}}{2}\left(\partial_{i}\chi\!+\!qA_{i}\right)^{2}+\frac{1}{2e^{2}}(\epsilon_{ijk}\partial_{j}a_{k})^{2} \\
&& +C\left(\epsilon_{ijk}\partial_{j}A_{k}-\frac{1}{2}\epsilon_{ijk}\epsilon^{abc}\hat{n}^{a}(\partial_{j}\hat{n}^{b})(\partial_{k}\hat{n}^{c})\right)^{2} \ .  \nonumber
\end{eqnarray}
The particles are coupled as usual to the electromagnetic gauge field ${\bf a}$, and the spins experience Zeeman coupling to the magnetic field $b_i=\epsilon_{ijk}\partial_j a_k$. We introduced an auxiliary gauge field ${\bf A}$ and pinned it to the spin chirality with the $C$ term in order to track the Hopf index of the spin configuration. ${\bf A}$ also couples to the charged particles in order to capture topological Hall effect (THE) and the chiral spin interaction induced by the magnetic field. THE arises because electrons carry both charge and spin; we assume adiabatic limit in which the electron spin is strictly aligned with the background magnetization $\hat{\bf n}$ by a strong Kondo-type coupling, so that a non-zero spin chirality (\ref{Chirality}) generates an effective Lorentz force which acts exactly as an external magnetic field. The factor $q=1/2$ converts the $4\pi$ skyrmion quantum of the chirality flux to the corresponding $2\pi$ magnetic flux quantum. Integrating out ${\bf A}$ reveals the chiral interaction $(\boldsymbol{\nabla}\times {\bf a}) {\bf J}$ between the external magnetic field and spin chirality ${\bf J}$. Together, these physical effects are a topological interaction between magnetic moments and the electromagnetic field. By the virtue of THE, the charged matter field $\theta$ could play the role of a framing or twist field for skyrmion flux loops. Nevertheless, we must couple ${\bf A}$ to a separate dedicated twist field $\chi$ in order to provide independent framing in the charge and spin sectors; $\chi$ physically implements a restoring torsion force against the skyrmion twist. Hopfions in the charge sector are indirectly governed by the Maxwell term for ${\bf a}$.

The Lagrangian density in real time is obtained by promoting the spatial indices $i,j,k$ into upper and lower space-tine indices $\mu,\nu,\lambda$ and relating them via a metric tensor $g=\textrm{diag}(1,-1,-1,-1)$ which implements a sign reversal of all Hamiltonian terms:
\begin{eqnarray}\label{Lagrangian3D}
\mathcal{L} &=& \frac{\kappa_{c}}{2}\left(\partial_{\mu}\theta\!+\!a_{\mu}+qA_{\mu}\right)^{2}+\frac{\kappa_{s}}{2}(\partial_{\mu}\hat{n}^{a})^{2}+\mu b_a\hat{n}^a + \mathcal{L}_{\textrm{B}} \nonumber \\
&& +\frac{\kappa_{t}}{2}\left(\partial_{\mu}\chi\!+\!qA_{\mu}\right)^{2}+\frac{1}{4e^{2}}(\epsilon^{\mu\nu\alpha\beta}\partial_{\alpha}a_{\beta})^{2} \\
&& +\frac{C}{2}\left(\epsilon^{\mu\nu\alpha\beta}\partial_{\alpha}A_{\beta}-\frac{1}{2}\epsilon^{\mu\nu\alpha\beta}\epsilon_{abc}\hat{n}^{a}(\partial_{\alpha}\hat{n}^{b})(\partial_{\beta}\hat{n}^{c})\right)^{2} \nonumber
\end{eqnarray}
Some coupling constants have been adjusted in order to retain the original spatial-index content, and it is understood that the unpaired indices are contracted by squaring, e.g. $(f_{\mu\nu\lambda}g^{\lambda})^{2}\equiv f_{\mu\nu\alpha}g^{\alpha}f^{\mu\nu\beta}g_{\beta}$. Ferromagnetic spin dynamics requires special care as it involves a Berry phase term $\mathcal{L}_{\textrm{B}}$; analogous theory can be written for antiferromagnetic spins represented by a smooth ``staggered magnetization'' $\hat{\bf n}$, but then the Berry phase, Zeeman interaction and the coupling of ${\bf A}$ to charged currents would all be gone. We will derive the field equations from the stationary action condition and express them using the current densities of charge $j_\mu^{\phantom{x}}$, twist $J_\mu^{\phantom{x}}$ and spin $J_\mu^a$
\begin{eqnarray}\label{Currents}
j_{\mu} &=& \partial_{\mu}\theta+a_{\mu}+qA_{\mu} \\
J_{\mu} &=& \partial_{\mu}\chi+qA_{\mu} \nonumber \\
J_{\mu}^{a} &=& \epsilon_{abc}^{\phantom{x}}\hat{n}^{b}\partial_{\mu}^{\phantom{x}}\hat{n}^{c} \ , \nonumber
\end{eqnarray}
as well as the topological currents
\begin{eqnarray}\label{TopCurrents}
J^{\mu\nu} &=& \frac{1}{2}\epsilon^{\mu\nu\alpha\beta}\epsilon_{abc}^{\phantom{x}}\hat{n}^{a}(\partial_{\alpha}^{\phantom{x}}\hat{n}^{b})(\partial_{\beta}^{\phantom{x}}\hat{n}^{c}) \\
\mathcal{I}^{\mu}_{\textrm{s}} &=& \frac{1}{q\phi_{0}^{2}}\epsilon^{\mu\nu\alpha\beta}(\partial_\nu^{\phantom{x}}\chi+qA_{\nu}^{\phantom{x}})\partial_{\alpha}A_{\beta}^{\phantom{x}} \nonumber \\
\mathcal{I}^{\mu}_{\textrm{c}} &=& \frac{1}{(2\pi)^{2}}\epsilon^{\mu\nu\alpha\beta}(\partial_\nu^{\phantom{x}}\theta+a_\nu^{\phantom{x}}+qA_{\nu}^{\phantom{x}})\partial_{\alpha}a_{\beta}^{\phantom{x}} \ . \nonumber
\end{eqnarray}
The spin chirality tensor $J^{\mu\nu}$ characterizes topological textures of the spins in space-time, such as skyrmions, while $\mathcal{I}^\mu_{\textrm{s}}$ and $\mathcal{I}^\mu_{\textrm{c}}$ are the current densities of the Hopf index in the spin and charge sectors respectively. Generally, there is one independent Hopf index for every U(1) gauge transformation in the theory. We will also use the field tensors
\begin{equation}
f_{\mu\nu}=\partial_{\mu}a_{\nu}-\partial_{\nu}a_{\mu} \quad,\quad F_{\mu\nu}=\partial_{\mu}A_{\nu}-\partial_{\nu}A_{\mu} \ .
\end{equation}

Without the Hopf currents, the Lagrangian density constructed so far makes no connection to the Hopf index. Nevertheless, its classical dynamics is very rich and it is useful to explore it before introducing additional couplings to $\mathcal{I}^\mu_{\textrm{s}}$ and $\mathcal{I}^\mu_{\textrm{c}}$. The field equations obtained from the stationary action include current conservation laws ($\theta$, $\chi$ variations)
\begin{equation}\label{ConservEq3D}
\partial_{\mu}j^{\mu}=0 \quad,\quad \partial_{\mu}J^{\mu}=0 \quad ,
\end{equation}
Maxwell equations (${\bf a}$, ${\bf A}$ variations)
\begin{equation}\label{ChargeEq3D}
\kappa_{c}j^{\mu}-\frac{1}{e^{2}}\partial_{\nu}f^{\mu\nu}=0 \ ,
\end{equation}
\begin{equation}\label{HopfEq3D}
\kappa_{c}\,q\,j^{\mu}-2C\left(\partial_{\nu}F^{\mu\nu}+\frac{1}{2}\epsilon^{\mu\nu\alpha\beta}\partial_{\nu}J_{\alpha\beta}\right) = 0 \ ,
\end{equation}
and the spin wave equation ($\hat{\bf n}$ variations)
\begin{eqnarray}\label{SpinEq3D}
&& S\epsilon^{abc}\hat{n}^{b}\partial_{0}\hat{n}^{c}+(\delta^{ab}-\hat{n}^{a}\hat{n}^{b})\biggl\lbrace -\kappa_{s}^{\phantom{x}}\partial_{\mu}\partial^{\mu}\hat{n}^{b}+\mu b^{b} \\
&& \qquad +\frac{3C}{2}\left(F^{\alpha\beta}+\frac{1}{2}\epsilon^{\mu\nu\alpha\beta}J_{\mu\nu}^{\phantom{x}}\right)(\partial_{\alpha}^{\phantom{x}}J_{\beta}^{b}-\partial_{\beta}^{\phantom{x}}J_{\alpha}^{b})\biggr\rbrace =0 \nonumber
\end{eqnarray}
The first term comes from the Berry phase $\mathcal{L}_{\textrm{B}}$, where $S$ is the quantum spin magnitude in the units of $\hbar$. This captures the ferromagnetic spin precession in the internal and external magnetic field.

Perhaps the most interesting aspect of the dynamics in this system is the transfer of gauge flux between the charge and spin sectors. If we express the charge current curl $\epsilon^{\mu\nu\alpha\beta}\partial_{\alpha}j_{\beta}$ with (\ref{ChargeEq3D}) and substitute in it the $j_\mu$ definition (\ref{Currents}), we reveal a correlation between the electromagnetic and spin-chirality gauge fields (assuming the absence of monopoles)
\begin{equation}\label{GaugeTransf}
f_{\mu\nu}+qF_{\mu\nu} = -\frac{1}{\kappa_{c}e^{2}}\partial_{\lambda}\partial^{\lambda}f_{\mu\nu} \approx 0 \ .
\end{equation}
We will mainly consider slowly and smoothly varying electromagnetic fields, and neglect the right-hand side in this formula. At the same time, $F_{\mu\nu}$ shapes the spin chirality $J_{\mu\nu}$. Defining
\begin{equation}
W_{\mu\nu}=F_{\mu\nu}+\frac{1}{2}\epsilon_{\mu\nu\alpha\beta}J^{\alpha\beta} \ ,
\end{equation}
the field equation (\ref{HopfEq3D})
\begin{eqnarray}\label{HopfEq3Db}
\partial^{\nu}W_{\mu\nu} = \frac{q\kappa_{c}}{2C}j_{\mu}
\end{eqnarray}
can be easily solved in the absence of charge currents with an arbitrary vector field $w_\mu$:
\begin{equation}
j_\mu=0 \quad\Rightarrow\quad W_{\mu\nu}=\epsilon_{\mu\nu\alpha\beta}\partial^{\alpha}w^{\beta} \ .
\end{equation}
In these conditions ($q=1/2$),
\begin{equation}
J_{\mu\nu} = -\frac{1}{2}\epsilon_{\mu\nu\alpha\beta}F^{\alpha\beta}+\partial_{\mu}w_{\nu}-\partial_{\nu}w_{\mu}
\end{equation}\vspace{-0.25in}
\begin{equation}
\frac{1}{2}\hat{n}^{a}(\partial_{\mu}^{\phantom{x}}J_{\nu}^{a}-\partial_{\nu}^{\phantom{x}}J_{\mu}^{a}) = -F_{\mu\nu}^{\phantom{x}}+\epsilon_{\mu\nu\alpha\beta}^{\phantom{x}}\partial^{\alpha}w^{\beta} \nonumber
\end{equation}
relate the spin currents and chirality to the internal electromagnetic field $F_{\mu\nu}$, or equivalently to the external electromagnetic field $f_{\mu\nu}$ by the virtue of (\ref{GaugeTransf}). Evidently, $w_\mu$ represents the independent spin wave content. The dynamics of spin waves is governed in the field equation (\ref{SpinEq3D}). After some algebra, one finds that spin waves in the absence of charge currents propagate according to:
\begin{eqnarray}\label{SpinWave3D}
&& S\epsilon^{abc}\hat{n}^{b}\partial_{0}\hat{n}^{c}-(\delta^{ab}-\hat{n}^{a}\hat{n}^{b})(\kappa_{s}^{\phantom{x}}\partial_{\mu}\partial^{\mu}\hat{n}^{b}-\mu b^{b}) = \nonumber \\
&& \qquad = -3CF^{\mu\nu}\left(\partial_{\mu}^{\phantom{x}}J_{\nu}^{a}-\frac{1}{2}\hat{n}^{a}\epsilon_{\mu\nu\alpha\beta}J^{\alpha\beta}\right) \ .
\end{eqnarray}
The term $\epsilon_{\mu\nu\alpha\beta}F^{\mu\nu}J^{\alpha\beta}$ on the right-hand side is the chiral interaction in disguise,
\begin{eqnarray}
\frac{1}{4}\epsilon_{\mu\nu\alpha\beta}F^{\mu\nu}J^{\alpha\beta} &=& B^{i}\frac{1}{2}\epsilon_{ijk}^{\phantom{x}}\epsilon^{abc}\hat{n}^{a}(\partial_{j}^{\phantom{x}}\hat{n}^{b})(\partial_{k}^{\phantom{x}}\hat{n}^{c}) \nonumber \\
&& -E^{i}\epsilon^{abc}\hat{n}^{a}(\partial_{0}^{\phantom{x}}\hat{n}^{b})(\partial_{i}^{\phantom{x}}\hat{n}^{c}) \ , \nonumber
\end{eqnarray}
where $E^i$ and $B^i$ are the spin-chirality electric and magnetic fields respectively, correlated with the external electromagnetic field by (\ref{GaugeTransf}). Consequently, the spin chirality can be induced by an external magnetic field and moved around by an electric field. This affects the dispersion of spin waves in the complicated manner captured by (\ref{SpinWave3D}). Qualitatively, the propagation of spin waves can mimic the features of particle propagation in magnetic fields. The presence of charge density and currents further complicates the equations by introducing sources to $W_{\mu\nu}$ in (\ref{HopfEq3Db}). 

In order to explore the effects of hopfion dynamics in the field equations, we need additional real-time Lagrangian density terms related to hopfions. When topological order is possible, a Lagrangian term of the form $j_\mu \mathcal{J}^\mu$ implements the conservation $\partial_\mu \mathcal{J}^\mu = 0$ of the topological defect current at the level of field equations. The topological conservation law formally arises due to abundant phase $\theta$ fluctuations in the particle current $j_\mu \sim \partial_\mu\theta + a_\mu$, with a remnant $a_\mu \mathcal{J}^\mu$ given by the Chern-Simons or generalized background-field coupling in the context of $\pi_{d-1}(S^{d-1})$ homotopy groups. A consequence of such a coupling is the attachment of charge to topological defects, characteristic for topologically ordered phases. A singular topological defect of a spin vector field $\hat{\bf n}$ in $d=3$ dimensions is a hedgehog, and indeed topologically ordered quantum liquids of hedgehogs are possible. However, hopfions are $d=3$ classical topological defects without singularities, similar to skyrmions in $d=2$. A topological current of hopfion singularities can be constructed only in $d=4$ spatial dimensions, i.e. $D=d+1=5$ dimensional space-time:
\begin{equation}
\mathcal{J}^{\mu} \sim \epsilon^{\mu\nu\alpha\beta\gamma}\partial_{\nu}(A_{\alpha}\partial_{\beta}A_{\gamma}) \equiv \partial_\nu \mathcal{I}^{\mu\nu} \xrightarrow{D=4}{} \partial_\nu \mathcal{I}^\nu \ .
\end{equation}
Its projection to the $D=4$ real-world, with one index stripped, is the divergence $\partial_\nu \mathcal{I}^\nu$ of the Hopf index current density (\ref{TopCurrents}). We cannot construct a topological Lagrangian term $j_\mu \mathcal{J}^\mu$ in $D=4$, but
\begin{equation}
\mathcal{L}_{i}^{\phantom{x}} = K_{i}^{\textrm{c}}\, (\partial_\mu^{\phantom{x}} \mathcal{I}^\mu_{\textrm{c}})^2 + K_{i}^{\textrm{s}}\, (\partial_\mu^{\phantom{x}} \mathcal{I}^\mu_{\textrm{s}})^2
\end{equation}
is permissible and shapes the dynamics of instantons. Hopf index conservation amounts to $\partial_\mu \mathcal{I}^\mu=0$, so $\mathcal{L}_i$ regulates the amount of fluctuations in which the topological charge conservation is locally violated.

Unfortunately, the instanton Lagrangian has no impact on the field equations. The $A_{\mu}$ variations of $\mathcal{L}_{i}$ cannot introduce any correction to (\ref{HopfEq3D}) because $\partial_{\mu}\mathcal{I}^{\mu}$ is a real scalar lacking the structure that could support topological singularities. For example, in the spin sector with $a_\mu=0$:
\begin{eqnarray}
\frac{\delta\mathcal{L}_{i}}{\delta A_{\sigma}} &=& \frac{2K_{i}}{\phi_{0}^{4}}(\partial_{\mu}\mathcal{I}^{\mu})\frac{\delta}{\delta A_{\sigma}}\epsilon^{\alpha\beta\gamma\delta}\Bigl(\partial_{\alpha}(A_{\beta}\partial_{\gamma}A_{\delta})\Bigr) \nonumber \\
&\to& \frac{4K_{i}}{\phi_{0}^{4}}\epsilon^{\sigma\alpha\beta\gamma}A_{\alpha}(\partial_{\beta}\partial_{\gamma}\partial_{\mu}\mathcal{I}^{\mu})=0
\end{eqnarray}
(note that integrations-by-parts were carried out assuming the absence of monopoles, $\epsilon^{\mu\alpha\beta\gamma} \partial_\alpha\partial_\beta A_\gamma = 0$). While instantons apparently do not influence the classical dynamics, they do leave a signature in the quantum noise.

Is there any other hopfion mechanism for the phenomenology of topological order? A method for analyzing the topological ground state degeneracy, presented in Section \ref{secTopDeg}, finds that no topological order of hopfions is possible in $d=3$ spatial dimensions. Nevertheless, we can attempt to obtain some form of charge-hopfion attachment in the $d=3$ field equations by considering another addition to the Lagrangian density,
\begin{equation}
\mathcal{L}_{t}^{\phantom{x}} = -K_{t}^{\textrm{s}}\rho\, j_\mu^{\phantom{x}} \mathcal{I}^\mu_{\textrm{s}} -K_{t}^{\textrm{c}}\rho (\partial_\mu^{\phantom{x}}\chi+qA_\mu^{\phantom{x}}) \mathcal{I}^{\mu}_{\textrm{c}} \nonumber \ .
\end{equation}
This introduces various complications in the field equations, and a notable correction to (\ref{ChargeEq3D}) which takes a simple form in the limit $a_\mu\to 0$:
\begin{equation}
j^{\mu}=\frac{K_{t}^{\textrm{s}}\rho}{\kappa_{c}}\mathcal{I}^{\mu}_{\textrm{s}} \ .
\end{equation}
This is problematic for several reasons. The charge density $j^0$ is effectively bound to the Hopf index density $\mathcal{I}_0$ and spread-out over an extended region in space where $\mathcal{I}_0\neq 0$. There is no singular point to which the charge can attach. Since fluctuations can smoothly vary the spatial distribution of $\mathcal{I}_0$, no local quasiparticles can carry a quantized fractional amount of charge despite the quantization of the total Hopf index. Hence, we cannot expect topological order from $\mathcal{L}_{t}$. We also do not expect the appearance of $\mathcal{L}_{t}$ in the effective theories of realistic systems. Topological terms are normally generated by a microscopic Berry curvature, which inserts the static topological current $\mathcal{J}_\mu$ into the Hamiltonian instead of $\mathcal{I}_\mu$. If, alternatively, a certain two-particle interaction were able to generate $\mathcal{L}_{t}$, then the amount $\nu$ of attached charge per topological flux quantum would scale in proportion to the particle density $\rho$; this is also incompatible with topological order because the co-mobility of particles and their topological defects requires an incompressible state with a constant and quantized filling factor $\nu$.

\subsection{Thermodynamic signatures and the chiral anomaly}\label{secHopfCorr}

The confined phase of hopfion instantons can be physically identified despite lacking obvious signatures in classical observables. The key to this is quantum noise. Instanton confinement is sharply characterized by a generalization of the Wilson loop operator
\begin{equation}\label{Wilson1}
\mathcal{C}(S^3) = \oint\limits_{S^3} d^3 x\, \hat{\eta}_\mu \mathcal{I}^\mu \ ,
\end{equation}
which is defined on a 3-sphere $S^3$ embedded in $D=4$ space-time; $\hat{\eta}_\mu$ is the unit-vector locally perpendicular to the sphere. According to the definition of the Hopf index (\ref{Hopf2}) and its current density $\mathcal{I}_\mu$ (\ref{TopCurrents}), this counts the Hopf-instanton charge, i.e. the number of instantons minus the number of anti-instantons inside the space-time volume $B^4$ bounded by the closed ``surface'' $S^3$. The variance of the random outcomes in the quantum measurements of $\mathcal{C}(S^3)$ is a thermodynamic characterization of the instanton phase \cite{Nikolic2023a}:
\begin{equation}\label{Wilson2}
\textrm{Var}\,\mathcal{C}(S^3) \xrightarrow{S^3\to\infty}{}
  \begin{cases}
    a S^3 &,\quad \textrm{confined phase} \\
    b B^4 & ,\quad \textrm{deconfined phase}
\end{cases} \ .
\end{equation}
If the instantons are confined, then they contribute to (\ref{Wilson1}) only within a finite distance from the ``surface'' $S^3$ given by their confinement length $\lambda$; every instanton is compensated further away, so that $\textrm{Var}\,\mathcal{C}(S^3)$ grows in proportion to the ``area'' of the boundary $S^3$. Otherwise, topological instanton fluctuations matter in the entire space-time and $\textrm{Var}\,\mathcal{C}(S^3)$ scales in proportion to the bounded volume $B^4$.

The Hopf index is conserved in the confined instanton phase. This means in practice that the noise spectrum of the Hopf index fluctuations is depleted at low frequencies \cite{Nikolic2023a}. Quantum noise quickly fills up the low-frequency spectrum upon entering the deconfined phase. The main challenge for the detection of this spectrum change is whether one can indirectly measure the Hopf index of the system. In the worst case scenario, we don't have any other probe but the heat capacity. The onset of low-frequency noise is naively expected to boost the heat capacity across the critical temperature for instanton deconfinement \cite{Nikolic2023a}. In some cases, it might be possible to indirectly probe the Hopf index fluctuations with quantum noise measurements. Such experiments would be certainly difficult for a variety of reasons, but they could, for example, exploit the chiral quantum anomaly in materials with a Dirac and Weyl electron spectrum. There is also a general possibility that a confined instanton phase could be identified by unconventional transport properties or observable local fluctuations, but it is not presently clear how and it could be system-specific.

The chiral anomaly of quantum electrodynamics is, actually, closely related to the Hopf index physics. It relates the chiral current non-conservation to the electromagnetic field fluctuations:
\begin{equation}\label{Anomaly}
\partial_{\mu}j^{5\mu}=\frac{1}{16\pi^{2}}\,\epsilon^{\mu\nu\alpha\beta}f_{\mu\nu}f_{\alpha\beta}
\end{equation}
when the Dirac quasiparticle mass $m$ vanishes \cite{Adler1969, Bell1969, Fujikawa1980, Nielsen1981, Nielsen1983, Peskin1995, ZinnJustin2001}. This result was originally obtained in the perturbation theory by introducing gauge-invariant regularizations of three-vertex one-loop Feynman diagrams \cite{Adler1969, Bell1969}. Different regularization schemes, each complicated in its own way, point to the same conclusion. The physics of this is peculiar because a massless Dirac theory possesses the chiral symmetry and conserves the chiral current $j^{5\mu}$ at the level of field equations. It is the quantum fluctuations which necessarily violate the chiral symmetry, for the reasons which are not made very transparent in the perturbation theory. For our purposes, we immediately recognize the hopfion current
\begin{equation}
\mathcal{I}^\mu = \frac{1}{(2\pi)^2}\,\epsilon^{\mu\nu\alpha\beta} a_{\nu}\partial_{\alpha}a_{\beta}
\end{equation}
of the electromagnetic field in the charge sector ($\phi_0=2\pi$) on the right-hand side of (\ref{Anomaly}):
\begin{equation}\label{Anomaly2}
\partial_{\mu}j^{5\mu} = \frac{1}{16\pi^{2}}\,\epsilon^{\mu\nu\alpha\beta}f_{\mu\nu}f_{\alpha\beta} = \partial_\mu \mathcal{I}^\mu \ .
\end{equation}
This provides new insight: the local events which violate the chiral conservation law are bound to the Hopf instantons. While particles are attached to topological defects in topologically ordered phases, here we have attachment of particle creation/annihilation events to the analogous events for topological defects. Many interesting questions arise from this similarity between topological order and quantum anomaly. Are both of them manifestations of correlations caused by instanton confinement? If yes, then can a quantum anomaly be fractionalized? Is the chiral anomaly a relativistic variant of the non-relativistic topological order?

We can gain valuable insight into these questions from the following transparent physical picture of the chiral anomaly. For simplicity, and in order to relate to the traditional context, let us scrutinize quantum electrodynamics whose real-time Lagrangian density is
\begin{equation}\label{QEDLagrangian}
\mathcal{L}=\psi^{\dagger}(i\mathcal{D}-\gamma^{0}m)\psi-\frac{1}{4}f_{\mu\nu}f^{\mu\nu} \ .
\end{equation}
$\gamma^\mu$ are the standard Dirac matrices. We must regularize this theory on a lattice in order to provide flux quantization for the gauge field $a_\mu$, a prerequisite for the Hopf index quantization. We first introduce a regulator which breaks the Lorentz invariance at high energies,
\begin{equation}\label{CLreg}
\mathcal{D} = D_{0}+\frac{i\gamma^0}{2M}(D_{i}D^{i}+\mathcal{A}_i\mathcal{A}^i) \quad,\quad D_{\mu}=\partial_{\mu}+ia_{\mu}+i\mathcal{A}_{\mu}
\end{equation}
featuring a large mass $M$ and a static background SU(2) gauge field
\begin{equation}
\mathcal{A}_{0}=0 \quad,\quad \mathcal{A}_{i}=-M\gamma_{i} \ .
\end{equation}
The usual relativistic Dirac theory $\mathcal{L} = \bar{\psi}(i\cancel{D}-m)\psi$ with $\bar{\psi}=\psi^{\dagger}\gamma^{0}$ and $\cancel{D}=\gamma^{\mu}(\partial_{\mu}+ia_{\mu})$ is recovered in the $M\to\infty$ limit. This regularization is physically transparent and converts the QED into the continuum limit description of a condensed matter system with a Berry curvature. It is now easy to construct a lattice Hamiltonian
\begin{equation}
H = -t \sum_{\bf r}\sum_{\mu} \psi_{\bf r}^{\dagger} e^{i(a_{{\bf r},\mu} + \mathcal{A}_{{\bf r},\mu})} \psi_{{\bf r}+\hat{\boldsymbol{\mu}}}^{\phantom{\dagger}} + h.c. + \Delta H \ ,
\end{equation}
where ${\bf r}$ are lattice sites and $\mu$ are the directions to nearest-neighbor sites. The continuum limit of $H$ reproduces the Hamiltonian of the theory (\ref{QEDLagrangian}). Some spurious two-fold degenerate Dirac nodes might arise at the corners of the first Brillouin zone due to time-reversal and lattice symmetries; these are easily gapped-out to high energies by appropriate chirality-altering hopping terms $\Delta H$, without disturbing the Dirac node at the $\Gamma$ point.

In the pristine QED with $M\to\infty$, the structure:
\begin{eqnarray}
\beta &=& \gamma^{0}=\gamma^{0\dagger}=\left(
  \begin{array}{cc}
    1 & 0\\
    0 & -1
  \end{array}\right) = \tau^{z}\otimes 1
\\
\alpha^{k} &=& \alpha^{k\dagger}=\gamma^{0}\gamma^{k}=\left(
  \begin{array}{cc}
    0 & \sigma^{k}\\
    \sigma^{k} & 0
  \end{array}\right) = \tau^{x}\otimes\sigma^{k} \nonumber
\end{eqnarray}\vspace{-0.21in}
\begin{equation}
\gamma^{5}=\gamma^{5\dagger}=i\gamma^0\gamma^1\gamma^2\gamma^3=\left(
  \begin{array}{cc}
    0 & 1\\
    1 & 0
  \end{array}\right)=\tau^{x}\otimes1 \nonumber
\end{equation}
expressed in terms of Pauli matrices $\sigma^a$ and $\tau^a$ produces the Hamiltonian
\begin{equation}\label{DiracHamiltonian}
H=-i\alpha^{k}(\partial_{k}+ia_{k})+\gamma^{0}m \;\xrightarrow[a_{\mu}\to0]{m\to0}\; \alpha^{k}p_{k} \ .
\end{equation}
Massless $m=0$ Dirac electrons have the spectrum
\begin{equation}\label{DiracSpectrum}
E_{\tau\sigma{\bf p}}=\tau\sigma|{\bf p}| \ .
\end{equation}
Since the matrices $\sigma^a$ couple to spin, the eigenvalue $\sigma=\pm1$ of $\boldsymbol{\sigma}\hat{{\bf p}}$ is helicity (alignment or anti-alignment between the electron's spin and momentum). The $\tau=\pm1$ eigenvalue of $\tau^{x}$ (as well as $\gamma^{5}$) is chirality. The band index $s=\tau\sigma$ is the product of helicity and chirality. The continuum-limit regularization (\ref{CLreg}) only adds
\begin{equation}
\delta H = \gamma^{0}\frac{(p_{i}+a_{i})^{2}}{2M}
\end{equation}
to the Dirac Hamiltonian (\ref{DiracHamiltonian}). This preserves the particle-hole symmetry, but explicitly breaks the chiral symmetry of QED at high energies. Photon absorption and emission that violates chiral current conservation is also made possible, with an amplitude proportional to $M^{-1}$. Note that removing the factor of $\gamma^0$ to restore the chiral symmetry is not permissible because, without it, the spectrum would possess a large Fermi surface that coexists with the Dirac node.

The regularization of QED plays an important role in the emergence of the chiral anomaly, and simultaneously reveals the origin of the anomaly in condensed matter systems. However, the physical picture which we will now construct invokes the regularization only in very subtle ways. The chiral anomaly is a correlation between the changes of chirality and the topological Hopf index. The classical field equations are derived from small and smooth variations of the fields, which by definition cannot capture the changes of a topological invariant. This is why the field equations fail to reveal the quantum anomaly. We will show next that as long as there is some microscopic mechanism (provided by the regularization) for the electromagnetic field fluctuations to alter electrons' chirality, the topological considerations alone link the chirality fluctuations to the Hopf index.

\begin{figure}
\includegraphics[width=2.0in]{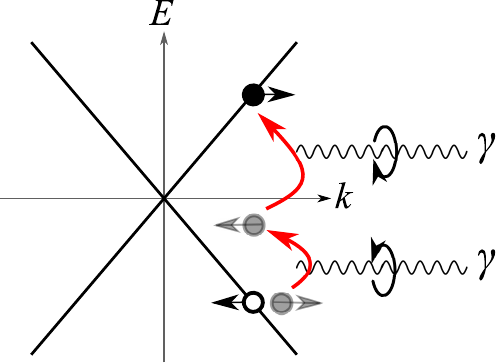}
\caption{\label{ChiralAnomaly}Two-photon absorption behind the chiral quantum anomaly, analogous (after time reversal) to the neutral pion decay $\pi^0\to\gamma+\gamma$. A massless Dirac electron in the occupied valence band has a well-defined chirality, say $\tau=-1$. This fixes its helicity to $\sigma=1$. By absorbing a photon ($\gamma$) with right-handed circular polarization, the electron flips its spin and gains some energy, leaving a hole in the valence band. This removal of $\tau=-1$ chirality from the valence band is equivalent to the injection of $\tau=1$. The absorption of a second right-handed photon, moving in the opposite direction from the first photon, flips the electron's spin again. The final particle excitation has chirality $\tau=1$ because it has the same helicity $\sigma=1$ (same spin and momentum) as the original electron but lives in a different band ($\sigma\tau$ is the band index, see Eq.\ref{DiracSpectrum}). Therefore, the particle-hole pair, created by absorbing a spinless photon pair, has no net angular momentum and carries the net chirality $\tau=2$. The pristine massless QED forbids the above second photon process in which the electron's chirality changes. However, the required physical regularization necessarily introduces a chirality-changing photon absorption in order to protect the gauge invariance.}
\end{figure}

The particle-hole excitations that carry a neutral chiral current
\begin{equation}
j^\mu = \bar{\psi}\gamma^{\mu}\psi = 0 \quad,\quad
j^{5\mu} = \bar{\psi}\gamma^{5}\gamma^{\mu}\psi \neq 0
\end{equation}
necessarily have a zero net spin due to the spin-momentum locking in the conduction and valence bands. Therefore, the conservation of charge, momentum and angular momentum permits only the zero-spin photon pairs to create such excitations, assuming that the chirality need not be conserved, see Fig.\ref{ChiralAnomaly}. It turns out that the zero-spin photon fluctuations also change the Hopf index of the electromagnetic field. This is easy to show by expressing the gauge field operator ${\bf a}$ in the Coulomb gauge using the creation $a_{{\bf k},p}^\dagger$ and annihilation $a_{{\bf k},p}^{\phantom{\dagger}}$ operators of photons with momentum ${\bf k}$ and linear polarization $p=\pm$:
\begin{equation}
{\bf a}({\bf r},t)=\sum_{p}\!\int\!\frac{d^{3}k}{(2\pi)^{3}}\frac{\hat{\boldsymbol{\epsilon}}_{1,{\bf k}}^{p}}{\sqrt{2\omega_{k}}}\Bigl\lbrack a_{{\bf k},p}^{\phantom{x}}e^{i({\bf k}{\bf r}-\omega_{k}t)}+h.c.\Bigr\rbrack
\end{equation}
The two orthogonal polarization vectors $\hat{\epsilon}_{1,{\bf k}}^p$ and $\hat{\epsilon}_{2,{\bf k}}^p$ satisfy $\hat{\boldsymbol{\epsilon}}_{1,{\bf k}}^{p}\times\hat{\boldsymbol{\epsilon}}_{2,{\bf k}}^{p}=\hat{{\bf k}}$ and $\hat{\boldsymbol{\epsilon}}_{i,{\bf k}}^{+} \hat{\boldsymbol{\epsilon}}_{i,{\bf k}}^{-}=0$, and the photon dispersion is $\omega_{k}=|{\bf k}|$. Substituting into (\ref{Anomaly}) gives us:
\begin{eqnarray}\label{TwoPhotons}
&& \frac{1}{16\pi^{2}}\int d^{3}r\,\epsilon^{\mu\nu\alpha\beta}f_{\mu\nu}f_{\alpha\beta} = \frac{\partial N_{\textrm{Hopf}}}{\partial t} \\
&& ~ \to -\frac{1}{2\pi^{2}}\!\sum_{p,p'}\!\int\!\!\frac{d^{3}k}{(2\pi)^{3}}\frac{\omega_{k}}{2}\hat{\boldsymbol{\epsilon}}_{1,{\bf k}}^{p}\hat{\boldsymbol{\epsilon}}_{2,-{\bf k}}^{p'}a_{{\bf k},p}^{\phantom{x}}a_{-{\bf k},p'}^{\phantom{x}}e^{-2i\omega_{k}t}\!+\!h.c. \nonumber \\
&& ~ = \frac{1}{8\pi^{2}}\frac{\partial}{\partial t}\sum_{\sigma}\!\int\!\!\frac{d^{3}k}{(2\pi)^{3}}\,\sigma\,b_{{\bf k},\sigma}^{\phantom{x}}b_{-{\bf k},\sigma}^{\phantom{x}}e^{-2i\omega_{k}t}\!+\!h.c. \nonumber \\
&& ~ \propto \frac{\partial N^{5}}{\partial t} \ . \nonumber
\end{eqnarray}
Behind the arrow, we retained only the terms which describe the photon absorption and emission. At the end, we switched to the circular polarization field operators $b_{{\bf k},\sigma}$, where $\sigma$ is photon's helicity: $\sigma=+1$ indicates that photon's spin is aligned with its momentum ${\bf k}$, and $\sigma=-1$ means anti-alignment. The final proportionality follows from the fact that whenever two photons of zero net momentum and spin are annihilated with $b_{{\bf k},\sigma}^{\phantom{x}}b_{-{\bf k},\sigma}^{\phantom{x}}$, i.e. successfully absorbed, their energy $2\omega_k$ is used to create an electron-hole pair with net spin zero, which carries a non-zero chirality in the massless Dirac spectrum. Therefore, the processes of quantum electrodynamics make it impossible to change the electron chirality without changing the Hopf index of the gauge field.

\begin{figure}
\subfigure[{}]{\includegraphics[width=2.2in]{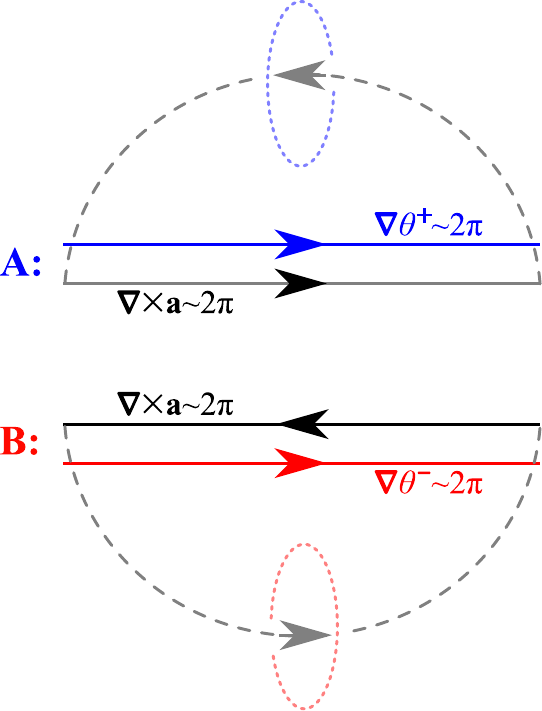}}\hspace{0.1in}
\raisebox{0.7in}{\subfigure[{}]{\includegraphics[width=1.0in]{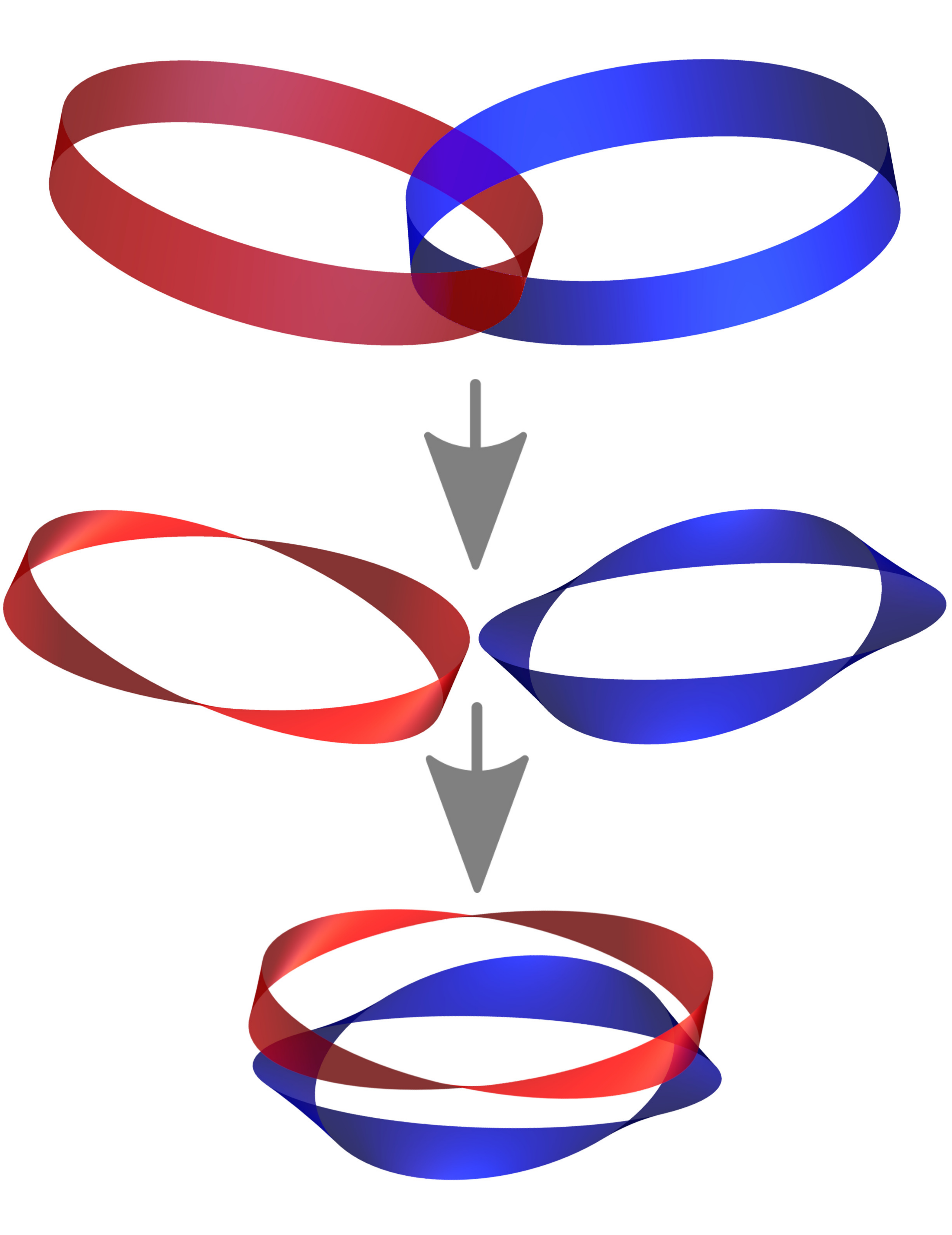}}}
\caption{\label{chiralHopf}The Hopf index and chiral anomaly. (a) Create two unlinked flux-quantum loops A and B (black lines, $\boldsymbol{\nabla}\times{\bf a}$) and accompany each by a $2\pi$ phase twist. Let the twists of A and B carry particle (blue) and hole (red) current respectively governed by the Lagrangian (\ref{QED0}); both currents have the same chirality. Dashed and dotted segments are out of sight, e.g. at infinity. The dotted colored loops represent far-away vortex singularities of $\theta^+$ (blue) and $\theta^-$ (red); they are interlinked with one flux loop each. In this configuration, each twisted flux loop contributes 1 to the Hopf index, totaling the change $\Delta N_{\textrm{Hopf}}=2$ according to (\ref{QEDHopf}). When the straight flux segments are smoothly deformed and pushed to overlap, their gauge flux quanta and charge currents cancel out. What remains is two units of the chiral current, $\boldsymbol{\nabla}\phi=\boldsymbol{\nabla}(\theta^++\theta^-)\sim 2\times 2\pi$. Therefore, $\Delta N^5 = \Delta N_{\textrm{Hopf}} = 2$. (b) An equivalent local process. Creating a pair of interlinked flux loops gives $\Delta N_{\textrm{Hopf}}=2$. A chiral current $\Delta N^5=2$ is left behind at the end of the process.}
\end{figure}

In order to find the exact proportionality constant in (\ref{TwoPhotons}), consider a qualitative description of massless Dirac particles coupled to a gauge field,
\begin{eqnarray}\label{QED0}
\mathcal{L}_{\textrm{QED}_{0}} &=& \frac{\kappa}{2}(\partial_{\mu}\theta^{+}+a_{\mu})^{2}+\frac{\kappa}{2}(\partial_{\mu}\theta^{-}-a_{\mu})^{2} \\
&=& \frac{\kappa}{4}(\partial_{\mu}\theta+2a_{\mu})^{2}+\frac{\kappa}{4}(\partial_{\mu}\phi)^{2} \ . \nonumber
\end{eqnarray}
It is not crucial to capture the fermionic statistics of charged matter fields $\theta^\pm$, we only must consider them uncondensed. By construction, $\theta^\pm$ carry opposite charge $\pm 1$ with respect to the U(1) gauge field $a_\mu$, and will represent the massless Dirac particles and holes which carry the same helicity $\tau$. By particle-hole symmetry, $\theta^\pm$ simultaneously represent particle (charge 1) currents of opposite helicity. Defining
\begin{equation}\label{ChargeHelicity}
\theta=\theta^{+}-\theta^{-}\quad,\quad\phi=\theta^{+}+\theta^{-}
\end{equation}
allows us to separately track the charge $j^\mu$ and chiral $j^{5\mu}$ currents
\begin{equation}
j_\mu^{\phantom{x}} \sim \partial_\mu^{\phantom{x}}\theta+2a_\mu^{\phantom{x}} \quad,\quad j^{5}_{\mu} \sim \partial_\mu^{\phantom{x}}\phi \ .
\end{equation}
This theory has only one gauge transformation
\begin{equation}
\theta^{\pm}\to\theta^{\pm}\pm\lambda\quad,\quad a_{\mu}\to a_{\mu}-\partial_{\mu}\lambda \ ,
\end{equation}
so we expect only one Hopf index:
\begin{equation}\label{QEDHopf}
N_{\textrm{Hopf}}(t) = \frac{1}{(2\pi)^{2}}\oint\limits _{S^{3}}d^{3}x\,\epsilon^{ijk}(\partial_{i}\theta+2a_{i})\partial_ja_{k} \ .
\end{equation}
The normalization is fixed by the requirement that the $2\pi$ winding of either $\theta^+$ or $\theta^-$ along a flux loop give a unit Hopf index. According to (\ref{HopfLink}), the inter-linking of flux loops gives only even values of the Hopf index. Therefore, changing $N_{\textrm{Hopf}}$ by $\pm1$ necessarily excites charge (twist) currents. A neutral chiral current ($j^{\phantom{x}}_\mu=0$, $j^{5}_\mu\neq 0$) carries at least $N^5=2$ helicity which has to be matched to some $N_{\textrm{Hopf}}\neq 0$; it turns out this is $N_{\textrm{Hopf}}=2$. Fig.\ref{chiralHopf} explains the basic process of creating a particle-hole excitation characterized by $\Delta N^5= \Delta N_{\textrm{Hopf}} = 2$, while all other processes can be derived from it by adding low-energy Hopf-preserving fluctuations. This reproduces the chiral anomaly
\begin{equation}\label{Anomaly3}
\Delta N^{5} = \Delta N_{\textrm{Hopf}} \ .
\end{equation}

From the current perspective, chiral anomaly requires the conservation of the combined Hopf index (\ref{Hopf2}) of the gauge and twist (matter) fields. This is putatively realized below the instanton deconfinement temperature $T_{\textrm{c}2}$. The correlation (\ref{Anomaly3}) is lost at $T>T_{\textrm{c}2}$, but reflects via quantum noise the deconfinement of pure gauge field instantons at a lower critical temperature $T_{\textrm{c}1}$. Hence, the Hopf index (\ref{Hopf1}) of the gauge field alone is conserved at $T<T_{\textrm{c}1}$, and its fluctuations are mirrored by chiral currents at $T_{\textrm{c}1}<T<T_{\textrm{c}2}$. The latter can be further modified if interactions produce additional effective gauge fields \cite{Galitski2021} which carry a Hopf index. Note that fermionic matter cannot screen out the magnetic flux loops, but it can affect the action cost of instantons by suppressing their electric field fluctuations at sufficiently low frequencies. The regime $T<T_{\textrm{c}1}$ can be probably accessed with a sufficiently low fermion density (i.e. chemical potential close to the Dirac node) because fermions can maintain phase coherence across the mean inter-particle distance (as in quantum Hall states).

\begin{figure}
\subfigure[{}]{\includegraphics[width=3.4in]{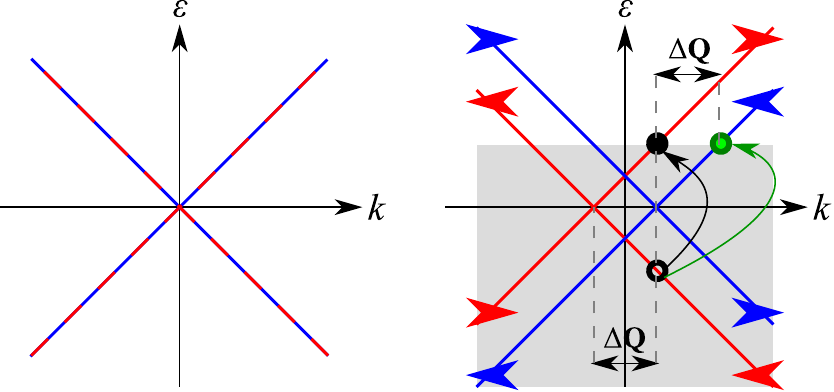}}
\subfigure[{}]{\includegraphics[width=3.4in]{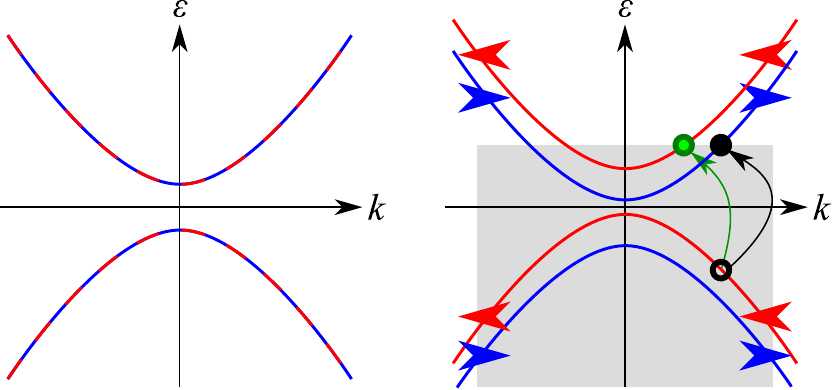}}
\caption{\label{Qspectra}Particle-hole excitations with and without net spin are distinguished by different energy/momentum transfers from photons when the spin degeneracy (left) is lifted (right) by a magnetic field or spin-orbit coupling. (a) The case of Dirac or Weyl spectrum: A spin-flip particle-hole excitation with the minimum threshold energy $2\epsilon_{\textrm{f}}$ on the Fermi sea (shaded) occurs with no momentum transfer, while the chiral excitations created from the quantum anomaly fluctuations require momentum transfer equal to the momentum separation $\Delta{\bf Q}$ between the intrinsic or generated Weyl nodes. The relativistic spin-momentum-locked spectrum makes $\Delta{\bf Q}$ independent of the Fermi energy $\epsilon_{\textrm{f}}$, but sensitive to a magnetic field. (b) The conventional bands without spin-momentum locking: The same-spin threshold transitions still require momentum transfers, but the transferred momentum also depends on the energy transfer.}
\end{figure}

In principle, the correlation between currents and the Hopf index can be exploited to detect instanton deconfinement. The chiral anomaly is just a specific manifestation of this correlation by Dirac electrons, least difficult to detect, but converting chiral to charge currents requires finite momentum transfers between photons and electrons. Since the total momentum of the electron gas is not conserved, the measurements of charge currents will exhibit quantum randomness which is correlated with the gauge field dynamics. We are interested in the resulting quantum noise spectrum. The noise contributed by virtual zero-spin particle-hole pairs is suppressed at low frequencies in a confined instanton phase. Fig.\ref{Qspectra} illustrates that zero-spin particle-hole fluctuations can be distinguished from the non-zero-spin fluctuations by momentum and energy transfer when the magnetic field or spin-orbit coupling lifts the spin degeneracy of the electron spectrum. This is, of course, a very difficult measurement for several reasons. It requires enough sensitivity and a subtraction of other well-understood sources of noise. And, the probability of photon-electron momentum transfer is suppressed by a negative power of the speed of light. Natural photons carry very small momenta, so it is really an emergent gauge field (with a speed of light comparable to a typical electron speed) that would give this method a better chance. One could devise a dielectric constant superlattice to impart umklapp scattering on the (emergent) photons that matches the needed momentum transfers. A layered Weyl heterostructure \cite{Burkov2011a} might just serve this purpose, and tuning the magnetic field can perhaps bring the momentum transfer into resonance. While measuring the noise spectrum would be the least ambiguous detection method, there are other options. An instanton deconfinement transition should theoretically boost the specific heat \cite{Nikolic2023a}, and it may also be accompanied by other observable properties (transport, for example).

We end this discussion by envisioning a mechanism for the fractional quantization of the chiral anomaly,
\begin{equation}\label{Anomaly4}
\Delta N^{5} = \nu\, \Delta N_{\textrm{Hopf}} \quad,\quad \nu=\frac{p}{q} \ .
\end{equation}
This requires a fractionalization of chirality without spoiling its classical conservation. Since massless Dirac electrons carry both charge and chirality, charge fractionalization will generally imply some form of chirality fractionalization. Charge fractionalization in $d=3$ spatial dimensions manifestly occurs in nuclear matter (irrespective of the microscopic mechanism), and can occur by the topological order involving hedgehogs or monopoles \cite{Nikolic2019}. An effective theory that captures the dynamics of fractional quasiparticles (partons) always requires an emergent gauge field $\widetilde{\bf a}$. If $\widetilde{\bf a}$ experiences extremely strong fluctuations, it necessarily recombines the fractional quasiparticles into electrons and stabilizes a conventional phase of matter. But, a Coulomb phase of $\widetilde{\bf a}$ allows the partons to be deconfined because they carry unit-charge relative to $\widetilde{\bf a}$ (note that the physical U(1) gauge field may be gapped in order to accommodate monopoles for the topological order). Since the Coulomb phase hosts flux loops characterized by a conserved Hopf index of $\widetilde{\bf a}$, the fluctuations of this Hopf index are correlated with the fluctuations of the \emph{fractionalized} quasiparticle's chirality. A rational quantization of the ``filling factor'' $\nu$ is inherited from the parent topological order. Disregarding questions about the feasibility of $d=3$ fractionalization in materials, the main theoretical issue is whether the gap of the topologically ordered phase can be disconnected from the usual chirality-mixing mechanism that gives Dirac electrons a mass. Naively, the classical chirality conservation should be able to survive because the energy gap can be produced purely from the destructive quantum interference caused by mobile topological defects.

\section{Hopfions in four dimensions: Topological order}\label{sec4D}

Interlinked loops of quantized flux can form topological defects with point singularities only in $d=4$ spatial dimensions. The topological charge of such a singularity is given by the Hopf index on a 3-sphere which encloses the singularity. Neither smooth transformations nor local quantum tunneling can change the Hopf topological charge of the field configuration. The latter protects Hopf singularities from quantum fluctuations and enables topological order with fractionalized charge and braiding statistics. Here we explore the Hopf topological order for the purposes of adding the $\pi_3(S^2)$ homotopy group to the general topological order classification. Higher order homotopy groups $\pi_{2n-1}(S^n)$ which generalize the Hopf singularities to $d=2n$ spatial dimensions (for even $n$) are also able to support topological order, but instead of analyzing this in detail we only construct the corresponding homotopy invariants in Appendices \ref{app2} and \ref{app3} for spin and charge sectors respectively.

\subsection{Charge and angular momentum fractionalization}\label{secFract}

A method for analyzing the topologically ordered states of spinor fields was discussed in Ref.\cite{Nikolic2019}. We first need to construct a ``singularity gauge field'' whose flux is computed as the Hopf index on an $S^3$ manifold surrounding a Hopf singularity. Then, we can formulate a field theory which captures the dynamics of this gauge field even when its flux diffuses as a result of quantum fluctuations. The topological protection mechanism will be implemented with a topological Lagrangian density term involving the singularity gauge field.

We already have the basic setup. The fundamental fields carry charge ($\theta$) and magnetic moments ($\hat{\bf n}$) in $d=4$ dimensional space. Both degrees of freedom can be packed into an SU(2) spinor $\psi({\bf x})$ whose magnitude $|\psi|^2=\rho\neq 0$ is fixed:
\begin{equation}
\psi = \psi_0 e^{i\theta} \quad,\quad \hat{\bf n}=\frac{1}{\rho}\psi^\dagger \boldsymbol{\sigma}\psi \ .
\end{equation}
There are two relevant gauge fields, one $a_\mu=\partial_\mu\theta$ extracted from the U(1) phase winding of $\psi$ and another $A_\mu$ representing the topological Hall effect from the skyrmion textures of $\hat{\bf n}$. To be concrete, we will focus on the spin sector ($\phi_0=4\pi$) and construct the Hopf index on a closed $S^3$ spatial manifold which encloses a volume $B^4$,
\begin{equation}\label{QuantFlux4D}
N=\frac{1}{\phi_{0}^{2}}\oint\limits _{S^{3}}d^{3}x\,\epsilon_{ijk}A_{i}\partial_{j}A_{k}=\frac{1}{\phi_{0}^{2}}\oint\limits _{S^{3}}d^{3}x\,\epsilon_{ijk}\mathcal{A}_{ijk} \ .
\end{equation}
The antisymmetric singularity gauge field can be expressed as
\begin{eqnarray}\label{SingGauge}
\mathcal{A}_{ijk} &=& \frac{1}{6}(A_{i}\partial_{j}A_{k}+A_{j}\partial_{k}A_{i}+A_{k}\partial_{i}A_{j} \\
&& -A_{k}\partial_{j}A_{i}-A_{j}\partial_{i}A_{k}-A_{i}\partial_{k}A_{j}) \ . \nonumber
\end{eqnarray}
The density $\mathcal{J}_{0}$ of Hopf singularities is revealed by representing the Hopf index as a $B^4$ volume integral using Stokes-Cartan theorem:
\begin{equation}
N=\frac{1}{\phi_{0}^{2}}\int\limits _{B^{4}}d^{4}x\,\mathcal{J}_{0} \ .
\end{equation}
From this, we can identify the current density of Hopf singularities
\begin{eqnarray}\label{HopfCurrent}
\mathcal{J}^{\mu} &=& \epsilon^{\mu\nu\alpha\beta\gamma}\partial_{\nu}\mathcal{A}_{\alpha\beta\gamma} = \epsilon^{\mu\nu\alpha\beta\gamma}\partial_{\nu}(A_{\alpha}\partial_{\beta}A_{\gamma}) \\
&=& \epsilon^{\mu\alpha\beta\gamma\delta}(\partial_{\alpha}A_{\beta})(\partial_{\gamma}A_{\delta})-\epsilon^{\mu\alpha\beta\gamma\delta}A_{\alpha}\partial_{\beta}\partial_{\gamma}A_{\delta} \ . \nonumber
\end{eqnarray}
The last term containing an antisymmetrized second derivative responds only to the monopole singularities of the gauge field $A_{\mu}$, which are line defects in $d=4$. We can omit this term if the monopoles have decidedly higher energy than other excitations. For our purposes, the presence of both charge and spin degrees of freedom introduces two gauge fields and two gauge transformations. We will show in Section \ref{secTopSpinor} that the gauge invariant topological current density has an expanded form
\begin{equation}
\mathcal{J}^{\mu} = \frac{1}{q^2} \epsilon^{\mu\alpha\beta\gamma\delta}\Bigl(\partial_{\alpha}(a_{\beta}+qA_{\beta})\Bigr)\Bigl(\partial_{\gamma}(a_{\delta}+qA_{\delta})\Bigr) \ .
\end{equation}

The topological Lagrangian density term has the purpose to implement the conservation of topological charge despite the defect delocalization. We can construct it in real time by coupling the charge current $j_\mu$ to the Hopf singularity current
\begin{equation}
\mathcal{L}_{t} = - K_{t}\rho \, j_{\mu}\mathcal{J}^{\mu} \ .
\end{equation}
Substituting $j_\mu = \partial_\mu\theta + a_\mu + qA_\mu$ with $q\phi_0=2\pi$ and integrating out the phase fluctuations $\theta$ gives us the topological conservation law $\partial_\mu\mathcal{J}^\mu = 0$ (see Appendix \ref{app1}). What remains is a generalization of the Chern-Simons coupling to the $\pi_3(S^2)$ homotopy group,
\begin{eqnarray}
\mathcal{L}_{t} &=& -\frac{K_{t}\rho}{q^2} \, \epsilon^{\mu\alpha\beta\gamma\delta}(a_{\mu}+qA_\mu) \\
&& \quad\times \Bigl(\partial_{\alpha}(a_{\beta}+qA_{\beta})\Bigr)\Bigl(\partial_{\gamma}(a_{\delta}+qA_{\delta})\Bigr) \nonumber \ .
\end{eqnarray}
We can alternatively implement the topological current conservation with the twist field $\chi$ minimally coupled to $A_\mu$; this allows the $a_\mu$ and $A_\mu$ gauge fields to have independent couplings in $\mathcal{L}_{t}$.  

The rest of the real-time Lagrangian density is a generalization of (\ref{Lagrangian3D}) to $d=4$:
\begin{eqnarray}\label{Lagrangian4D}
\mathcal{L} &=& \frac{\kappa_{c}}{2}\left(\partial_{\mu}\theta\!+\!a_{\mu}+qA_{\mu}\right)^{2}+\frac{\kappa_{s}}{2}(\partial_{\mu}\hat{n}^{a})^{2}+\mu b_a\hat{n}^a + \mathcal{L}_{\textrm{B}} \nonumber \\
&& +\frac{\kappa_{t}}{2}\left(\partial_{\mu}\chi\!+\!qA_{\mu}\right)^{2}+\frac{1}{12e^{2}}(\epsilon^{\mu\nu\lambda\alpha\beta}\partial_{\alpha}a_{\beta})^{2} \\
&& +\frac{C}{6}\left(\epsilon^{\mu\nu\lambda\alpha\beta}\partial_{\alpha}A_{\beta}-J^{\mu\nu\lambda}\right)^{2} + \mathcal{L}_{t} \ . \nonumber
\end{eqnarray}
The featured currents are (\ref{HopfCurrent}) and
\begin{eqnarray}
j_{\mu} &=& \partial_{\mu}\theta+a_{\mu}+qA_{\mu} \\
J_{\mu}^{a} &=& \epsilon^{abc}\hat{n}^{b}\partial_{\mu}^{\phantom{x}}\hat{n}^{c} \ . \nonumber
\end{eqnarray}
The spin chirality is a rank-3 tensor in $D=d+1=5$ space-time,
\begin{equation}
J_{\mu\nu\lambda}^{\phantom{x}}=\frac{1}{2}\epsilon_{\mu\nu\lambda\alpha\beta}^{\phantom{x}}\epsilon^{abc}\hat{n}^{a}(\partial_{\alpha}^{\phantom{x}}\hat{n}^{b})(\partial_{\beta}^{\phantom{x}}\hat{n}^{c})=\frac{1}{2}\epsilon_{\mu\nu\lambda\alpha\beta}^{\phantom{x}}\hat{n}^{a}\partial_{\alpha}^{\phantom{x}}J_{\beta}^{a} \nonumber
\end{equation}
The field equations reproduce the same kinds of phenomena as in $d=3$, but with a new important feature attributed to the topological term. Quantized charge becomes attached to quantized Hopf singularities,
\begin{equation}
\kappa_{c}j^{\mu}-\frac{1}{e^{2}}\partial_{\nu}f^{\mu\nu}-3K_{t}\rho\mathcal{J}^{\mu}=0 \ .
\end{equation}
The system can respond to the presence of charge currents by binding topological currents to them
\begin{equation}\label{ChargeFract}
j_\mu = \frac{\nu}{\phi_0^2} \mathcal{J}_\mu
\quad;\quad
\frac{\nu}{\phi_0^2} = \frac{3K_t\rho}{\kappa_c}
\end{equation}
instead of generating the electromagnetic field $f_{\mu\nu}$. The proportionality constant, or the ``filling factor'', is rationally quantized  $\nu=p/q$ in any incompressible quantum liquid. Otherwise, fluctuations can push some frustrated defects through the space occupied by particles (or vice versa) and hence localize the particles (or defects respectively) by destructive quantum interference.

The equation (\ref{ChargeFract}) with a rational $\nu$ describes charge fractionalization because topological defects carry an integer-quantized Hopf index. This is a hallmark of topological order, seen in all $\pi_{d-1}(S^{d-1})$ homotopy groups and now in $\pi_3(S^2)$ as well. The flux-loop structure of the Hopf topological defects provides another mechanism for fractionalization which is not present in the $\pi_{d-1}(S^{d-1})$ homotopies. Consider a hopfion made from U(1) flux loops in the charge sector. The fractional charge $\nu$ attached to the topological defect experiences a system of inter-linked or self-linked flux loops on the $S^{3}$ manifolds that enclose the singularity. The relevant manifolds are those with a radius smaller than the particle-defect binding length. Within such an $S^{3}$ manifold, quantum fluctuations of the topologically ordered state explore many locations and orientations of the flux loops. This frustrates the motion of the attached charge $\nu$ and localizes it to a flux loop, but gives it freedom to move along the loop (parallel to the magnetic field). Other inter-linked loops are seen as the source of Aharonov-Bohm flux by this charge. Consequently, the charge is encouraged to move around its loop and produce a circulating charge current with non-zero orbital angular momentum and magnetic moment. For a circular loop $C$ of radius $r$ and perimeter $l=2\pi r$, interlinked with $n$ quantized flux loops (including self), the resulting orbital magnetic moment amplitude is (in the units $c=1$)
\begin{equation}
\mu=IS=\nu e\frac{2\pi n}{2\pi r}\pi r^{2}=\frac{\nu en}{2}l \ .
\end{equation}
The magnetic moment is itself an antisymmetric rank-2 tensor in $d=4$ dimensions. In this derivation, we interpreted the gauge field $a_{i}$ as the background charge current $I/\nu e\equiv j_{i}=a_{i}$ on the loop, normalized to the unit particle number and then divided $\oint dx_{i}a_{i}=2\pi n$ with the loop perimeter to extract the average current along the loop. The orbital angular momentum is
\begin{equation}
L=\nu m|2\pi rj_{i}|r=\nu m\cdot2\pi r^{2}\cdot\frac{2\pi n}{2\pi r}=\nu mnl \ ,
\end{equation}
(note that the quasiparticle mass $\nu m$ is effectively fractionalized via momentum conservation), so that the Bohr magneton
\begin{equation}
\mu_{\textrm{B}}=\frac{\mu}{L}=\frac{e}{2m}
\end{equation}
still has the standard non-fractionalized value (in the $\hbar=1$ units) associated with orbital motion in quantum mechanics. Nevertheless, the magnetic moment and angular momentum are both fractionalized according to the filling factor $\nu$. We ought to recover the proper angular momentum units by comparing with a non-fractionalized particle ($\nu=1$) whose angular momentum quantization $L=n\hbar\to n$ is clear. The fractionalization by hopfion loops effectively reduces the microscopic angular momentum quantum $\hbar$ to a fractional value $\nu\hbar$, and the magnetic moment to
\begin{equation}
\mu=\nu\mu_{\textrm{B}}\times\textrm{integer} \ .
\end{equation}

It should be noted that the total fractional angular momentum is conserved in the presence of rotational symmetry, and restricted to small magnitudes due to the short length-scale of particle-defect binding (the flux loops in this analysis live on microscopic $S^3$ manifolds). But, there are further complications which we will not attempt to analyze. Even though the fractional charge $\nu$ is indivisible, it can tunnel from one linked loop to another and form superposition states which are still deeply quantized due to the microscopic binding to the topological defect. This drives the twist-exchange processes shown in Fig.\ref{HopfPreservation}. Other Hopf index preserving processes which link or unlink flux loops also tend to preserve the Aharonov-Bohm flux seen by the fractional charge.

\subsection{Fractional braiding statistics}\label{secBraid}

\begin{figure}
\includegraphics[height=0.9in]{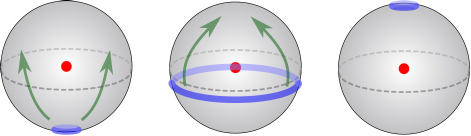}
\caption{\label{Braid4D}A three-dimensional counterpart of the four-dimensional braiding. Set up a sphere around a fractional quasiparticle that contains a topological defect (a monopole in this example). Create an infinitesimal loop at the south pole of the sphere, then stretch and sweep it across the sphere, and finally collapse it into a point to annihilate at the north pole. All entities in this example acquire one additional dimension for the braiding with a Hopf singularity.}
\end{figure}

The topological order of Hopf singularities is equipped with a topologically protected braiding operation. This cannot be a particle exchange like in the fractional quantum Hall state because no topological invariant can be defined on contractible loop-paths. Instead, the braiding operation must be carried out on a 3-sphere path that can serve as a manifold for the computation of the Hopf index. A fractional quasiparticle is a point-like object which combines a fractional charge $\nu$ with a unit Hopf singularity. The mentioned 3-sphere manifold has to enclose the quasiparticle and reflect on its ``surface'' the Hopf charge $N=1$ of the singularity. The lower-dimensional object we braid should traverse the path comprised of this 3-sphere manifold. Therefore, the braided object will be a 2-sphere, and the braiding operation will be a $d=4$ analogue of the process depicted in Fig.\ref{Braid4D}. After a braiding operation, the many-body wavefunction of the system acquires an Abelian fractionally quantized ``Aharonov-Bohm phase''. No other braiding operations have topologically protected outcomes. Specifically, the braiding of flux loops does not enjoy topological protection because loops are volatile under the Hopf index preserving dynamics (see Fig.\ref{HopfPreservation}).

For simplicity, let us neglect any Hopf structure in the charge sector. A hopfion singularity is surrounded by the gauge field $\mathcal{A}_{\mu\nu\lambda}$ defined in (\ref{SingGauge}). The topological charge of the singularity determines the quantized flux (\ref{QuantFlux4D}) of $\mathcal{A}_{\mu\nu\lambda}$ on any 3-sphere $S^{3}$ that encloses the singularity. We wish to assign to this an Aharonov-Bohm ``phase''
\begin{equation}
\phi=\frac{q_{3}}{\phi_{0}^{2}}\oint d^{3}x\,\epsilon_{ijk}A_{ijk}=q_{3}N
\end{equation}
accumulated by a 2-sphere that we create at the ``south'' pole point of $S^{3}$, continuously stretch and sweep across the $S^{3}$ manifold, then finally shrink into a point and consume at the ``north'' pole. Alternatively, we can consider an $S^{2}\times S^{1}$ manifold to capture a braiding operation between a 2-sphere and a quasiparticle in which a quasiparticle is pushed through a 2-sphere on a closed loop. The current $J_{\mu\nu\lambda}$ of the 2-sphere objects minimally couples to $\mathcal{A}_{\mu\nu\lambda}$ and generates these braiding operations via the operator $\exp(i\,d^{3}x\,\epsilon_{ijk}J_{ijk})$. The bare current is the topologically trivial ``pure gauge'' content of $\mathcal{A}_{\mu\nu\lambda}$, which integrates out to $N=0$:
\begin{equation}\label{PureGauge}
J_{\mu\nu\lambda}=\partial_{\mu}\theta_{\nu\lambda}-\partial_{\nu}\theta_{\mu\lambda}+\partial_{\lambda}\theta_{\mu\nu} \ .
\end{equation}
The matter field $\theta_{\mu\nu}$ introduced here is an antisymmetric tensor. The full gauge-invariant current we need is
\begin{equation}
J_{\mu\nu\lambda} = \partial_{\mu}\theta_{\nu\lambda}-\partial_{\nu}\theta_{\mu\lambda}+\partial_{\lambda}\theta_{\mu\nu}+\mathcal{A}_{\mu\nu\lambda} \ .
\end{equation}

We must construct the braiding operator $\mathcal{B}$ using the gauge-invariant generator $\mathcal{J}_{\mu\nu\lambda}$, but the matter fields in its definition immediately drop out because they are integrated over a manifold without a boundary. Hence,
\begin{equation}
\mathcal{B} = \exp\left(iq_{3}\oint\limits _{S^{3}}d^{3}x\,\epsilon_{ijk}\mathcal{A}_{ijk}\right) = e^{iq_{3}^{\phantom{x}}\phi_{0}^{2}N}
\end{equation}
should create a 2-sphere excitation $J_{0ij}$ at the ``south'' pole of $S^{3}$, drive it around a stationary fractionalized quasiparticle at the origin, and then annihilate it at the ``north'' pole of $S^{3}$. If the quasiparticle at the origin carried $N$ hopfion quanta, then $iq_3^{\phantom{x}}\phi_0^2 N$ would be the Aharonov-Bohm phase acquired in the braiding operation. There is a problem, however. The charge $q_{3}$ associated with the current $J_{\mu\nu\lambda}$ should be an integer in any conventional state of matter, yet the acquired phase proportional to $\phi_{0}^{2}$ would be non-trivial due to not being an integer multiple of $2\pi$. Analogous constructions of braiding operations for vortices and monopoles are not plagued by this problem. We must renormalize the generating current operator to fix this problem.

Consider the analogy with particle braiding, where $\exp(\oint dx_i A_i)$ creates a unit dipole, drives the particle of the dipole around a loop and then annihilates it with the antiparticle. The dipole creation step,
\begin{equation}
e^{dx_i a_i} = e^{dx_i \partial_i \theta} = e^{i\theta({\bf x}+{\bf dx})} e^{-i\theta({\bf x})} \ ,
\end{equation}
manifestly involves the particle creation $e^{i\theta}$ and annihilation $e^{-i\theta}$ operators. The phase operator $\theta$ is canonically conjugate to the number operator $n$. Charge quantization implies integer eigenvalues for $n$, making $\theta$ an angle from a $2\pi$ interval. This determines the normalization of the current operator $j_\mu = \partial_\mu \theta + a_\mu$. Going back to our problem, the initial creation $\exp(-iJ^{\mu\nu\lambda}\epsilon_{\mu\nu\lambda}d^{3}x)$ of a braided 2-sphere ought to utilize the properly quantized ``phase'' operators $\theta_{\mu\nu}$ within (\ref{PureGauge}). We can reveal this quantization by constructing everything from the singular vortex-loop configurations of the twist field $\chi$. We begin by applying a singular gauge transformation to extract vorticity from $\chi$ into the gauge field
\begin{equation}
A_{\mu}=\frac{1}{q}\partial_{\mu}\chi \ .
\end{equation}
Then, by comparing (\ref{PureGauge}) and (\ref{SingGauge}) through a similar singular gauge transformation $\mathcal{J}_{\mu\nu\lambda}\to\mathcal{A}_{\mu\nu\lambda}$, we identify
\begin{equation}
\theta_{\mu\nu}=\frac{1}{3q^{2}}\chi\partial_{\mu}\partial_{\nu}\chi
\end{equation}
with an implicit index antisymmetrization. An underlying assumption is that the $\chi$ configuration does not contain monopoles, i.e. $\epsilon_{\cdots\mu\nu\lambda}\partial_\mu\partial_\nu\partial_\lambda\chi = 0$. The rank-3 current currently has the form
\begin{equation}
J_{\mu\nu\lambda}=\frac{1}{3q^{2}}\Bigl\lbrack\partial_{\mu}(\chi\partial_{\nu}\partial_{\lambda}\chi)-\partial_{\nu}(\chi\partial_{\mu}\partial_{\lambda}\chi)+\partial_{\lambda}(\chi\partial_{\mu}\partial_{\nu}\chi)\Bigr\rbrack \nonumber
\end{equation}
where the antisymmetrization of indices is implicitly understood. The $\chi$ field carries $2\pi$-quantized vorticity, so that
\begin{equation}
\frac{1}{q^{2}}\partial_{\mu}(\chi\partial_{\nu}\partial_{\lambda}\chi)\,d^{3}x\to\frac{2\pi}{q^{2}}(\partial_{\mu}\chi)dx^{\mu} \ .
\end{equation}
The omitted index antisymmetrization keeps the $\mu$ direction orthogonal to the directions $\nu,\lambda$. Then, the $\partial_{\mu}$ derivative acts along the singular flux line and sees $\epsilon_{\cdots\mu\nu\lambda}\partial_\nu\partial_\lambda\chi$ as a constant. By definition, the twist field $\chi$ is a proper angle, so $\partial_\mu\chi$ will integrate to an integer multiple of $2\pi$. Evidently, the extra factor $2\pi/q^2$ must be divided out in order to get an adequately normalized rank-3 current:
\begin{equation}\label{Rank3Current}
\widetilde{J}_{\mu\nu\lambda}=\frac{q^{2}}{2\pi}\Bigl( \partial_{\mu}\theta_{\nu\lambda}-\partial_{\nu}\theta_{\mu\lambda}+\partial_{\lambda}\theta_{\mu\nu}+\mathcal{A}_{\mu\nu\lambda}\Bigr) \ .
\end{equation}
Now,
\begin{equation}
\widetilde{\mathcal{B}} = \exp\left(iq_{3}\oint\limits _{S^{3}}d^{3}x\,\epsilon_{ijk}\widetilde{J}_{ijk}\right) = e^{iq_{3}^{\phantom{x}}(q^2\phi_{0}^{2}/2\pi)N}
\end{equation}
correctly braids a quantized 2-sphere around a point Hopf singularity because $q\phi_0=2\pi$ in both charge and spin sectors.

The last task is to determine the ``charge'' $q_{3}$ associated with the current $J_{\mu\nu\lambda}$. The U(1) flux quantum carried by $F_{\mu\nu}$ remains $\phi_0$ in the fractionalized phases, but the charge unit carried by $J_{\mu}$ becomes a rational number $\nu$. We do not want to embed a factor of $\nu$ into the operator $\widetilde{J}_{\mu\nu\lambda}$, which ought to create unit ``charges''. Instead, we explicitly provide a factor of $q_3=\nu$ in the braiding generator, to be associated with the $J_{\mu}$ operators within (\ref{Rank3Current}). In this sense, the braiding operator will braid the smallest possible bundle of charge and flux comprising $\widetilde{J}_{\mu\nu\lambda}$ that the incompressible quantum liquid can support. The resulting topologically-protected hopfion braiding statistics (phase) between a 2-sphere and a point fractional quasiparticle made of $N$ Hopf quanta is:
\begin{equation}
\widetilde{\mathcal{B}} = e^{2\pi i\nu N} \ .
\end{equation}
Note that further induced field corrections to braiding have not been considered here.

\subsection{Topological ground state degeneracy}\label{secTopDeg}

The incompressible quantum liquid of Hopf singularities exhibits topological order, i.e. a ground state degeneracy on topologically non-trivial manifolds. We will demonstrate this on the four-dimensional spatial manifold $\mathcal{M}=S^{1}\otimes M^3$ with $M^3=S^1\otimes\mathbb{R}$. Quantum fluctuations on such manifolds, which one might call vacuum instantons, can convert a classical ground state from one topological sector with a given Hopf index to another. Our goal here is to demonstrate that vacuum instantons lift the classical ground state degeneracy down to a finite residual degeneracy. We will consider only a simple subset of the full problem in order to exhibit topological order at zero temperature. Vacuum instantons should not be confused with defect instantons which were mentioned in earlier sections. Defect instantons cost more action, and create or annihilate local topological defect excitations. Their confinement, measured with a separate long-range correlation function, is a prerequisite for topological order. 

For our purposes, a classical ground state is characterized by two integer topological invariants, a Hopf index $N$ on $M^3$ and a vortex winding number $N'$ on the complement $S^1$ submanifold of $\mathcal{M}$. The gauge field configurations in these states are
\begin{eqnarray}\label{ClassGS}
a_{i}({\bf x}) &=& 2\pi\frac{N'}{L}\delta_{i,4} \\
A_{i}({\bf x}) &=& \frac{1}{q}\left(\frac{x_2\delta_{i,1}-x_1\delta_{i,2}}{x_1^{2}+x_2^{2}}n+\frac{2\pi N}{L}\delta_{i,3}\right) \ , \nonumber
\end{eqnarray}
where $x_1,x_2\in\mathbb{R}^2$, $x_3\in S^1\subset M$, $x_4\in S^1\subset \mathcal{M}$, and both $S^1$ submanifolds have the perimeter $L$. A vortex line 
\begin{equation}
\epsilon_{ijk}\partial_{j}A_{k}\to\frac{2\pi n}{q}\delta(x_1)\delta(x_2)\delta_{i,3}
\end{equation}
is threaded along the $x_3$ axis through the origin of $\mathbb{R}$; this singularity will need to be regularized later on. We mention in passing that the quantization of $N$ and $N'$ is enforced by the presence of matter fields that couple to $a_i$ and $A_i$; for example, the integer winding of $\theta$ over $S^1$ can be compensated without a high gradient $(\partial_\mu\theta+a_\mu)^2$ price only by quantizing the appropriate gauge field components. The configurations (\ref{ClassGS}) are classically degenerate in the $L\to\infty$ limit. We implicitly integrate out all other configurations which cost more energy, and explore how the remaining macroscopic degeneracy can be lifted by quantum fluctuations. The relevant effective Lagrangian density has the form
\begin{eqnarray}
\mathcal{L} &=& \frac{1}{12e^{2}}(\epsilon_{\mu\nu\lambda\alpha\beta}\partial_{\alpha}a_{\beta})^{2}+\frac{C}{6}(\epsilon_{\mu\nu\lambda\alpha\beta}\partial_{\alpha}A_{\beta})^{2} \nonumber \\
&& -\frac{q^{2}\nu}{(2\pi)^{2}}\epsilon^{\mu\nu\alpha\beta\gamma}a_{\mu}\partial_{\nu}(A_{\alpha}\partial_{\beta}A_{\gamma}) \ ,
\end{eqnarray}
where we have retained only a part of the full topological term for the demonstration of topological order. Choosing Coulomb gauge $a_0=0$, $A_0=0$ and allowing time variations of the topological invariants, we substitute (\ref{ClassGS}) into the Lagrangian density and get
\begin{eqnarray}
\mathcal{L} &=& \frac{(2\pi)^{2}}{2e^{2}L^{2}}(\partial_{0}N')^{2}+\frac{(2\pi)^{2}C}{q^{2}}\left\lbrack \delta(x_{1})\delta(x_{2})\right\rbrack ^{2}n^{2} \nonumber \\
&& +\frac{C}{q^{2}}\left\lbrack \frac{(\partial_{0}n)^{2}}{x_{1}^2+x_{2}^2}+\frac{(2\pi)^{2}(\partial_{0}N)^{2}}{L^{2}}\right\rbrack \nonumber \\
&& -4\pi\nu\,\frac{n}{L^{2}}\delta(x_1)\delta(x_2)N'\partial_{0}N \ .
\end{eqnarray}
The Lagrangian $L_i$ of vacuum instantons obtains by integrating out ${\bf x}\in\mathcal{M}$. We are forced to regulate the infra-red and ultra-violet divergence by reducing the $\mathbb{R}^2$ submanifold to a ring with inner radius $r\to 0$ and outer radius $R\to\infty$:
\begin{eqnarray}
L_i &=& \frac{(2\pi)^{2}}{2e^{2}}\pi R^{2}(\partial_{0}N')^{2}+\frac{(2\pi)^{2}C}{q^{2}}\frac{L^{2}}{\pi r^{2}}\,n^{2} \nonumber \\
&& +\frac{C}{q^{2}}\left\lbrack 2\pi L^{2}\log\left(\frac{R}{r}\right)(\partial_{0}n)^{2}+(2\pi)^{2}\pi R^{2}(\partial_{0}N)^{2}\right\rbrack \nonumber \\
&& -4\pi n\nu\,N'\partial_{0}N \ .
\end{eqnarray}
The ultraviolet cut-off distance $r$ appears only in the terms that contain the vorticity $n$, and these terms are relevant only if the vorticity can fluctuate. We may neglect such fluctuations here because their energy cost is logarithmically enhanced in comparison to the cost of all other fluctuations. So, if we assume that the fluctuations of $n$ are frozen, i.e. $\partial_{0}n\to0$, the vacuum instanton Lagrangian becomes
\begin{eqnarray}
L_i &=& \frac{(2\pi)^{2}}{2e^{2}}\pi R^{2}(\partial_{0}N')^{2}+\frac{(2\pi)^{2}C}{q^{2}}\pi R^{2}(\partial_{0}N)^{2} \nonumber \\
&& -4\pi n\nu\,N'\partial_{0}N \ .
\end{eqnarray}
up to a constant. We can now treat $N$ and $N'$ as canonical coordinates for instanton fluctuations. The corresponding canonical momenta are
\begin{equation}
P=\frac{\delta L_i}{\delta\partial_{0}N}=\frac{2(2\pi)^{2}C}{q^{2}}\pi R^{2}\,\partial_{0}N-\frac{4\pi n\nu}{m}\,N'
\end{equation}\vspace{-0.2in}
\begin{equation}
P'=\frac{\delta L_i}{\delta\partial_{0}N'}=\frac{(2\pi)^{2}}{e^{2}}\pi R^{2}\partial_{0}N' \ .
\end{equation}
The instanton Hamiltonian is then
\begin{eqnarray}
H_i &=& P\partial_{0}N+P'\partial_{0}N'-L_i \\
&=& \frac{1}{2M}\left(P+4\pi n\nu N'\right)^{2}+\frac{1}{2M'}P'^{2} \ , \nonumber
\end{eqnarray}
where the masses
\begin{equation}
M=\frac{(2\pi)^{3}R^{2}C}{q^{2}} \quad,\quad M'=\frac{(2\pi)^{3}R^{2}}{2e^{2}}
\end{equation}
are macroscopically large due to $R\to\infty$.

As in the treatment of fractional quantum Hall states, we imagine that $N,N'\in\mathbb{Z}$ are integer-valued coordinates of sites on a fictitious square lattice. Their canonical conjugates $P,P'$ must be angles, so it is necessary to regularize the instanton Hamiltonian into a compact form
\begin{equation}\label{InstantonHamiltonian}
H_i \to -t\cos(P+4\pi n\nu N')-t'\cos(P') \ .
\end{equation}
This is Hofstadter model expressed in Landau gauge, with $2n\nu$ flux quanta per lattice plaquette. Its ground states capture the fate of the classical degeneracy after taking vacuum instantons into account. If the filling factor $\nu$ is rational so that $2n\nu=p/q$ with mutually prime integers $p,q$ ($n\in\mathbb{Z}$), then the ground state degeneracy is $q$ for any finite $t,t'$. Under current assumptions, $t,t'\propto R^{-2}\to 0$ both vanish in the thermodynamic $R\to\infty$ limit, and hence preserve the infinite degeneracy of the classical topological sectors. However, this is a consequence of suppressing the local flux-changing events in our formalism. The U(1) flux lines which determine $N$ and $N'$ do not live in the physical space $\mathcal{M}$, so no energy cost is associated with their presence or absence. One can imagine these flux lines protruding through $\mathcal{M}$ due to occasional fluctuations, and temporarily creating visible magnetic dipoles of any size. This dynamics is governed only by the local energy of the visible dipoles, but may result with changes of the total flux $N$ and $N'$ threaded through the openings of $\mathcal{M}$. As a consequence, we should indeed assume that the hopping integrals $t,t'>0$ are finite.

This establishes topological order at zero temperature, with ground state degeneracy on $\mathcal{M}$ determined by the filling factor $\nu$. The vacuum instanton Hamiltonian (\ref{InstantonHamiltonian}) is ultimately analogous to that of a fractional quantum Hall liquid in two dimensions, although the ground state degeneracy is slightly different for the same filling factor. This topological order does not survive at finite temperatures. However, the defect instanton events which create and annihilate the Hopf singularities in the $D=d+1=5$ dimensional space-time are strongly confined by a linear potential in the action, so a confined instanton phase survives at finite low temperatures. When defect instantons are confined, the topological charge of Hopf singularities is conserved regardless of whether topological order is present. This opens a possibility of other correlated phases at low temperatures.

It is interesting to point out that the presented method can be also applied to the more physical $d=3$ models of hopfion dynamics. The result is, however, trivial. The ground state of a putative incompressible quantum liquid of conserved hopfions is non-degenerate in three spatial dimensions. The vacuum instanton Hamiltonian equivalent to (\ref{InstantonHamiltonian}) would be one-dimensional because a $d=3$ spatial manifold leaves no room for the $N'$ variable.

\subsection{Spinor representation of the topological term}\label{secTopSpinor}

Electrons are naturally represented by a spinor field $\psi$ that carries information about charge and spin. The charge content is handled by the phase factor $\psi\propto e^{i\varphi}$ and the spin content $\hat{{\bf n}}(\theta,\phi)$ is given by the spinor structure in some representation of the Spin(3)=SU(2) group:
\begin{equation}\label{Spinor}
\psi(\hat{{\bf n}})=e^{-iJ_{3}\phi}e^{-iJ_{2}\theta}e^{i\varphi}\psi_{0} \ ,
\end{equation}
where $J^a$ are the spin projection operators. The reference spinor $\psi_0$ is constant and all dynamics is expressed with the space-time variations of $\varphi,\theta,\phi$. Here we construct the topological Lagrangian density directly in terms of the spinor field, and study its properties. This provides the most accurate formulation of the topological field theory which resolves some issues with the previous more effective approach.

In the previous discussions, we normalized the charge current $j_\mu \sim \partial_\mu\varphi +a_\mu +qA_\mu$ in a manner that emphasizes its relationship to the phase gradient $\partial_\mu\varphi$ of the matter field. We now wish to use the standard normalization of the current extracted from a spinor,
\begin{equation}\label{SpinorCurrent}
\rho=\psi^{\dagger}\psi\quad,\quad j_{\mu}=-\frac{i}{2}\Bigl(\psi^{\dagger}(\partial_{\mu}\psi)-(\partial_{\mu}\psi^{\dagger})\psi\Bigr) \ .
\end{equation}
The corresponding imaginary-time topological Lagrangian density with the same coupling constant $K_t$ is
\begin{equation}\label{TopologicalTerm}
\mathcal{L}_{t}=iK_{t}\,j_{\mu}\mathcal{J}_{\mu} \ .
\end{equation}
Our goal is to reproduce this form and the Hopf singularity current
\begin{equation}
\mathcal{J}_{\mu}=\epsilon_{\mu\nu\alpha\beta\gamma}\partial_{\nu}(A_{\alpha}\partial_{\beta}A_{\gamma})
\end{equation}
by applying a certain singular derivative operator $\psi^{\dagger}\mathcal{D}\psi$ to the spinor field $\psi$. This operator will contain antisymmetrized derivatives which automatically give zero at every point $x_\mu$ where $\psi(x_\mu)$ is analytic; we want $\mathcal{D}$ to pick only the singularities of $\psi$, and for that purpose a mathematical rule for taking derivatives of singular functions will be implicitly provided. The structure of the Hopf index suggests that we should explore the Lagrangian density
\begin{equation}
\mathcal{L}_{t}\propto\epsilon_{\mu\nu\alpha\beta\gamma}\psi^{\dagger}\partial_{\mu}\partial_{\nu}\partial_{\alpha}\partial_{\beta}\partial_{\gamma}\psi \ ,
\end{equation}
with understanding that the correct type of topological defects will be detected when present. The gradient of (\ref{Spinor}) is
\begin{equation}\label{PsiGradient}
\partial_{\mu}\psi=i\left\lbrack \partial_{\mu}\varphi-J_{3}\partial_{\mu}\phi-e^{-iJ_{3}\phi}J_{2}e^{+iJ_{3}\phi}\partial_{\mu}\theta\right\rbrack\psi \ .
\end{equation}
We will at first concentrate on the charge sector and consider $\theta,\phi\to\textrm{const}$. After applying a singular gauge transformation to transfer the gauge field $a_\mu$ into $\partial_\mu\varphi$, we get
\begin{equation}
\theta,\phi\to\textrm{const} \quad\Rightarrow\quad \partial_\mu\psi\to ia_{\mu}\psi \ .
\end{equation}
Then, the formal application of the product rule for derivatives gives us
\begin{eqnarray}
&& \epsilon_{\mu\nu\alpha\beta\gamma}\partial_{\mu}\partial_{\nu}\partial_{\alpha}\partial_{\beta}\partial_{\gamma}\psi
= i\epsilon_{\mu\nu\alpha\beta\gamma}\partial_{\mu}\partial_{\nu}\partial_{\alpha}\partial_{\beta}a_{\gamma}\psi \\
&& \qquad = i\epsilon_{\mu\nu\alpha\beta\gamma}\partial_{\mu}\partial_{\nu}\partial_{\alpha}(\partial_{\beta}a_{\gamma})\psi-\epsilon_{\mu\nu\alpha\beta\gamma}\partial_{\mu}\partial_{\nu}\partial_{\alpha}a_{\gamma}a_{\beta}\psi \nonumber \\
&& \qquad = i\epsilon_{\mu\nu\alpha\beta\gamma}\partial_{\mu}\partial_{\nu}(\partial_{\alpha}\partial_{\beta}a_{\gamma})\psi-\epsilon_{\mu\nu\alpha\beta\gamma}\partial_{\mu}\partial_{\nu}a_{\alpha}(\partial_{\beta}a_{\gamma})\psi \nonumber \\
&& \qquad = i\epsilon_{\mu\nu\alpha\beta\gamma}\partial_{\mu}(\partial_{\nu}\partial_{\alpha}\partial_{\beta}a_{\gamma})\psi-\epsilon_{\mu\nu\alpha\beta\gamma}\partial_{\mu}a_{\nu}(\partial_{\alpha}\partial_{\beta}a_{\gamma})\psi \nonumber \\
&& \qquad\quad -\partial_{\mu}^{\phantom{x}}\mathcal{J}_{\mu}^{\textrm{c}}\psi \nonumber \ ,
\end{eqnarray}
where
\begin{equation}
\mathcal{J}_{\mu}^{\textrm{c}} = \epsilon_{\mu\nu\alpha\beta\gamma}^{\phantom{x}}\partial_{\nu}^{\phantom{x}}(a_{\alpha}^{\phantom{x}}\partial_{\beta}^{\phantom{x}}a_{\gamma}^{\phantom{x}})
\end{equation}
is the Hopf index current density in the charge sector. The first term in the final expression describes $d=4$ monopoles; we will drop this term because monopoles are homotopically distinct topological defects from hopfions. The second term similarly captures the line singularities of $a_{\mu}$ in $d=4$, which would be point-like monopoles in $d=3$; hopfions are not equipped with such singularities. So, for the purposes of hopfions in the charge sector, the topological Lagrangian density (\ref{TopologicalTerm}) should be
\begin{equation}\label{TopologicalTerm2}
\mathcal{L}_{t}=K_{t}\,\psi^{\dagger}\partial_{\mu}\mathcal{J}_{\mu}\psi=-K_{t}\,\psi^{\dagger}\epsilon_{\mu\nu\alpha\beta\gamma}\partial_{\mu}\partial_{\nu}\partial_{\alpha}\partial_{\beta}\partial_{\gamma}\psi \ .
\end{equation}
Antisymmetrized derivatives were retained in various steps of this derivation. The implicit rule for calculating such derivatives of singular functions is given by the Stokes-Cartan theorem and similar topological relationships. For example, a complex field $\psi=\psi_0 e^{i\varphi}$ with non-zero magnitude $\rho=|\psi_0|^2$ in two dimensions may have a vortex singularity at the origin; then, the ``singularity gauge field'' $a_i = \partial_i \varphi$ has a singular flux $\epsilon_{ijk} \partial_j a_k \equiv -i\rho^{-1} \psi^\dagger \epsilon_{ijk} \partial_j \partial_k \psi = 2\pi \delta_{i,0} \delta({\bf x})$ because $\oint dx_\mu a_\mu = 2\pi$, has a fixed well-defined value on any loop that encloses the origin. This specifies the basic rule for taking antisymmetric derivatives, and its extension to higher orders can be constructed recursively \cite{Nikolic2019}.

We will show next that the same antisymmetric derivative extracts the Hopf singularity in the spin sector as well. The $\phi\in(0,2\pi)$ angle in (\ref{Spinor}) microscopically provides both the twist field $\chi\sim\phi$ and the spin chirality gauge field $qA_\mu\sim\partial_\mu\phi$. Consider the following integral over a spatial volume $B^4$ bounded by $S^3$:
\begin{eqnarray}\label{I4a}
I_{4} &=& \int\limits_{B^{4}}d^{4}x\,\psi^{\dagger}\epsilon_{ijkl}\partial_{i}\partial_{j}\partial_{k}\partial_{l}\psi \\
&=& I_3 - \int\limits_{B^{4}}d^{4}x\,\epsilon_{ijkl}(\partial_{i}\psi^{\dagger})(\partial_{j}\partial_{k}\partial_{l}\psi) \nonumber \\
&=& I^{\phantom{x}}_{3}+I'_3 +\int\limits_{B^{4}}d^{4}x\,\epsilon_{ijkl}(\partial_{j}\partial_{i}\psi^{\dagger})(\partial_{k}\partial_{l}\psi) \nonumber \ .
\end{eqnarray}
Two integrations by parts and Stokes-Cartan theorem produce the boundary integrals
\begin{eqnarray}
I_3 &=& \oint\limits _{S^{3}}d^{3}x\,\psi^{\dagger}\epsilon_{ijk}\partial_{i}\partial_{j}\partial_{k}\psi \\
I'_3 &=& \oint\limits _{S^{3}}d^{3}x\,\epsilon_{ijk}(\partial_{i}\psi^{\dagger})(\partial_{j}\partial_{k}\psi) \nonumber
\end{eqnarray}
which add up to zero because $S^3$ itself has no boundary:
\begin{equation}
I^{\phantom{x}}_3+I'_3 = \oint\limits _{S^{3}}d^{3}x\,\epsilon_{ijk}\partial_{i} (\psi^{\dagger}\partial_{j}\partial_{k}\psi) = 0 \ .
\end{equation}
The remaining volume integral still responds only to the singularities of $\psi({\bf x})$ which arise in (\ref{Spinor}) due to the spin texture $\hat{\bf n}$ or charge current vortices. The spin-rotation generators $J_{i}$ place the spin in the unique direction $\hat{{\bf n}}(\theta,\phi)$, but also introduce a sign-changing branch-cut in the spinor for fermionic particles due to having eigenvalues equal to odd integer halves. This is formally corrected by providing a U(1) rotation:
\begin{equation}\label{Spinor2}
\psi(\hat{{\bf n}})=e^{-iJ_{3}\phi}e^{-iJ_{2}\theta}e^{i(\varphi+s\phi/2)}\psi_{0} \ ,
\end{equation}
and amounts to the shift $J_{3}\to J_{3}-s/2$ of the spin projection operator. The value of $s$ must be an odd integer for fermions and an even integer for bosons, unless $\phi$ winds by an integer multiple of $4\pi$. Looking at the two equivalent minimal representations $\psi_\pm = e^{\pm i\phi/2} \psi$ of (\ref{Spinor}) with $\psi_0 = |\uparrow\rangle$,
\begin{equation}
\psi_{+}=\left(\begin{array}{c}
\cos\left(\frac{\theta}{2}\right)\\
e^{i\phi}\sin\left(\frac{\theta}{2}\right)
\end{array}\right)
\quad,\quad
\psi_{-}=\left(\begin{array}{c}
e^{-i\phi}\cos\left(\frac{\theta}{2}\right)\\
\sin\left(\frac{\theta}{2}\right)
\end{array}\right) \quad ,
\end{equation}
we see that the choices $s=1$ and $s=-1$ minimally regularize the vortex singularity at $\theta=0$ and $\theta=\pi$ respectively. This generalizes to $s=2S$ and $s=-2S$ respectively in higher spin representations. The values $s=2(S+n)$ at $\theta=0$ and $s=2(-S+n)$ at $\theta=\pi$ leave a $2\pi n$ vortex singularity of $\phi$ in the spinor $\psi$. We will pick $s=2(S+1)$ in order to capture the $2\pi$ vorticity of $\phi$ at ``skyrmion centers'',
\begin{equation}\label{SChoice}
S' = S-\frac{s}{2} \xrightarrow{s\to 2(S+1)} -1 \ .
\end{equation}
This amounts to the choice of an arbitrary spin quantization axis $\hat{\bf z}$ such that $\hat{\bf n}=\hat{\bf z}$ ($\theta=0$) at the skyrmion centers and $\hat{\bf n}\to-\hat{\bf z}$ ($\theta=\pi$) in the far-away zero-vorticity ferromagnetic regions. A skyrmion center defined hereby will coincide with the symmetry center when the spin texture is rotationally symmetric, but generally cannot be unambiguously identified because skyrmions are not singular spin configurations. Since we pinned the skyrmion centers to $\theta=0$, we always have $J_{3}\psi=S\psi$ there given that (\ref{Spinor2}) is an eigenvector of $J_3$ at $\theta=0$. So, starting from (\ref{Spinor2}), we find
\begin{equation}
\partial_{j}\psi=i\left\lbrack \partial_{j}\varphi-\left(J_{3}-\frac{s}{2}\right)\partial_{j}\phi-e^{-iJ_{3}\phi}J_{2}e^{iJ_{3}\phi}\partial_{j}\theta\right\rbrack \psi \nonumber \ ,
\end{equation}
and after some algebra
\begin{equation}\label{CurlPsi}
\epsilon_{\cdots ij}\partial_{i}\partial_{j}\psi = i(\epsilon_{\cdots ij}\partial_{i}\partial_{j}\varphi)\psi-i(\epsilon_{\cdots ij}\partial_{i}\partial_{j}\phi)\left(J_{3}-\frac{s}{2}\right)\psi \ .
\end{equation}
The derivation discards $\epsilon_{\cdots ij}\partial_{i}\partial_{j}\theta$ because $\theta\in\lbrack0,\pi\rbrack$ cannot wind and support a vortex singularity. We point out in passing that the statistics correction $s$ has no impact on the dynamics of spin-hedgehog topological defects \cite{Nikolic2019} because hedgehogs introduce $\phi$ vorticity at both north $\theta=0$ and south $\theta=\pi$ poles, and the effects of $s$ at both poles cancel out.

Let us simplify the further analysis by temporarily neglecting charge currents, i.e setting $\varphi=\textrm{const}$. When we substitute (\ref{CurlPsi}) in (\ref{I4a}) with $\varphi=\textrm{const.}$, we get
\begin{eqnarray}\label{I4b}
I_4 &=& -\int\limits _{B^{4}}d^{4}x\,\epsilon_{ijkl}(\partial_{i}\partial_{j}\phi)(\partial_{k}\partial_{l}\phi)\,\psi^{\dagger}\left(J_{3}-\frac{s}{2}\right)^2\psi \nonumber \\
&=& -S'^{2}|\psi|^{2}\int\limits _{B^{4}}d^{4}x\,\epsilon_{ijkl}(\partial_{i}\partial_{j}\phi)(\partial_{k}\partial_{l}\phi) \ .
\end{eqnarray}
If we introduce a parameter $\xi\in\mathbb{R}$ which specifies the points ${\bf r}^{(n)}(\xi)$ on the $n^{\textrm{th}}$ skyrmion's center, then we can express the singular vorticity of $\phi$ in a $d=3$ manifold as
\begin{eqnarray}
\epsilon_{ijk}^{\phantom{x}}\partial_{j}^{\phantom{x}}\partial_{k}^{\phantom{x}}\phi &=& \sum_{n}\int dr_{i}^{(n)}\,2\pi\delta\bigl({\bf r}-{\bf r}^{(n)}(\xi)\bigr) \\
&=& \int d\xi\,\frac{\partial r_{i}^{(n)}}{\partial\xi}\,2\pi\delta\bigl({\bf r}-{\bf r}^{(n)}(\xi)\bigr) \ , \nonumber
\end{eqnarray}
where the $\delta$-function is 3-dimensional. The parametrization of a $\phi$-vortex in $d=4$ must take the form of a rank-2 tensor because a skyrmion center is a sheet of points ${\bf r}^{(n)}(\xi_{1},\xi_{2})$. Summing over all skyrmions ($n$), we obtain
\begin{eqnarray}
&& \epsilon_{ijkl}^{\phantom{x}}\partial_{k}^{\phantom{x}}\partial_{l}^{\phantom{x}}\phi=\sum_{n}\int d\xi_{1}^{\phantom{x}}d\xi_{2}^{\phantom{x}}\left(\frac{\partial r_{i}^{(n)}}{\partial\xi_{1}}\frac{\partial r_{j}^{(n)}}{\partial\xi_{2}}-\frac{\partial r_{i}^{(n)}}{\partial\xi_{2}}\frac{\partial r_{j}^{(n)}}{\partial\xi_{1}}\right) \nonumber \\
&& \qquad \times 2\pi\,\delta\bigl({\bf r}-{\bf r}^{(n)}(\xi_{1}^{\phantom{x}},\xi_{2}^{\phantom{x}})\bigr) \ .
\end{eqnarray}
The $\delta$-function is four-dimensional here. Substituting the last expression into (\ref{I4b}) and using
\begin{equation}
\epsilon_{pqij}\epsilon_{pqkl}=2(\delta_{ik}\delta_{jl}-\delta_{il}\delta_{jk})
\end{equation}
yields
\begin{eqnarray}\label{I4c}
I_{4} &=& -(2\pi S')^{2}|\psi|^{2}\sum_{n,m}\,\int\limits_{B^{4}}d^{4}r\,\int d\xi_{1}^{(m)}d\xi_{2}^{(m)}d\xi_{1}^{(n)}d\xi_{2}^{(n)} \nonumber \\
&& \times \epsilon_{ijkl}\,\frac{\partial r_{i}^{(m)}}{\partial\xi_{1}^{(m)}}\frac{\partial r_{j}^{(m)}}{\partial\xi_{2}^{(m)}}\frac{\partial r_{k}^{(n)}}{\partial\xi_{1}^{(n)}}\frac{\partial r_{l}^{(n)}}{\partial\xi_{2}^{(n)}} \\&& \times \delta\bigl({\bf r}-{\bf r}^{(n)}(\xi_{1}^{(n)},\xi_{2}^{(n)})\bigr)\,\delta\bigl({\bf r}-{\bf r}^{(m)}(\xi_{1}^{(m)},\xi_{2}^{(m)})\bigr) \ . \nonumber
\end{eqnarray}

Let us assume that a Hopf singularity of spins is at the origin. We can enclose it by concentric $S^{3}$ 3-spheres that each carry the same Hopf index. The inter-linked skyrmions which materialize the Hopf index are loops in each $S^{3}$ manifold, so every pair of coordinates $r_{i}$ and $r_{j}$ in the last integral lives on a cone-shaped skyrmion sheet embedded in the $d=4$ space; one coordinate (parametrized by $\xi_1$) is bound to the skyrmion loop on the 3-sphere and the other (parametrized by $\xi_2$) to the ``radial'' direction which emanates from the Hopf singularity. All skyrmions need to meet at the singularity in order for $I_{4}$ to capture them; the skyrmion lines do not intersect on any $S^{3}$ away from the singularity, so only the singularity itself can contribute to $I_{4}$. We will extract the skyrmion linking number from $I_{4}$. The first step is to integrate out ${\bf r}$:
\begin{eqnarray}
I_{4} &=& -(2\pi S')^{2}|\psi|^{2}\sum_{n,m}\int d\xi_{1}^{(m)}d\xi_{2}^{(m)}d\xi_{1}^{(n)}d\xi_{2}^{(n)} \nonumber \\
&& \quad \times \epsilon_{ijkl}\,\frac{\partial r_{i}^{(m)}}{\partial\xi_{1}^{(m)}}\frac{\partial r_{j}^{(m)}}{\partial\xi_{2}^{(m)}}\frac{\partial r_{k}^{(n)}}{\partial\xi_{1}^{(n)}}\frac{\partial r_{l}^{(n)}}{\partial\xi_{2}^{(n)}} \\
&& \quad \times \delta^{4}\bigl({\bf r}^{(m)}(\xi_{1}^{(m)},\xi_{2}^{(m)})-{\bf r}^{(n)}(\xi_{1}^{(n)},\xi_{2}^{(n)})\bigr) \nonumber
\end{eqnarray}
We now distinguish the loop $\xi_{1}$ and radial $\xi_{2}$ parameters for each pair $n,m$ of skyrmion cones. The loop parameter is integrated over a one-dimensional manifold without a boundary (the perimeter of the 2-cone base), and the radial parameter is integrated out on a line such that ${\bf r}(\xi_{1},\xi_{2})$ is carried from the singularity to the final $S^{3}$ shell of some radius $R$ (the boundary of $B^{4}$). Then, the four-dimensional Dirac function can be broken up into the radial and $S^{3}$ factors
\begin{equation}
\delta^{4}(\Delta{\bf r})=\delta(\Delta r_{r})\delta^{3}(\Delta{\bf r}_{S^{3}})
\end{equation}
where $\Delta r_{r}$ and $\Delta{\bf r}_{S^{3}}$ are the projections of $\Delta{\bf r}={\bf r}^{(m)}-{\bf r}^{(n)}$ onto the radial direction and the $S^{3}$ manifold orthogonal to it respectively. We immediately observe that, by construction, $\Delta{\bf r}_{S^{3}}^{\phantom{x}}=\Delta{\bf r}_{S^{3}}^{\phantom{x}}(\xi_{1}^{(m)},\xi_{1}^{(n)})\propto R$ is parametrized only by the loop coordinates $\xi_{1}$ of the two skyrmions, provided that both live on the same shell, $\xi_{2}^{(m)}=\xi_{2}^{(n)}$, of radius $R$. The latter condition is enforced by the first $\delta$-function factor. Now we can integrate out one of the radial $\xi_{2}$ coordinates and rename the other one into $\eta$:
\begin{eqnarray}
I_{4} &=& (2\pi S')^{2}|\psi|^{2}\sum_{n,m}\int d\xi_{1}^{(m)}d\xi_{1}^{(n)}d\eta \\
&& \times \epsilon_{rijk}\,\frac{\partial r_{i}^{(m)}}{\partial\xi_{1}^{(m)}}\frac{\partial r_{j}^{(n)}}{\partial\xi_{1}^{(n)}}\left(\frac{\partial r_{k}^{(n)}}{\partial\eta}-\frac{\partial r_{k}^{(m)}}{\partial\eta}\right)\,\delta^{3}(\Delta{\bf r}_{S^{3}}) \ . \nonumber
\end{eqnarray}
We tacitly assumed $\partial r_{r}^{(m)}/\partial\xi_{2}^{(m)}>0$, $\partial r_{r}^{(n)}/\partial\xi_{2}^{(n)}>0$, as well as $\partial r_{r}^{(m)}/\partial\xi_{1}^{(m)}=\partial r_{r}^{(n)}/\partial\xi_{1}^{(n)}=0$, which can be always arranged by the choice of parametrization (e.g., the $S^{3}$ shells are indeed the equal-radius 3-spheres, etc.). Next, we rewrite $\delta^{3}({\bf r}-{\bf r}')$ as
\begin{equation}
\delta^{3}({\bf r}-{\bf r}')=-\frac{1}{4\pi}\nabla^{2}\frac{1}{|{\bf r}-{\bf r}'|}=\frac{1}{4\pi}\boldsymbol{\nabla}\frac{{\bf r}-{\bf r}'}{|{\bf r}-{\bf r}'|^{3}} \ .
\end{equation}
This allows us to apply Gauss' theorem to the last integral over a 3-dimensional domain. The boundary of this 3-dimensional manifold is precisely the 2-dimensional manifold of $\Delta{\bf r}_{S^{3}}\subset S^{3}$ probed by the linking number integral. We have:
\begin{eqnarray}\label{I4d}
I_{4} &=& -\frac{(2\pi S')^{2}}{4\pi}|\psi|^{2}\sum_{n,m}\!\int\!\epsilon_{ijk}\,dr_{i}^{(m)}dr_{j}^{(n)}d\eta\,\frac{\partial(r_{k}^{(m)}\!-\!r_{k}^{(n)})}{\partial\eta} \nonumber \\
&& \quad\times \partial_{p}\frac{r_{p}^{(m)}-r_{p}^{(n)}}{\left\vert {\bf r}^{(m)}-{\bf r}^{(n)}\right\vert ^{3}} \\
&=& -\frac{(2\pi S')^{2}}{4\pi}|\psi|^{2}\sum_{n,m}\int\!\epsilon_{ijk}\,dr_{i}^{(m)}dr_{j}^{(n)}\,\frac{r_{k}^{(m)}-r_{k}^{(n)}}{\left\vert {\bf r}^{(m)}-{\bf r}^{(n)}\right\vert ^{3}} \ . \nonumber
\end{eqnarray}
Note that only the upper bound of the $\eta$ integral
\begin{eqnarray}
I_{\eta} &=& \int d\eta\frac{\partial\delta r_{k}}{\partial\eta}\partial_{k}\frac{\delta r_{k}}{|\delta{\bf r}|^{3}}=\int d\delta r_{k}\frac{\partial}{\partial\delta r_{k}}\frac{\delta r_{k}}{|\delta{\bf r}|^{3}} \nonumber \\
&=& \frac{\delta r_{k}}{|\delta{\bf r}|^{3}}\Biggr\vert_{\eta}-\frac{\delta r_{k}}{|\delta{\bf r}|^{3}}\Biggr\vert_{0}
\end{eqnarray}
contributes to (\ref{I4d}), while the lower bound disappears for physical reasons. At the singularity ($\eta=0$), we have $\delta{\bf r}={\bf r}^{(m)}-{\bf r}^{(n)}\to0$, so $\delta r_{k}/|\delta{\bf r}|^{3}$ is not convergent or well-defined. However, this divergence must be regularized away upon the coarse-graining of microscopic degrees of freedom, which goes into the construction of the field theory. Placing the lower bound of $I_{\eta}$ at some small non-zero value of $\eta\to\epsilon>0$ would ensure an exact cancellation with the upper bound, i.e. $I_{4}=\int d^{2}r\,I_{\eta}\to 0$; each $\eta$ shell picks the same topologically protected Hopf index $N$. However, the lower bound is strictly at the $\eta\to0$ singularity since we formulated the original integral over the 4-dimensional volume $B^{4}$ which includes the singularity (the purpose of this was precisely to develop a prescription for integrating out a singular Hopf index density in $d=4$ dimensions). It is not possible to extract the value of the Hopf index by integrating out the skyrmion loop structure within a single point in space. The lower bound of $I_\eta$ should average out to zero because the Hopf index transforms non-trivially under spatial inversions and no information about such transformations can be packed into a point using the existing degrees of freedom.

Since (\ref{I4d}) reduces the integral (\ref{I4a}) to the Gauss' linking number for skyrmions, we can read the Hopf index $N\to N^{\textrm{s}}$ of the spin texture directly from the spinor field $\psi$. Taking (\ref{SChoice}) into account, 
\begin{equation}
I_4 \xrightarrow{\varphi\to\textrm{const.}} -(2\pi)^{2}|\psi|^{2}N^{\textrm{s}} .
\end{equation}
The same holds in the charge sector,
\begin{equation}
I_4 \xrightarrow{\theta,\phi\to\textrm{const.}} -|\psi|^{2}\!\int\limits_{S^{3}}\! d^{3}x\,\epsilon_{ijk}a_{i}\partial_{j}a_{k} = -(2\pi)^2 |\psi|^2 N^{\textrm{c}} \ .
\end{equation}
The Hopf invariants of charge (c) and spin (s) currents are represented by different gauge fields
\begin{eqnarray}\label{HopfGaugeSectors}
N^{\textrm{c}} &=& \frac{1}{(2\pi)^{2}}\int\limits _{S^{3}}d^{3}x\,\epsilon_{ijk}a_{i}\partial_{j}a_{k} \\
N^{\textrm{s}} &=& \frac{1}{(4\pi)^{2}}\int\limits _{S^{3}}d^{3}x\,\epsilon_{ijk}A_{i}\partial_{j}A_{k} \ , \nonumber
\end{eqnarray}
and define separate Hopf index current densities
\begin{eqnarray}\label{HopfCurrentSectors}
\mathcal{J}_{\mu}^{\textrm{c}} &=& \epsilon_{\mu\nu\alpha\beta\gamma}\partial_{\nu}(a_{\alpha}\partial_{\beta}a_{\gamma}) \\
\mathcal{J}_{\mu}^{\textrm{s}} &=& \epsilon_{\mu\nu\alpha\beta\gamma}\partial_{\nu}(A_{\alpha}\partial_{\beta}A_{\gamma}) \nonumber
\end{eqnarray}
normalized to the corresponding flux quanta. But, it is easy to see from (\ref{I4a}) and (\ref{CurlPsi}) that combining the charge ($\varphi$) and spin ($\phi$) vorticity embeds into the $I_4$ integral the gauge invariant combination $a_\mu + q A_\mu$ of the two gauge fields, with $q=1/2$. Hence,
\begin{equation}\label{I4e}
I_{4} = -|\psi|^{2}\int\limits _{S^{3}}d^{3}x\,\epsilon_{ijk}(a_{i}+qA_{i})\partial_{j}(a_{k}+qA_k) \ .
\end{equation}
The topological Lagrangian density (\ref{TopologicalTerm2}) can be related to (\ref{I4a}) with
\begin{eqnarray}
\int\limits_{-\infty}^{\tau_{0}}d\tau\int\limits_{B^{4}}d^{4}x\,\mathcal{L}_{t} &\propto& \int\limits _{-\infty}^{\tau_{0}}d\tau\int\limits _{B^{4}}d^{4}x\,\psi^{\dagger}\epsilon_{0ijkl}\partial_{0}\partial_{i}\partial_{j}\partial_{k}\partial_{l}\psi \nonumber \\
&=& \int\limits_{B^{4}}d^{4}x\,\psi^{\dagger}\epsilon_{ijkl}\partial_{i}\partial_{j}\partial_{k}\partial_{l}\psi = I_{4} \nonumber \ ,
\end{eqnarray}
so we can use the structure of (\ref{I4e}) and (\ref{HopfCurrentSectors}) to deduce
\begin{equation}\label{TopologicalTerm3}
\mathcal{L}_{t} = iK_{t}\,j_{\mu} \mathcal{J}_{\mu}
\end{equation}
in imaginary time, where the combined Hopf index current is
\begin{equation}\label{HopfCurrent2}
\mathcal{J}_\mu = \frac{1}{q^2}\, \epsilon_{\mu\nu\alpha\beta\gamma}\partial_{\nu}(a_{\alpha}+qA_{\alpha})\partial_{\beta}(a_{\gamma}+qA_{\gamma}) \ .
\end{equation}

The full Lagrangian density of non-relativistic particles may be
\begin{eqnarray}\label{SpinorLagrangian}
\mathcal{L}&=&\psi^{\dagger}D_{0}\psi+\frac{1}{2m}|D_{i}\psi|^{2}-\mu|\psi|^{2}+u|\psi|^{4} \\
&& + \frac{1}{12e^{2}}(\epsilon^{\mu\nu\lambda\alpha\beta}\partial_{\alpha}a_{\beta})^{2} + \mathcal{L}_{\textrm{int}} + \mathcal{L}_t \ , \nonumber
\end{eqnarray}
where $D_{\mu}=\partial_{\mu}+ia_{\mu}$. Some additional interactions $\mathcal{L}_{\textrm{int}}$, such as the spin-orbit coupling or a Kondo coupling to localized magnetic moments, might be needed to stabilize the phases with a non-trivial Hopf index dynamics. The topological Lagrangian density $\mathcal{L}_t$ is written in its purely spinor form (\ref{TopologicalTerm2}), so there is no need to include the auxiliary spin-chirality gauge field $A_\mu$ in this Lagrangian; the spinor kinematics automatically implements a topological Hall or magnetoelectric effect. This theory is a continuum-limit effective theory where the topological term arises from some microscopic Berry's phase. The earlier Lagrangian (\ref{Lagrangian4D}) is the next low-energy approximation obtained from (\ref{SpinorLagrangian}) when the spinor retains a sufficient phase stiffness, i.e. the particles are not localized. Apart from driving topological order, $\mathcal{L}_t$ also implements a Hopf variant of the topological magnetoelectric effect. Since $\mathcal{L}_t\to0$ in conventional phases, the resulting relationship $a_\mu+qA_\mu\to 0$ neutralizes any spin Hopf singularity by binding a charge Hopf singularity to it. Then, electrons moving through a spin-texture background that features interlinked skyrmions would exhibit charge current flow as if they were moving through the equivalent set of interlinked magnetic field loops.

The accurate picture of charge fractionalization needs to be derived from the spinor theory like (\ref{SpinorLagrangian}). The charge current of non-relativist particles is given by (\ref{SpinorCurrent}) with $j^0=\rho$. When we extract the U(1) gauge field from $\psi\sim e^{i\varphi}$ by a singular gauge transformation $a_{\mu}\to\partial_{\mu}\varphi$, the Lagrangian density variation with respect to $a_\mu$ gives
\begin{equation}\label{ChargeFieldEq}
j^{\mu}-\frac{1}{e^{2}}\partial_{\nu}f^{\mu\nu}=3K_{t}|\psi|^2\mathcal{J}^{\mu} \ .
\end{equation}
A finite density of Hopf singularities in the topologically ordered ground state needs to be driven by a Berry curvature field
\begin{equation}\label{TopOrderDrive}
\mathcal{B} = \langle \mathcal{J}^0 \rangle \neq 0
\end{equation}
which can be generated with a combination of the external electromagnetic field and spin texture according to (\ref{HopfCurrent2}). This field, and not the particle density, determines the coupling constant of the topological Lagrangian density (\ref{TopologicalTerm2})
\begin{equation}\label{TopCoupling}
K_{t} = \frac{1}{3\mathcal{B}}
\end{equation}
according to (\ref{ChargeFieldEq}) and $j^0=\rho=|\psi|^2$. At the same time, the Hopf index of the ground state is quantized
\begin{equation}
N=\frac{1}{(4\pi)^{2}}\int\limits_{B^{4}}d^{4}x\,\mathcal{J}^{0} \in \mathbb{Z}
\end{equation}
and captures the mutual inter-linking of both magnetic and skyrmion flux loops in three-dimensional manifolds. Since $\mathcal{B}/(4\pi)^2$ is the density of topological defects, the number of particles per topological defect is
\begin{equation}
\nu = \frac{|\psi|^2}{\mathcal{B}/(4\pi)^2} \ ,
\end{equation}
The imaginary-time topological Lagrangian density of the effective theory can now be expressed as
\begin{eqnarray}
\mathcal{L}_{t} &=& i\frac{\nu}{3(2\pi)^2} \epsilon_{\mu\alpha\beta\gamma\delta} (\partial_\mu\theta+a_{\mu}+qA_{\mu}) \\
&& \quad\times \partial_{\alpha}(a_{\beta}+qA_{\beta})\partial_{\gamma}(a_{\delta}+qA_{\delta}) \ . \nonumber
\end{eqnarray}
This is tailored to the particle density $\rho=|\psi|^2$ of a particular incompressible quantum liquid, since the density determines the filling factor $\nu$. In contrast, the master theory (\ref{SpinorLagrangian}) contains the topological term (\ref{TopologicalTerm2}) whose coupling constant (\ref{TopCoupling}) does not depend on the density.

Apart from the Hopf singularities, spinor fields in $d=4$ spatial dimensions can support monopoles and hedgehogs as topological defects. Hedgehogs in $d=4$ require a spinor representation of the Spin(4) group instead of Spin(3)=SU(2), but monopoles are derived from the same charge currents as the hopfions. The topological Lagrangian density
\begin{equation}\label{TopologicalTerm4}
\mathcal{L}_{t}=-K_{t}\,\psi^{\dagger}\epsilon_{\mu\nu\alpha\beta\gamma}\partial_{\mu}\partial_{\nu}\partial_{\alpha}\partial_{\beta}\partial_{\gamma}\psi
\end{equation}
is sensitive to all of these topological defects, and all can be handled with a complex coupling $K_t$ in imaginary time. The real part $\textrm{Re}(K_t)$ is the coupling for hopfions as discussed here; apart from generating a current $j_\mu$ factor, two partial derivatives are converted into gauge fields as $\partial_\mu \sim ia_\mu$. The imaginary part $\textrm{Im}(K_t)$ is the coupling for monopoles or hedgehogs; beside the current factor, three derivatives are converted into the antisymmetric rank-3 gauge field $\partial_\mu \partial_\nu \partial_\lambda \to i\mathcal{A}_{\mu\nu\lambda}$ whose flux over a 3-sphere is the enclosed monopole/hedgehog topological invariant in the $\pi_3(S^3)$ homotopy group. The mismatch between the collected factors of $i$ is compensated by the corresponding real and imaginary parts of $K_t$ to give the proper (real) field equations in real time. Note that the symmetry transformations and even the Hermitian property of (\ref{TopologicalTerm4}) depend on the character of extracted singularities on top of the transformations of $\psi$. It is unlikely that monopoles and hedgehogs could coexist with Hopf singularities, but the ability to construct the same topological term for both implies the possibility of driving phase transitions between different types of topological order by adjusting the particle density $|\psi|^2$. Phase transitions between incompressible quantum liquids can also be independently driven by the Hopf singularities of charge or spin currents.

Physical models which could possibly host incompressible quantum liquids of hopfions need to break the mirror inversion symmetries and remain invariant under time reversal. This is how the Hopf index, the topological Lagrangian density, and the driving agent (\ref{TopOrderDrive}) of Hopf topological orders transform,
\begin{equation}
\mathcal{B}\xrightarrow{x_{0}\to-x_{0}}\mathcal{B}\quad,\quad \mathcal{B}\xrightarrow{x_{i}\to-x_{i}}-\mathcal{B} \ .
\end{equation}
This is immediately compatible with transformations of the U(1) gauge field $a_\mu$. However, the equivalent transformations of the spin-chirality gauge field $A_\mu$
\begin{eqnarray}
A_{0}\xrightarrow{x_{0}\to-x_{0}}A_{0} &\quad,\quad& A_{0}\xrightarrow{x_{i}\to-x_{i}}A_{0} \\
A_{j}\xrightarrow{x_{0}\to-x_{0}}-A_{j} &\quad,\quad& A_{j}\xrightarrow{x_{i}\to-x_{i}}(-1)^{\delta_{ij}}A_{j} \nonumber
\end{eqnarray}
impose a constraint on the transformations of the spin vector $\hat{\bf n}$. The relationship
\begin{equation}
\epsilon_{\cdots ij}\partial_{i}A_{j}\sim\epsilon_{\cdots ij}\epsilon^{abc}\hat{n}^{a}(\partial_{i}^{\phantom{x}}\hat{n}^{b})(\partial_{j}^{\phantom{x}}\hat{n}^{c})
\end{equation}
implies
\begin{eqnarray}
\partial_{i}A_{j}\xrightarrow[i\neq j]{x_{0}\to-x_{0}}-\partial_{i}A_{j}&\quad\Rightarrow\quad& \hat{n}^{a}\xrightarrow{x_{0}\to-x_{0}}-\hat{n}^{a} \nonumber \\
\partial_{i}A_{j}\xrightarrow[i\neq j]{x_{k}\to-x_{k}} (-1)^{\delta_{ik}+\delta_{jk}}\partial_{i}A_{j} &\quad\Rightarrow\quad& \hat{n}^{a}\xrightarrow{x_{k}\to-x_{k}}+\hat{n}^{a} \nonumber
\end{eqnarray}
This is reminiscent of the pseudovector behavior in $d=3$, but the transformation under mirror inversion is trivial and insensitive to the spin direction $a$. In fact, the non-trivial $d=3$ pseudovector behavior under inversion cannot be naturally generalized to $d=4$ because the spin indices $a\in\lbrace 1,2,3\rbrace$ cannot be correlated with the spatial directions $i\in\lbrace 1,2,3,4\rbrace$.

\section{Conclusions}\label{secConclusions}

We developed a field theory description of generic systems whose states can be characterized by a Hopf index. The basic ingredient is a real or emergent vector gauge field, generally coupled to matter in $d=3$ or $d=4$ spatial dimensions. Extensions to higher dimensions are possible but not of interest in the present study. We observed that electrons can host independent Hopf invariants in their charge and spin sectors, and that the dynamics of the two is normally correlated via a type of magnetoelectric effect. Furthermore, the coupling of matter to an independent gauge field promotes by itself two distinct Hopf invariants, one per degree of freedom. The conservation of a Hopf index means that closed loops of vorticity or gauge flux can maintain their topologically non-trivial pattern of interlinks despite quantum and thermal fluctuations.

One of our main findings is that a ``pseudogap'' phase of interlinked loops can be thermodynamically distinguished from trivial disordered states in $d=3$ dimensions with a generalization of the Wilson loop correlator. This phase can exist at finite temperatures below a loop-unlinking transition or an alternative monopole-driven transition in which the loops fall apart. It need not spontaneously break a symmetry or have topological order. Instead, it features a sharp suppression of the topological index fluctuations at low frequencies,  which is expected to leave an imprint in the specific heat as well as the quantum noise of the currents coupled to the Hopf index. The latter is manifested through the chiral quantum anomaly in systems with a Dirac or Weyl electron spectrum.

In $d=4$ spatial dimensions, the Hopf index conservation can produce topological order. Even though it is irrelevant to materials, we characterized its basic properties as a part of the effort to classify topological orders. This Hopf topological order can exist only at zero temperature and has many similarities to the fractional quantum Hall effect. However, it differs from the Hall effect by the precise topological ground state degeneracy for the same filling factor. It also features a direct mechanism for angular momentum fractionalization, due to the ability of fractional charge to move along a flux loop which is interlinked with another loop. Furthermore, a fractional braiding statistics is topologically protected only between point-quasiparticles and 2-sphere excitations. Loop-loop braiding is not quantized. Interestingly, the classical Hopf index can be conserved on a topologically trivial manifold at finite temperatures, so additional ``pseudogap'' states can exist as in $d=3$ dimensions.

This work is meant only to open the problem of topological Hopf dynamics for further investigation. Many interesting questions remain. The detailed nature of the correlated phases with conserved Hopf index is not yet known. We observed that the lowest-temperature phases of this kind could perhaps protect some other knot invariants, but this requires a comprehensive further study because quantum fluctuations have a general tendency to create and annihilate small loops by tunneling. Field theory can be used to calculate the correlation functions for charge and spin currents in these phases, and also characterize the loop-unlinking phase transition. It would be also important to analyze microscopic models where the loop-unlinking transitions captured by the field theory can be explicitly demonstrated.

In terms of relating to experiments, one should look for materials which host emergent U(1) gauge fields and admit correlated low-temperature phases without obvious long-range order. Correlated superconductors such as cuprates and chiral quantum magnets such as Pr$_2$Ir$_2$O$_7$ are perhaps currently the best candidates. The Nernst effect of cuprates \cite{Ong2001, Wang2006, Li2010} hints that quantized dynamical vortex loops may exist in the pseudogap state even without an external magnetic field; such loops can inter-link, and their dynamics may be slowed-down by correlation effects which are often theoretically modeled with emergent gauge fields \cite{Sachdev2016x}. The quantum magnets of interest are U(1) spin liquid candidates, among which Pr$_2$Ir$_2$O$_7$ is an excellent example \cite{Machida2007, Machida2010, Armitage2017b}; its properties are not completely favorable, but spins on the pyrochlore lattice can generate useful effective U(1) gauge fields \cite{Hermele2004a}. If strong interactions can be induced in a Dirac or Weyl semimetal, then such a system might also be able to realize a Hopf ``pseudogap'' state and provide a window into its dynamics via the chiral anomaly.

It is known that various topological states, quantum Hall liquids in particular \cite{Thouless1982, Niu1985}, can be equivalently described in real and momentum spaces. The momentum space description is built upon a Berry curvature, while the real space description is conveniently given by a topological Lagrangian term. Both predict the same response functions. A similar relationship exists for the Hopf invariant, but should be studied further. We analyzed the Hopf index dynamics in the real space, where the relevant topological Lagrangian term for $d=3$ spatial dimensions is the $\theta$-term of the chiral quantum anomaly \cite{Goswami2023}. We also constructed a more impactful topological term for $d=4$, which leads to topological order. The equivalent momentum-space description has been useful for exploring Hopf insulators and identifying their protected surface states \cite{Moore2008, Duan2013, Duan2014, Duan2016, Hasan2017, Volovik2017, Schuster2019, Slager2019, Soluyanov2021, Trifunovic2021, Piechon2022, Goswami2023}. Presumably, some phases discussed here are strongly correlated generalizations of Hopf band insulators. This suggests another approach to the possible realizations of the correlated Hopf quantum liquids -- by opening an interaction gap inside a Hopf insulator's band, and relating the Hopf quantum liquids to symmetry-protected topological (SPT) phases.

There are also some bigger theoretical questions. An emerging picture from this study is that topological orders are members of a broader family of correlated phases which are unified by instanton confinement, i.e. the conservation of a classical topological index. We developed a basic renormalization group assessment of instanton confinement in Ref.\cite{Nikolic2023a}. However, it is not clear yet if further universal characterization is possible, apart from the basic notion that delocalized topological defects play the leading role in the dynamics. Recent studies of superconductors \cite{Nikolic2011a, Nikolic2014, Nikolic2014a} and magnets \cite{Nikolic2019b} with strong spin-orbit coupling have found various mean-field phases with ordered arrays of topological defects and anti-defects (vortices and hedgehogs, respectively). Quantum and thermal fluctuations can in principle melt such lattices of topological defects, while the spin-orbit coupling protects against a massive defect annihilation. The spin-orbit coupling also provides an independent ``magnetic length'' scale, which can define the extent of short-range coherence for a quantum liquid with delocalized topological defects. Hence, this is a possible route for obtaining unconventional ``pseudogap'' or SPT phases. Instanton confinement here indicates that the total zero topological charge is conserved. There is evidence \cite{Nikolic2022b} that such phases could possess a topologically non-trivial excitation spectrum, featuring nodes in the bulk and gapless states on the system boundary.

The relationship between the Hopf index and chiral anomaly is also fundamentally interesting. The QED ground state on $\mathbb{R}^{4}$ is in the confined-instanton phase with a conserved Hopf index, while no topological order exists on non-simply connected manifolds. This is a hallmark of Abelian gauge fields; the action cost $\sim\log(R)$ of an uncompensated instanton which alters the Hopf index scales at least as a logarithm of the system size $R$. In contrast, non-Abelian gauge fields allow instantons which evolve into a pure-gauge configuration at infinite distances, and hence enable a finite action cost \cite{Belavin1975, tHooft1976, Shuryak1998, tHooft2000, Oglivie2012}. This leads to quark confinement, but removes the confined-instanton phase. If the spatial manifold is topologically non-trivial, even the Abelian gauge fields are not able to hold on to their Hopf index due to vacuum instantons. This translates into a non-degenerate ground state and the absence of hopfion topological orders in $d=3$. Nevertheless, the hopfion dynamics of Abelian gauge fields is not-trivial. The low-temperature confined-instanton phase is a subtle replacement for topological order, where the particle-flux attachment is replaced with the attachment of certain particle creation/annihilation events to the equivalent topological defect events, i.e. instantons. The latter is a quantum anomaly. We have envisioned a scenario in which the anomaly becomes fractionalized just like its topological order counterpart, and this idea can be explored further.

\section{Acknowledgements}\label{secAck}

This research was partly supported by the Department of Energy, Basic Energy Sciences, Materials Sciences and Engineering Award DE-SC0019331. Support was also provided by the Quantum Science and Engineering Center at George Mason University.

\appendix

\section{Topological current conservation from the topological Lagrangian term}\label{app1}

The Lagrangian density with a topological term couples the charge current $j_\mu \sim \partial_\mu \theta + a_\mu$ to the current $\mathcal{J}_\mu$ of topological defects. Here we show that abundant phase $\theta$ fluctuations lead to the conservation $\partial_\mu \mathcal{J}_\mu=0$ of the topological current. We will work in the continuum limit. The alternative would be to construct a compact lattice theory with $\theta\in(0,2\pi)$ and integer-valued topological current $\mathcal{J}_\mu$, but this would be plagued by an inconsistency of the topological coupling $\mathcal{L}_t \sim i j_\mu \mathcal{J}_\mu$ on the lattice. Since the charge current $j_\mu$ lives on the physical lattice and $\mathcal{J}_\mu$ lives on the dual lattice, a topological coupling would have to explicitly break some lattice symmetry or have a fairly complicated form. 

The insight from the continuum limit is still valuable even though we lose access to Mott insulator phases. Integrating out $\theta$ in the imaginary-time path integral yields
\begin{eqnarray}
&& \int \mathcal{D}\theta\,e^{-\int d^{d}x\left(\frac{\kappa_{c}}{2}j_{\mu}j_{\mu}+iK_{d}j_{\mu}\mathcal{J}_{\mu}\right)} \\
&& \qquad \propto \int \mathcal{D}\theta\,e^{-\int d^{d}x\left\lbrack\frac{\kappa_{c}}{2}(\partial_{\mu}\theta)^{2}+\kappa_{c}(\partial_{\mu}\theta)a_{\mu}+iK_{d}(\partial_{\mu}\theta)\mathcal{J}_{\mu}\right\rbrack} \nonumber \\
&& \qquad \to \int \mathcal{D}\theta\,e^{\int d^{d}x\,\left\lbrack \frac{\kappa_{c}}{2}\theta\partial_{\mu}\partial_{\mu}\theta-i\,\theta\,K_{d}\partial_{\mu}\mathcal{J}_{\mu}\right\rbrack}  \nonumber \\
&& \qquad = \int \mathcal{D}\theta\,\exp\biggl\lbrace \int d^{d}x \biggl\lbrack \frac{1}{2}\Bigl(\theta-i(K_{d}\partial_{\mu}\mathcal{J}_{\mu})(\kappa_{c}\partial_{\mu}\partial_{\mu})^{-1}\Bigr) \nonumber \\
&& \qquad\qquad\qquad \times (\kappa_{c}\partial_{\mu}\partial_{\mu})\Bigl(\theta-i(\kappa_{c}\partial_{\mu}\partial_{\mu})^{-1}(K_{d}\partial_{\mu}\mathcal{J}_{\mu})\Bigr) \nonumber \\
&& \qquad\qquad +(K_{d}\partial_{\mu}\mathcal{J}_{\mu})(\kappa_{c}\partial_{\mu}\partial_{\mu})^{-1}(K_{d}\partial_{\mu}\mathcal{J}_{\mu}) \biggr\rbrack \biggr\rbrace \nonumber \\
&& \qquad \propto \exp \left\lbrack \int d^{d}x\,(K_{d}\partial_{\mu}\mathcal{J}_{\mu})(\kappa_{c}\partial_{\mu}\partial_{\mu})^{-1}(K_{d}\partial_{\mu}\mathcal{J}_{\mu}) \right\rbrack \nonumber \ .
\end{eqnarray}
We gauged-away $\partial_{\mu}a_{\mu}\to0$. The indices in different brackets are independent even if the same symbol is used, and $\theta$ was integrated out naively on the $(-\infty,\infty)$ interval as appropriate for large $\kappa_{c}$. We see that the current non-conserving $\partial_{\mu}\mathcal{J}_{\mu} \neq 0$ fluctuations would be strongly suppressed in the resulting action if $\kappa_c$ were small (the operator $\partial_\mu\partial_\mu \to -k^2$ has only negative eigenvalues).

The last effective action has a more transparent form in momentum space,
\begin{equation}
 S_{\textrm{eff}} = \int\frac{d^d k}{(2\pi)^d}\, \frac{|\Delta({\bf k})|^2}{k^2}  \ ,
\end{equation}
where $\Delta({\bf k})$ is the Fourier transform of $K_d\partial_\mu\mathcal{J}_\mu$. This translates into an interaction potential $V(\delta r)$
\begin{equation}
 S_{\textrm{eff}} = \int d^dr\,d^dr'\, \Delta({\bf r})\Delta({\bf r}')\,V(|{\bf r}-{\bf r}'|)
\end{equation}
which naively behaves as $V(\delta r) \sim 1/\delta r^{d-2}$ by dimensional analysis. Its ultra-violet divergence in the $\delta r\to 0$ limit is the agent that suppresses the local fluctuations of $\Delta \propto \partial_\mu\mathcal{J}_\mu$. In practice, the amount of suppression is limited by an ultra-violet cut-off length $\delta r \to a$. The continuum limit regime holds as long as we are justified approximating $a\to 0$. Incompressible quantum liquids are still consistently captured in the continuum limit, requiring the topological current conservation despite the loss of long-range phase coherence (the stability of such phases is confirmed by the renormalization group of instanton confinement). If the extremely abundant fluctuations push the system into a Mott insulator phase, then we are no longer justified in regarding the cut-off length $a$ small; constructing a compact lattice theory with a lattice constant $a$ is in order, and the topological currents gain a new freedom to fluctuate without being conserved. This amounts to a dual picture featuring the condensation of topological defects.

\section{Topological invariant in the $\pi_{2n-1}(S^n)$ homotopy group}\label{app2}

A generalization of the Hopf index to the $\pi_{2n-1}(S^{n})$ homotopy groups involves a Spin($n+1$) unit-magnitude vector field $\hat{{\bf n}}({\bf x})=(\hat{n}^{a_{1}},\cdots,\hat{n}^{a_{n+1}})$ which lives in a $d=2n-1$ dimensional sphere manifold ${\bf x}\in S^{2n-1}$. The flux of the vector field chirality on any $n$-sphere $S^n$ embedded in $S^{2n-1}$ is a quantized topological invariant of the $\pi_n(S^n)$ homotopy group:
\begin{equation}
\frac{1}{n!\,S_{n}}\oint\limits_{S^{n}}d^{n}x\,\epsilon_{j_{1}\cdots j_{n}}\epsilon^{a_{0}\cdots a_{n}}\hat{n}^{a_{0}}(\partial_{j_{1}}\hat{n}^{a_{1}})\cdots(\partial_{j_{n}}\hat{n}^{a_{n}})\in\mathbb{Z} \ .
\end{equation}
$S_n$ is the ``area'' of the unit $n$-sphere. In our notation, if the Levi-Civita tensor has fewer indices than $d$ as in the last formula, it is understood that all directions spanned by the indices live on the integration manifold. The standard mathematical notation utilizes the wedge product:
\begin{equation}
\frac{1}{n!\,S_{n}}\oint\limits_{S^{n}}\epsilon^{a_{0}\cdots a_{n}}\hat{n}^{a_{0}}\,(d\hat{n}^{a_{1}})\wedge\cdots\wedge(d\hat{n}^{a_{n}})\in\mathbb{Z}
\end{equation}
where $d\hat{n}^{a}=dx^{i}\partial_{i}\hat{n}^{a}$, but we will emphasize all indices here because they are helpful in the bookkeeping of the multiplicity factors like $n!$. Let us define the chirality tensor
\begin{eqnarray}
\chi_{i_{1}\cdots i_{n-1}} &=& \frac{1}{n!}\epsilon_{i_{1}\cdots i_{n-1}j_{1}\cdots j_{n}}\epsilon^{a_{0}\cdots a_{n}} \\
&& \qquad\times \hat{n}^{a_{0}}(\partial_{j_{1}}\hat{n}^{a_{1}})\cdots(\partial_{j_{n}}\hat{n}^{a_{n}}) \ . \nonumber
\end{eqnarray}
Since $|\hat{{\bf n}}|=1$ and $\hat{\bf n}\in S^n$, all linearly independent directions that $\hat{\bf n}$ and its derivatives can span are exhausted in the chirality expression. Hence,
\begin{equation}
\partial_{i_{1}}\chi_{i_{1}\cdots i_{n-1}}=\cdots=\partial_{i_{n-1}}\chi_{i_{1}\cdots i_{n-1}}=0 \ .
\end{equation}
This can be generally solved by introducing a smooth antisymmetric tensor gauge field $A_{i_1\cdots i_{n-1}}$ of rank $n-1$ in our $d=2n-1$ space:
\begin{equation}\label{appMagnFlux}
\chi_{i_{1}\cdots i_{n-1}}=\epsilon_{i_{1}\cdots i_{n-1}jk_{1}\cdots k_{n-1}}\partial_{j}A_{k_{1}\cdots k_{n-1}} \ .
\end{equation}
Now, we set up a system of $n$-dimensional open manifolds $D_{p}$ whose boundaries are $n-1$ dimensional interlinked ``flux loop'' manifolds $C_{p}$. The cross-sections of $C_{p}$ ``tubes'' are also $n$-dimensional; this is intentional, and in fact fixes the space dimensionality to $d=2n-1$. The flux carried by each ``loop'' is a quantized ``current'' computed through the $S^{n}$ cross-section of the ``loop'',
\begin{equation}\label{appLoopCurrent}
I_{p}=n_{p}I_{0}=\frac{1}{(n-1)!}\oint\limits_{S^n} d^{n}x\,\hat{\eta}_{i_{1}\cdots i_{n-1}}\chi_{i_{1}\cdots i_{n-1}} \ ,
\end{equation}
where
\begin{equation}\label{appI0a}
I_{0}=S_{n}\quad,\quad n_{p}\in\mathbb{Z} \ .
\end{equation}
The last integral employs another convenient custom notation: $\hat{\eta}_{i_{1}\cdots i_{n-1}}({\bf x})$ is defined on the integration manifold ${\bf x}\in S^n$ as the unit antisymmetric tensor which is locally orthogonal to $S^{n}$, pointing ``outwards''. The integral (\ref{appLoopCurrent}) is then the chirality ``flux'' through the ``surface'' $S^n$, a multi-dimensional generalization of the electric flux $\oint d{\bf A}\, {\bf E} = \oint d^2x\, \hat{\eta}_i E_i$ from the $d=3$ Gauss' law of electrodynamics. The total ``current'' passing through the ``loop'' $C_{p}$ is:
\begin{eqnarray}
\sum_{q(C_{p})}I_{q} &=& \sum_{q(C_{p})}I_{0}n_{q} = \frac{1}{(n-1)!}\int\limits_{D_{p}}d^{n}x\,\hat{\eta}_{i_{1}\cdots i_{n-1}}\chi_{i_{1}\cdots i_{n-1}} \nonumber \\
&=& \oint\limits_{C_{p}}d^{n-1}x\,\epsilon_{k_{1}\cdots k_{n-1}}A_{k_{1}\cdots k_{n-1}} \ .
\end{eqnarray}
We used Stokes-Cartan theorem in the last step. Next, we multiply this by $I_{0}n_{p}$ and sum over $p$ (all ``loops''), noting that the cross-section $S^n(p)$ of a ``loop'' is orthogonal to its boundary $C_p$,
\begin{eqnarray}
&& \sum_{p,q}I_{0}^{2}n_{p}^{\phantom{x}}n_{q}^{\phantom{x}}
  = \frac{1}{(n-1)!}\sum_{p}\int\limits_{S^n_{\phantom{x}}(p)}\!\! d^{n}x\,(\hat{\eta}_{i_{1}\cdots i_{n-1}}\chi_{i_{1}\cdots i_{n-1}}) \nonumber \\
&& \qquad\qquad\qquad \times \oint\limits_{C_{p}}d^{n-1}x\,\epsilon_{k_{1}\cdots k_{n-1}}A_{k_{1}\cdots k_{n-1}} \nonumber \\
&& \qquad = \int\! d^{2n-1}x\,\epsilon_{i_{1}\cdots i_{n-1}jk_{1}\cdots k_{n-1}}A_{i_{1}\cdots i_{n-1}}\partial_{j}A_{k_{1}\cdots k_{n-1}} \ . \nonumber
\end{eqnarray}
Recalling $n_p,n_q\in\mathbb{Z}$ and (\ref{appI0a}), the last equation reveals the integer quantization $N\in\mathbb{Z}$ of the Hopf index
\begin{equation}\label{appHopf}
N = \frac{1}{I_{0}^2} \!\oint\limits_{S^{2n-1}}\!\!\!\! d^{2n-1}x\,\epsilon_{i_{1}\cdots i_{n-1}jk_{1}\cdots k_{n-1}}A_{i_{1}\cdots i_{n-1}}\partial_{j}A_{k_{1}\cdots k_{n-1}}
\end{equation}
This derivation parallels the one from Section \ref{secHopf}. It applies to the spin sector of the Spin($n+1$) group in $d=2n-1$ spatial dimensions, where the chirality gauge field is generally a higher-rank tensor derived from the vector field $\hat{\bf n}$ configuration
\begin{eqnarray}
&& \epsilon_{i_{1}\cdots i_{n-1}jk_{1}\cdots k_{n-1}}\partial_{j}A_{k_{1}\cdots k_{n-1}} = \\
&& \qquad = \frac{1}{n!}\epsilon_{i_{1}\cdots i_{n-1}j_{1}\cdots j_{n}}\epsilon^{a_{0}\cdots a_{n}} \hat{n}^{a_{0}}(\partial_{j_{1}}\hat{n}^{a_{1}})\cdots(\partial_{j_{n}}\hat{n}^{a_{n}}) \ , \nonumber
\end{eqnarray}
and $I_0 = S_n$.

The expression (\ref{appHopf}) reveals a constraint on $n$ which was not apparent before: if $n$ is odd, then an integration by parts yields $N=-N$, i.e. $N=0$. Non-trivial integer values of the Hopf index are obtained only for even $n$. Indeed, the homotopy groups $\pi_{2n-1}(S^n)$ have an integer-valued topological invariant, the Hopf index, only for even $n$, with an exception of $\pi_1(S^1)$ whose integer invariant is not the Hopf index; see Table \ref{appHomotopy}. Any additional structure of the homotopy invariants is not contained in (\ref{appHopf}).

\begin{table}
\centering
\begin{tabular}{|c|c|c|}
\hline
\; $k$ \quad & \quad $n=2k-1$ \quad & \quad $n=2k$ \quad \\
\hline
\; $1$ \quad & \quad $\pi_1(S^1) = \mathbb{Z}$ \quad & \quad $\pi_3(S^2) = \mathbb{Z}$ \quad \\
\; $2$ \quad & \quad $\pi_5(S^3) = \mathbb{Z}_2$ \quad & \quad $\pi_7(S^4) = \mathbb{Z} \times \mathbb{Z}_{12}$ \quad \\
\; $3$ \quad & \quad $\pi_9(S^5) = \mathbb{Z}_2$ \quad & \quad $\pi_{11}(S^6) = \mathbb{Z}$ \quad \\
\; $4$ \quad & \quad $\pi_{13}(S^7) = \mathbb{Z}_2$ \quad & \quad $\pi_{15}(S^8) = \mathbb{Z} \times \mathbb{Z}_{120}$ \quad \\
\hline
\end{tabular}
\caption{\label{appHomotopy}Homotopy groups of the $\pi_{2n-1}(S^n)$ kind.}
\end{table}

Based on this, we anticipate the existence of topological orders in $d=4k$ ($k\in\mathbb{N}$) spatial dimensions, supported by currents that transform according to a representation of the Spin($2k+1$) group. We will argue in the next Appendix that charge currents and other types of currents may also be able to support these generalized Hopf topological orders. Topological defects characterized by the $\pi_{4k-1}(S^{2k})$ homotopy invariant bind a fractional amount of charge to form elementary quasiparticles.

\section{Hopf index of the U(1) gauge field and the axial vector anomaly in higher dimensions}\label{app3}

The general Hopf index (\ref{appHopf}) is applicable to the charge sector as well. To begin with, the tensor gauge field $A_{i_1\cdots i_{n-1}}$ can directly describe the special monopoles of a U(1) gauge field $A_i$ whose quantized topological charge $I_0=2\pi$ obtains as the ``magnetic'' (\ref{appMagnFlux}) flux through an $n$-sphere submanifold of the $S^{2n-1}$ space,
\begin{equation}\label{appFlux}
\Phi = \frac{1}{(n-1)!}\oint\limits_{S^n} d^{n}x\,\hat{\eta}_{i_{1}\cdots i_{n-1}}\chi_{i_{1}\cdots i_{n-1}} = I_0 \times \textrm{integer} \ .
\end{equation}
These special monopoles are singular on $(d-1)-(n-1)=n-1$ submanifolds instead of being point singularities in the $d=2n-1$ dimensional space. This is determined by the rank of the gauge field. Nevertheless, the flux quantization is derived from the lowest rank of $A_i$ through a hierarchical construction \cite{Nikolic2019}. Other kinds of monopole structures, if featured in the dynamics, may also be able to carry a Hopf index.

Of particular interest are the elementary monopole structures, i.e. ``vortices''. These are $d-2$ dimensional singular domains in the $d$ dimensional space. Their topological charge is quantized in the units of $2\pi$, and obtains from the curl of the U(1) gauge field $A_i$ in the $2$-plane manifold orthogonal to the singularity. If no other monopole structures are present, then we can construct the tensor $A_{i_1\cdots i_{n-1}}$ only from the curls of $A_\mu$,
\begin{eqnarray}\label{appDecomp}
A_{i_{1}\cdots i_{n-1}} &=& \frac{1}{n!}\epsilon_{i_{1}\cdots i_{n-1}j_{1}\cdots j_{n}}\epsilon_{k_{1}\cdots k_{n-1}j_{1}\cdots j_{n}} \\
&& \times A_{k_{1}}(\partial_{k_{2}}A_{k_{3}})\cdots(\partial_{k_{n-2}}A_{k_{n-1}}) \ . \nonumber
\end{eqnarray}
Note that we are specializing to the cases of even $n$, i.e. odd $n-1$, which provide a non-trivial Hopf index. Using (\ref{appMagnFlux}), we find the specialized form of the flux (\ref{appFlux})
\begin{eqnarray}\label{appFlux2}
&& \Phi = \frac{1}{2^{n/2}}\int d^{n}x\,\hat{\eta}_{i_{1}\cdots i_{n-1}}\epsilon_{i_{1}\cdots i_{n-1}k_{1}\cdots k_{n}} \nonumber \\
&& \qquad\qquad \times F_{k_{1}k_{2}}F_{k_{3}k_{4}}\cdots F_{k_{n-1}k_{n}} \ ,
\end{eqnarray}
where
\begin{equation}
F_{ij} = \partial_i A_j - \partial_j A_i \ .
\end{equation}
The flux quantum $I_0$ is different than before and we must compute it from scratch. Consider a magnetic field with singular structures of quantized flux. These must be $d-2$ dimensional ``vortex'' manifolds $V_{p}$ of singular points in a $d$-dimensional world: an intersection between such a manifold and an orthogonal $2$-plane is a singular point, so that the line integral of $A_{i}$ on the closed loop on the $2$-plane around the singularity yields $2\pi$. If the $2$-plane is spanned by the coordinates $x_{i},x_{j}$ and the orthogonal singular $d-2$ manifold by $x_{k_{1}},\dots,x_{k_{d-2}}$, then 
\begin{equation}
F_{ij}({\bf x})=2\pi\sum_{p}\int\limits _{V_{p}}d^{d-2}y\,\epsilon_{ijk_{1}\cdots k_{d-2}}\delta^{d}({\bf x}-{\bf y}) \ .
\end{equation}
Substituting into (\ref{appFlux2}) tells us that only the $d-n=n-1$ dimensional overlaps of $n/2$ ``vortices'' matter; each overlap manifold is orthogonal to all $k_{1},\dots,k_{n}$, which means it is orthogonal to the integration manifold of ${\bf x}$. Therefore, the manifold of ${\bf x}$ intersects each $V_{p}$ overlap manifold at a single point. The product of all Dirac $\delta$-functions integrates out to $1$, and we have 
\begin{eqnarray}
&& \Phi = \frac{(2\pi)^{k}}{2^{k}}\sum_{p_{1}\cdots p_{k}} \; \int\limits_{V_{p_{1}}}\!d^{d-2}y_{1}\cdots \int\limits_{V_{p_{k}}}\!\! d^{d-2}y_{k}\int d^{n}x \nonumber \\
&& \quad \times \hat{\eta}_{i_{1}\cdots i_{n-1}}\epsilon_{i_{1}\cdots i_{n-1}k_{1}\cdots k_{n}} \epsilon_{k_{1}k_{2}l_{1}^{1}\cdots l_{d-2}^{1}}\cdots\epsilon_{k_{n-1}k_{n}l_{1}^{k}\cdots l_{d-2}^{k}} \nonumber \\
&& \quad \times\delta^{d}({\bf x}-{\bf y}_{1})\cdots\delta^{d}({\bf x}-{\bf y}_{k}) \nonumber
\\
&& \quad = (2\pi)^{n/2}(n-1)!\times\textrm{integer}
\end{eqnarray}
Note $n=2k$. The flux quantum is
\begin{equation}
I_0 = (2\pi)^{n/2}(n-1)!
\end{equation}
because the Levi-Civita tensors in the last integral make integer multiples of $2^{n/2}(n-1)!$, due to the $n/2$ pair-wise $k_{i}$ permutations and the permutations of $i_{1},\dots,i_{n-1}$.
The resulting Hopf index (\ref{appHopf}) is integer-valued:
\begin{eqnarray}\label{appHopf2}
N' &=& \left\lbrack \frac{(n-1)!}{I_{0}}\right\rbrack^{2} \!\!\oint\limits_{S^{2n-1}}\!\! d^{2n-1}x\,\epsilon_{i_{1}\cdots i_{2n-1}} \\
&& \qquad \times A_{i_{1}}(\partial_{i_{2}}A_{i_{3}})\cdots(\partial_{i_{2n-2}}A_{i_{2n-1}}) \nonumber \\
&=& \frac{2}{(4\pi)^{n}} \!\!\oint\limits_{S^{2n-1}}\!\! d^{2n-1}x\,\epsilon_{i_{1}\cdots i_{2n-1}}A_{i_{1}}F_{i_{2}i_{3}}\cdots F_{i_{2n-2}i_{2n-1}} \nonumber \ .
\end{eqnarray}

Unconstrained tensor gauge fields $A_{i_1\cdots i_{n-1}}({\bf x})$ split into equivalence classes which are enumerated by $N\in \mathbb{Z}$ in (\ref{appHopf}). However, the decomposition (\ref{appDecomp}) is a constraint on the tensor gauge field. One may expect that the integers $N'$ of (\ref{appHopf2}) enumerate the constrained equivalence classes by having a quantum $q$ larger than $1$, i.e. $N'/q\in \mathbb{Z}$. Let us examine this by scrutinizing the symmetries of $N$ and $N'$. Since $A_{i_{1}\cdots i_{n-1}}$ is an antisymmetric tensor, we can always reorder its indices as $i_{1}<i_{2}<\cdots<i_{n-2}<i_{n-1}$. Any resulting sign change is absorbed by the equivalent reordering of the indices in the Levi-Civita tensor. If we denote the ordered set of indices with angle brackets, $\langle i_{1}\cdots i_{n-1}\rangle$, and generate only non-equivalent sets of indices under ordering, then (\ref{appHopf}) becomes:
\begin{eqnarray}
N &=& \left\lbrack \frac{(n-1)!}{I_{0}}\right\rbrack^{2} \!\!\oint\limits_{S^{2n-1}}\!\! d^{2n-1}x\,\epsilon_{\langle i_{1}\cdots i_{n-1}\rangle j\langle k_{1}\cdots k_{n-1}\rangle} \nonumber \\
&& \qquad\qquad \times A_{\langle i_{1}\cdots i_{n-1}\rangle}\partial_{j}A_{\langle k_{1}\cdots k_{n-1}\rangle} \ .
\end{eqnarray}
Furthermore, we can exchange the $i$ and $k$ index sets; an integration by parts assures that the value of $N$ does not change. For a canonical ordering, we can always associate the smallest of the $i_{1},k_{1}$ indices to the gauge field factor without the derivative:
\begin{eqnarray}
N &=& 2\times\left\lbrack\frac{(n-1)!}{I_{0}}\right\rbrack^{2} \!\!\oint\limits_{S^{2n-1}}\!\! d^{2n-1} x\,\epsilon_{\langle\langle k_{1}\cdots k_{n-1}\rangle j\langle i_{1}\cdots i_{n-1}\rangle\rangle} \nonumber \\
&& \qquad\qquad \times A_{\langle i_{1}\cdots i_{n-1}\rangle}\partial_{j}A_{\langle k_{1}\cdots k_{n-1}\rangle} \equiv 2Q \ .
\end{eqnarray}
There are no further symmetries to exploit, so the quantity $Q$ represents the irreducible (essential) information carried by the gauge field which determines its topological equivalence class. The decomposition (\ref{appDecomp}) has more symmetry. Looking at the left-hand side of (\ref{appHopf2}), we see that we can freely permute $n-1$ index pairs of the $\partial_i A_j$ factors. Furthermore, we can carry out $n$ different integrations by parts to apply a derivative on the leading $A_{i_1}$ factor. Together, these symmetry transformations are equivalent to all $n!$ permutations among the indices of the gauge field factors. Hence, we have:
\begin{eqnarray}
N' &=& n!\times\left\lbrack\frac{(n-1)!}{I_{0}}\right\rbrack^{2} \!\!\oint\limits_{S^{2n-1}}\!\! d^{2n-1}x\, \epsilon_{\langle j(i_{2}i_{3})\cdots(i_{2n-2}i_{2n-1})\rangle} \nonumber \\
&& \quad \times A_{j}(\partial_{i_{2}}A_{i_{3}})\cdots(\partial_{i_{2n-2}}A_{i_{2n-1}})\equiv n!\,Q' \ .
\end{eqnarray}
This exhausts the symmetry of $N'$. The quantity $Q'$ contains irreducible information about the constrained field (\ref{appDecomp}) which determines its equivalence class. Now, both $Q$ and $Q'$ are quantized in some units; the most general Hopf index $N\in\mathbb{Z}$ requires $Q$ to be quantized in the units of $1/2$. However, $Q'$ cannot be equivalently quantized in the units of $1/n!$. If that quantization were possible, then every configuration of $A_{i_{1}\cdots i_{n-1}}$ could be smoothly deformed into a form given by (\ref{appDecomp}) because the two forms would have the same topological index. If we generate all possible $A_{i_{1}\cdots i_{n-1}}$ configurations, we will also cover all possible $A_{i}$ configurations by generating them from the decompositions (\ref{appDecomp}) whenever possible. The period of finding these decompositions relative to all possible $A_{i_{1}\cdots i_{n-1}}$ configurations matches the relative quantizations of $N'$ and $N$. This period is also reflected in the relationships $N=2Q$ and $N'=n!Q'$. Generating the ``essential'' information $Q,Q'$ at a fixed frequency gives us the frequency of symmetry-restricted values $N,N'$ in the respective symmetry classes. So, taking the same quantization (frequency) for $Q$ and $Q'$ yields:
\begin{equation}
Q=\frac{1}{2}q\quad,\quad Q'=\frac{1}{2}q \quad\Rightarrow\quad N=q\quad,\quad N'=\frac{n!}{2}q \nonumber
\end{equation}
with $q\in\mathbb{Z}$. Consequently, we can extract an integer factor $N''\equiv q = (2/n!)N' \in \mathbb{Z}$ from (\ref{appHopf2}) which depends on the field configuration:
\begin{equation}
N'' = \frac{4}{n!\,(4\pi)^{n}} \!\!\!\oint\limits_{S^{2n-1}}\!\!\!\! d^{2n-1}x\, \epsilon_{i_{1}\cdots i_{2n-1}}A_{i_{1}}F_{i_{2}i_{3}}\cdots F_{i_{2n-2}i_{2n-1}}
\end{equation}
This is the irreducible part of the constrained Hopf index. In the light of the quantum anomaly discussion from Section \ref{secHopfCorr}, one anticipates the non-conservation of chiral currents in the form of
\begin{equation}
\partial_{\mu}j_{c\mu}=\frac{2}{n!\,(4\pi)^{n}}\,\epsilon_{\mu_{1}\cdots\mu_{2n}}F_{\mu_{1}\mu_{2}}F_{\mu_{3}\mu_{4}}\cdots F_{\mu_{2n-1}\mu_{2n}} \ ,
\end{equation}
since the $\Gamma_{c}=i^{n-1}\Gamma_{0}\Gamma_{1}\cdots\Gamma_{d}$ matrix in $j_{c\mu} = \bar{\psi}\Gamma_{\textrm{c}}\Gamma_{\mu}\psi$ is constructed to have $\pm 1$ eigenvalues (i.e. we match integer changes of the chiral charge to the integer changes of the Hopf index). This is, in fact, obtained rigorously as the axial vector anomaly of the quantum electrodynamics generalized to $d$ dimensions \cite{Peskin1995}.

\bigskip\bigskip
\raggedbottom

\end{document}